\documentclass[11pt,a4paper]{article} 
\pdfoutput=1
\usepackage{jheppub}
\newcommand{\half}{\frac{1}{2}}

\newcommand\beq{\begin{equation}}
\newcommand\eeq{\end{equation}}
\newcommand\bea{\begin{eqnarray}}
\newcommand\eea{\end{eqnarray}}
\newcommand\bi{\begin{itemize}}
\newcommand\ei{\end{itemize}}
\newcommand\ben{\begin{enumerate}}
\newcommand\een{\end{enumerate}}
\newcommand{\stau}{\tilde{\tau}_{1}}
\newcommand{\chiz}{\tilde{\chi}_{1}^{0}}

\newcommand{\LP}{\Lambda_{+}}
\newcommand{\LM}{\Lambda_{-}}
\newcommand{\SP}{\Sigma_{+}}
\newcommand{\SM}{\Sigma_{-}}
\newcommand{\cw}{\cos\theta_{\rm W}}
\newcommand{\sw}{\sin\theta_{\rm W}}
\newcommand{\cb}{\cos\beta}
\newcommand{\mw}{M_{\rm W}}
\newcommand{\cwsq}{\cos^{2}\theta_{\rm W}}
\newcommand{\swsq}{\sin^{2}\theta_{\rm W}}
\newcommand{\cbsq}{\cos^{2}\beta}
\newcommand{\mwsq}{M^{2}_{\rm W}}
\newcommand{\cwfr}{\cos^{4}\theta_{\rm W}}
\newcommand{\swfr}{\sin^{4}\theta_{\rm W}}
\newcommand{\cbfr}{\cos^{4}\beta}
\newcommand{\mwfr}{M^{4}_{\rm W}}

\newcommand{\smu}{\tilde{\mu}}
\newcommand{\staur}{\tilde{\tau}}
\newcommand{\drr}{\delta_{RR}}
\newcommand{\Drr}{\Delta_{RR}}
\newcommand{\msmu}{m^{2}_{\tilde{\mu}_{R}}}
\newcommand{\mstau}{m^{2}_{\tilde{\tau}_{R}}}

\def\dfrac#1#2{{\displaystyle\frac{#1}{#2}}}
\newif\ifboo \boofalse

\def\lsim{\mathrel{\rlap{\lower4pt\hbox{\hskip1pt$\sim$}}
    \raise1pt\hbox{$<$}}}         
\def\gsim{\mathrel{\rlap{\lower4pt\hbox{\hskip1pt$\sim$}}
    \raise1pt\hbox{$>$}}}         

\newcommand{\lstau}{\tilde{l}_{1}}

\newcommand{\csttgl}{\chiz\;\lstau\rightarrow\gamma\;\tau}
\newcommand{\csttgm}{\chiz\;\lstau\rightarrow\gamma\;\mu}
\newcommand{\ststttt}{\lstau\;\lstau\rightarrow\tau\;\tau}
\newcommand{\ststttm}{\lstau\;\lstau\rightarrow\tau\;\mu}
\newcommand{\ststtmm}{\lstau\;\lstau\rightarrow\mu\;\mu}
\newcommand{\ststbtgg}{\lstau\;\lstau^{*}\rightarrow\gamma\;\gamma}
\newcommand{\csttzl}{\chiz\;\lstau\rightarrow Z\;\tau}
\newcommand{\csttzm}{\chiz\;\lstau\rightarrow Z\;\mu}
\newcommand{\ststbtzg}{\lstau\;\lstau^{*}\rightarrow Z\;\gamma}
\newcommand{\cctttb}{\chiz\;\chiz\rightarrow\tau\;\bar{\tau}}
\newcommand{\ccttmb}{\chiz\;\chiz\rightarrow\tau\;\bar{\mu}}
\newcommand{\cctmmb}{\chiz\;\chiz\rightarrow\mu\;\bar{\mu}}
\newcommand{\sigmav}{\langle \sigma v \rangle}

\title{Flavored Co-annihilations}
\author[a]{Debtosh Chowdhury,}
\author[b]{Raghuveer Garani }
\author[a]{and Sudhir K. Vempati}

\affiliation[a]{Centre for High Energy Physics, Indian Institute of Science, Bangalore 560 012, India}
\affiliation[b]{Department of Physics, University of Cologne, 50923 Cologne, Germany}
\emailAdd{debtosh@cts.iisc.ernet.in}
\emailAdd{veergarani@gmail.com}
\emailAdd{vempati@cts.iisc.ernet.in}

\abstract{
Neutralino dark matter in supersymmetric models is revisited in the presence of flavor violation in the soft supersymmetry breaking
sector.  We focus on flavor violation
in the sleptonic sector  and study the implications for the co-annihilation regions.  
Flavor violation is introduced by a single  $\tilde{\mu}_R-\tilde{\tau}_R$ insertion in the 
slepton mass matrix. Limits on this insertion from BR($\tau \to \mu + \gamma$) are weak in some regions of the parameter space
where cancellations happen within the amplitudes. We look for overlaps in parameter space where both the co-annihilation condition
as well as the cancellations within the amplitudes occur. In mSUGRA, such overlap regions are not existent, whereas they 
are present in  models with non-universal Higgs boundary conditions (NUHM). The effect of flavor violation is two fold: (a) it
shifts the co-annihilation regions towards lighter neutralino masses (b) the co-annihilation cross sections would be modified 
with the inclusion of flavor violating diagrams which can contribute significantly.    
Even if flavor violation is within the presently allowed limits, this is sufficient to modify the thermally averaged cross-sections by about 
(10-15)\% in mSUGRA and (20-30)\% in NUHM, depending on the parameter space.  In the overlap regions, the flavor violating cross sections
become comparable and in some cases even dominant to the flavor conserving ones. A comparative study of the channels is presented 
for mSUGRA and NUHM cases. 
} 

\arxivnumber{1104.4467}
\begin{document}
\maketitle

\section{Introduction}

\begin{figure}[ht]
\begin{center}
\includegraphics[width=1.0\textwidth,angle=0]{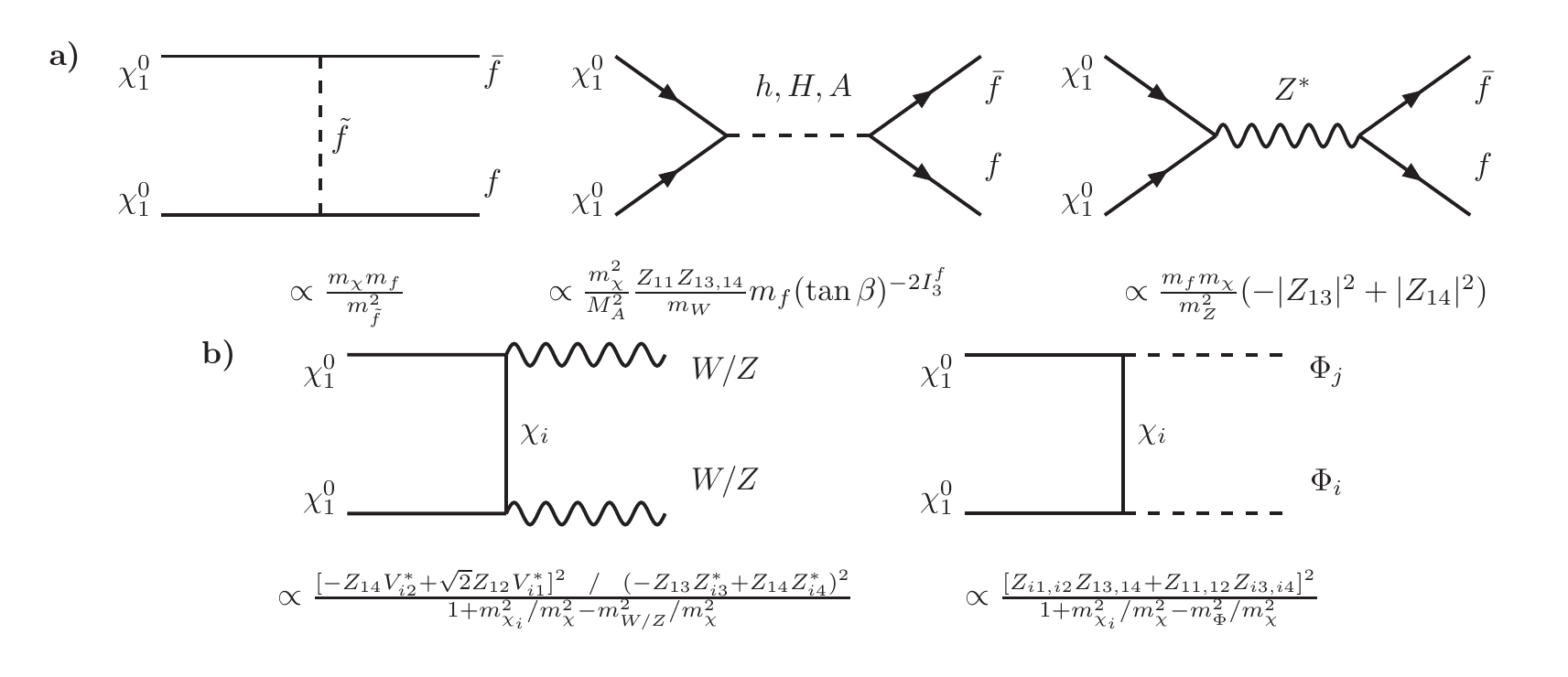}
\end{center}
\caption{{\bf Annihilation channels appearing in the $\Omega_{DM}$ calculation.} $V$ and $Z$ are the chargino and neutralino mixing matrices \cite{cmv}.}
\label{annhchannels}
\end{figure}

Supersymmetric standard models have a natural dark matter candidate namely, the lightest supersymmetric particle (LSP) if R-parity is conserved \cite{Jungman:1995df}. In mSUGRA/CMSSM models, the LSP typically is the lightest neutralino \cite{Goldberg:1983nd,Ellis:1983ew,Chankowski:1998za}. In most of  mSUGRA /CMSSM parameter space, the lightest neutralino is mostly a bino ($\widetilde{B^{0}}$); the bino component being close to 99\%. With the bino cross-section being small,  the neutralinos are overproduced resulting in a larger dark matter relic density compared to WMAP \cite{wmap7} allowed range. There are however, some {\it special} regions in the mSUGRA parameter space where the neutralino is able\footnote{See also Ref. \cite{ArkaniHamed:2006mb}} to satisfy the relic density limits \cite{Baer:2003wx,Djouadi:2006be}. 
These are the (i) Bulk region, (ii) Stop ($\tilde{t}$) co-annihilation region,  (iii) Stau ($\tilde\tau$) co-annihilation region, (iv) $A-$pole funnel region and (v) Focus point/ Hyperbolic branch regions. The various processes which play an important role in each of these sub-cases is shown in Fig. (\ref{annhchannels}). 

The stau--co-annihilation region requires the mass of the lightest stau, $\stau$ to be close to the mass of the LSP. The stop--co-annihilation is typically realized with large $A-$terms, which is also the case with the bulk region \cite{utpala-term}.
Among the above depicted regions, discounting the case of large $A-$terms, $\tilde{\tau}$--co-annihilation and the focus point regions are most sensitive to pre-GUT scale effects and the see-saw mechanism \cite{cmv,barger,gomez-lola-kang,Kang:2009pj,Biggio:2010me,Esteves:2010ff}. It has been shown that the co-annihilation region gets completely modified in the $SU(5)$ GUT theory and leads to upper bounds in the neutralino masses \cite{cmv}. Similarly, in the presence of type I, type II or type III see-saw mechanisms \cite{cmv,barger, Biggio:2010me,Esteves:2010ff} $\tilde{\tau}$--co-annihilation regions get completely modified. Strong implications can also be felt in the focus point regions unless the right handed neutrino masses are larger than the GUT scale \cite{barger}.  GUT scale effects can even revive no-scale models \cite{olivereview}.  It has also been shown that in the presence of large $A-$terms `new' regions with $\tilde{\tau}$--co-annihilation appear \cite{cmv,olive1}.

In the present work, we consider  flavor violation  in the sleptonic sector and study its implications for the co-annihilation regions. In generic MSSM, flavor violation can appear either in the
left handed slepton sector (LL), right handed slepton sector (RR) or left-right mixing sector (LR/RL) of the sleptonic mass matrix. However, we concentrate on the flavor violation in RR sector
as it has some interesting properties related to cancellations in the lepton flavor violating amplitudes as discussed below. 
Such flavor mixing is not difficult to imagine. It appears generically in most supersymmetric grand unified theories. 
A classic example is the  SUSY  SU(5)  GUT model.  If the supersymmetry breaking soft terms are considered universal at scales much above the gauge coupling unification scale ($M_{GUT}$),   typically the Planck scale,  then the running of the soft terms between the Planck scale and the GUT scale could generate the RR flavor 
violating entries in the sleptonic sector \cite{BHS,Calibbi:2006nq}.

 For demonstration purposes, lets consider the superpotential of the $SU(5)$ SUSY-GUT:
\begin{equation}
W  =  h^u_{ij} {\bf 10}_i {\bf 10}_j  {\bf \bar{5}}_H + h^d_{ij} {\bf 10}_i {\bf\bar{5}}_j {\bf 5}_H + \cdots 
\end{equation}
where ${\bf 10}$ contains $\{q,u^c,e^c\}$ and ${\bf \bar{5}}$ contains $\{d^c, l\}$. As supersymmetry
is broken above the GUT scale, the soft terms receive RG (renormalisation group) corrections between the high 
scale $M_X$ and $M_{GUT}$, which can be estimated using the leading log solution of the relevant RG equation.  
For example, the soft mass of ${\bf 10}$ would receive corrections:
\begin{equation}
\Delta^{RR}_{ij} = \left(m^{2}\right)_{{\bf\widetilde{10}}_{ij}} \approx - {3 \over 16 \pi^2}\, h^2_t\; V_{ti}\, V_{tj}\, \left( 3 m_0^2 + A_0^2\right)\, \log \left({M_X^2 \over M_{GUT}^2}\right),
\end{equation}
where $V_{ij}$ stands for the $ij^{th}$ element of the CKM matrix. Since ${\bf 10}$ contains $e^c$, the flavor violation in the
CKM matrix (in the basis where charged leptons and down quarks are diagonal) now appears
in the right handed slepton sector. Below the GUT scale, the RG scaling of the soft masses just
follows the standard mSUGRA evolution and no further flavor violation is generated in the
sleptonic sector in the absence of right handed neutrinos or any other seesaw mechanism. 
Assuming $M_X \approx 10^{18}$ GeV, the leading log estimates of the ratios of flavor violating
entries to the flavor conserving ones,  $\delta^{RR}_{ij} \equiv \Delta^{RR}_{ij}/m_{\tilde{l}}^2 $, are\footnote{$m_{\tilde{l}}^2 $ is the flavor conserving average slepton mass.} given in the Table \ref{leadinglogsu(5)}. We
have taken  $A_{0}=0$ and  $h_{t}\approx 1$. At 1-loop level $\delta$  it is roughly independent of $m_{0}$.

\begin{table}[htdp]
\caption{Flavor Violation generated in $SU(5)$ Model}
\begin{center}
\begin{tabular}{|c|c|}
\hline 
$|\delta|$ & Value  \\
\hline
\hline 
$\left|\delta_{\mu e}^{RR}\right|$ & $7.8\cdot 10^{-5}$\\
$\left|\delta_{\tau e}^{RR}\right|$ & $2.0 \cdot 10^{-3}$\\
$\left|\delta_{\tau \mu}^{RR}\right|$ & $1.4 \cdot 10^{-2}$\\
\hline
\end{tabular}
\end{center}
\label{leadinglogsu(5)}
\end{table}
From the Table \ref{leadinglogsu(5)}, we see that  the RG generated $\delta^{RR}_{ij} $ is typically of $\mathcal{O}(10^{-3} - 10^{-5})$. Such small values will not have any implications on the co-annihilation regions or rare flavor violating decays. While non-universality at the GUT scale in this case is RG induced, there are models where non-universal soft terms can arise from non-trivial K\"{a}hler metrics in supergravity, this could be the case in models with flavor symmetry at the high scale \`a la Froggatt-Nielsen models (see for example, discussions in \cite{fnsoftterms,Dudas:1996fe,Barbieri:1997tu,Kobayashi:2002mx,Chankowski:2005qp,Antusch:2008jf,Scrucca:2007pj}). In such cases, the $\delta_{RR}$'s could be much larger, even close to $\mathcal{O}(1)$.  These terms would then receive little corrections through RG as they are evolved from the GUT scale to the electroweak scale. Recently, in an interesting paper \cite{susylr}, supersymmetric models with Left-Right symmetry have been studied with particular emphasis on 
leptonic flavor violation. In these models, both left handed and right handed sleptonic sectors have flavor violation with the constraint
that $\delta_{RR} (\Lambda_r)\; = \; \delta_{LL} (\Lambda_r)$, where $\Lambda_r$ is the left right symmetry breaking scale. In such cases it could be possible\footnote{Subsequent to the appearance to this work on arXiv flavored co-annihilations have been studied by the group \cite{Esteves:2011gk}.} to generate $\delta_{RR} \sim \mathcal{O}(10^{-1})$.

In this present work, we will follow a model-independent approach and assume the presence of a single flavor violating parameter $\Delta^{RR}_{\mu\tau}$ and study the implications of it for the co-annihilation region. We will consider the simplistic case of universal soft-masses at the $M_{GUT}$ scale with non-zero
 $\delta_{23}^{RR}$  which is treated as  a free parameter. To distinguish from the standard mSUGRA model, we will call this model $\delta$-mSUGRA and similar nomenclature also holds for the other
 supersymmetry breaking models which we consider in this work.

While flavor violating entries in the sleptonic mass matrices are strongly constrained in general, the constraints on leptonic $\delta_{23}^{RR}$ entries are 
weak in some regions of the parameter space \cite{Hisano:1995nq,Masina:2002mv,Paradisi:2005fk}. This leads to the possibility that large flavor violation could
 be present in the sleptonic right handed sector.  In these regions cancellations happen between various contributions to the lepton flavor violating (LFV) amplitudes. 
 If such cancellation regions overlap with regions where sleptonic co-annihilations are important, flavor violation has
 to be considered in evaluating the co-annihilation cross-sections in the early universe. 
 This is the basic point of the paper where we show that flavor violating processes can play a dominant role  in the co-annihilation 
 regions of the supersymmetric breaking soft parameter space.  The processes contributing to relic density in these regions are called \textit{flavored co-annihilations}. 

It turns out that with mSUGRA/CMSSM boundary conditions, the parameter space where the flavor violating constraints are relaxed does not overlap with the $\stau$ co-annihilation regions unless one considers extremely large values of $\delta \geqslant 0.8$.  The overlap is not very significant and is mostly ruled out by other phenomenological constraints. However, if one relaxes the complete universality in the Higgs sector i.e., within non-universal Higgs mass models (NUHM), there is an overlap between these regions, paving way for large flavor violation to coexist with co-annihilation regions.

The fact that in $\delta$-NUHM these regions do overlap has already been observed independently by Hisano et al. \cite{Hisano:2002iy,Hisano:2008ng}. However, they have studied $\mu\,\rightarrow\,e\,\gamma$ transitions and their co-annihilating partner is not really a mixed flavor state. Further, they have not studied the relic density regions in detail.

In this present work we elaborate on these regions and study the consequences of it. The rest of the paper is organized as follows: In section [\ref{sec2}] we
discuss the effect on $\delta$ in the co-annihilation regions both in the mass of the co-annihilating partner and in the cross section. We also show that overlap between regions of LFV cancellations
and co-annihilations are not possible in $\delta$-mSUGRA. 
In section [\ref{sec3}] we show  that in $\delta$-NUHM regions do exist where flavored co-annihilations become important.  Relative importance of various cross-sections in the flavored co-annihilation regions  is elaborated in section [\ref{sec4}]. 
We close with a summary and brief
implications for LHC in  [\ref{sec5}]. In Appendix [\ref{appendix1}] we have written down the approximate expression of the soft-masses for mSUGRA and NUHM scenario for three different values to $\tan\beta$. 
In Appendix [\ref{appB}] we present $\delta$-mSUGRA in more detail using approximate results. Description of numerical packages used and numerical procedures followed are in Appendix [\ref{appendix2}]. In Appendix [\ref{loopfunc}], we present loop functions which are relevant to the discussion in the text. In Appendix [\ref{appD}] we present the analytic form of the cross-sections for some scattering processes relevant for the present discussions. 
\begin{figure}[htbp]
\begin{center}
\includegraphics[width=0.50\textwidth,angle=0]{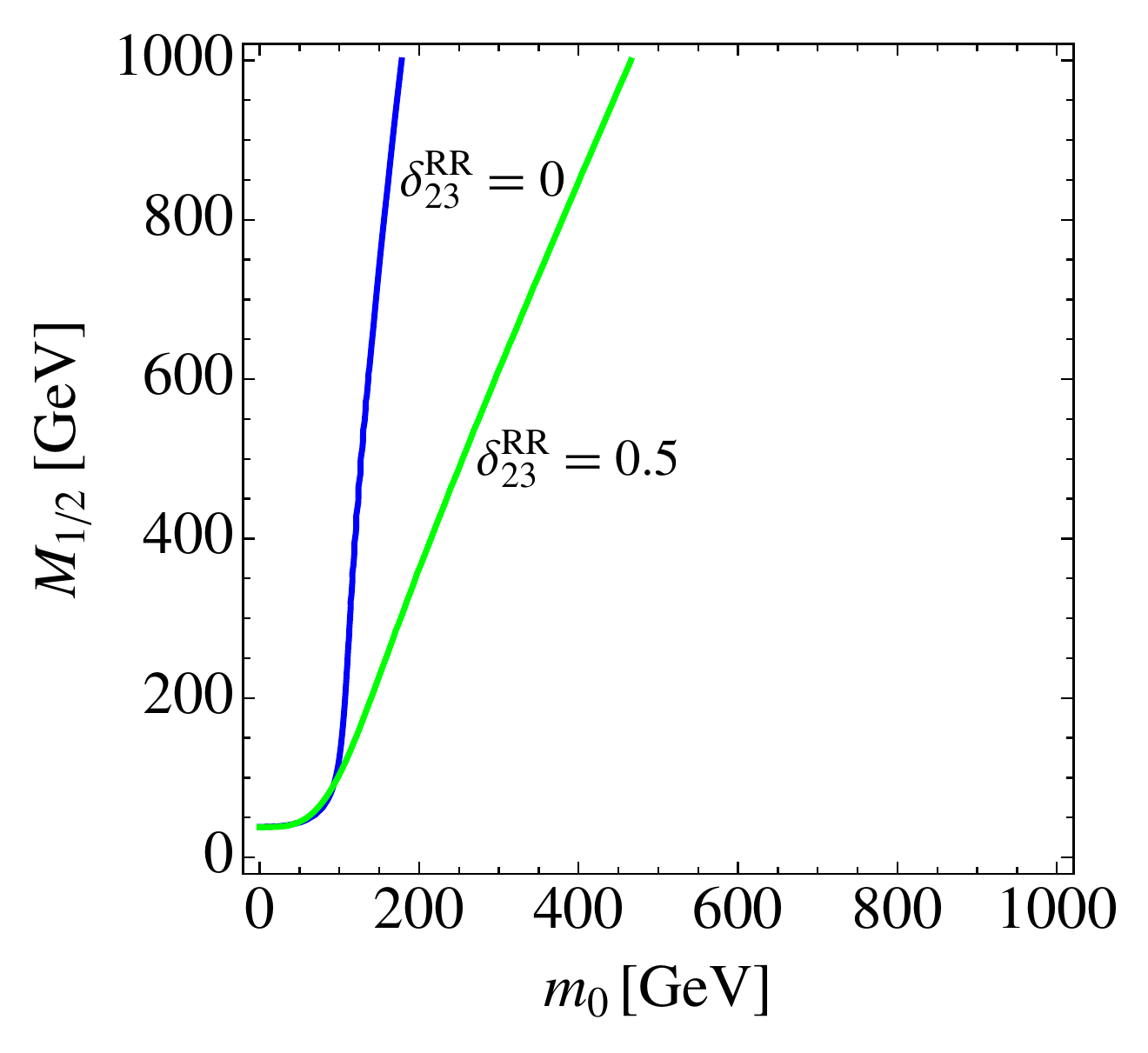}
\end{center}
\caption{{\bf The Co-annihilation region with and without flavor mixing.}
In the above figure we plot the condition $m_{\tilde\tau_1} - m_{\tilde\chi_1^0} = 0$ for $\delta =0$ (blue line) and for $\delta = 0.5$ (green line). Here we have chosen $\tan\beta = 5$ and $A_0 = 0$.}
\label{mom12}
\end{figure}

\section{Co-annihilation with Flavor Violation}
\label{sec2}
Co-annihilations play an important role in reducing the (relic) number density of the dark matter particle by increasing its interactions at the
decoupling point.  It requires having another particle which is almost degenerate in mass with the dark matter particle and should share
a quantum number with it \cite{Griest:1990kh}. In mSUGRA, $\tilde{\chi}_{1}^0$ can have co-annihilations with $\tilde{\tau}_1$ in regions of
the parameter space where $m_{\tilde{\tau}_1} \approx m_{\tilde{\chi}_1^0}$.  We will now generalize this condition\footnote{The condition
can be more accurately expressed as $m_{\tilde{\tau}_1}  = m_{\tilde{\chi}_1^0} + \delta m$, where $\delta m$  lies within 10-15 GeV. } in the presence of
flavor violation.   As discussed in the introduction,  we will   consider a single  $\mu-\tau$  flavor mixing term in the RR sector, 
$\Delta^{\mu \tau}_{RR}$ to be present at the weak scale. Similar analysis also holds for the $e-\tau$ flavor mixing. The slepton mass matrix is defined by
\begin{align}
&\qquad \qquad \mathcal{L}_{int}  \supset - \half\, \Phi^{T} \, \mathcal{M}^2_{\tilde{l}} \, \Phi \\
\intertext{where $\Phi^T = \Big\{\tilde{e}_L, \tilde{\mu}_L, \tilde{\tau}_L, \tilde{e}_R, \tilde{\mu}_R, \tilde{\tau}_R \Big\}$ and } 
\mathcal{M}^2_{\tilde{l}} &= \begin{pmatrix}
m_{\tilde{e}_{L}}^2 & 0 & 0 & m_{\tilde{e}_{LR}}^2 & 0 & 0 \\
0 & m_{\tilde{\mu}_{L}}^2 & 0 & 0 & m_{\tilde{\mu}_{LR}}^2 & 0 \\
0 & 0 &  m_{\tilde{\tau}_{L}}^2 & 0 & 0 & m^2_{\tilde{\tau}_{LR}} \\
m_{\tilde{e}_{LR}}^2 & 0 & 0 & m_{\tilde{e}_{R}}^2 & 0 & 0 \\
0 & m_{\tilde{\mu}_{LR}}^2 & 0 & 0 & m_{\tilde{\mu}_{R}}^2 & \Delta^{\mu \tau}_{RR} \\
0 & 0 &  m_{\tilde{\tau}_{LR}}^2 & 0 & \Delta^{\mu \tau}_{RR} & m^2_{\tilde{\tau}_{R}} \\
\end{pmatrix}, \label{slep6}
\end{align}
where, $m_{\tilde{f}_{LR}}^{2} = m_{f} \left(A_{f} - \mu \tan \beta \right)$'s are the flavor conserving left-right mixing term, $m_{\tilde{f}_{L}}^{2}$'s are the left handed slepton mass term and $m_{\tilde{f}_{R}}^{2}$'s denote the right handed slepton masses. In the limit of vanishing electron mass\footnote{In all our numerical calculations, we have used the full $6 \times 6$ mass matrix without any approximations. This approximation is valid only in models with universal scalar masses, like mSUGRA, NUHM etc.}
and zero flavor mixing in the selectron sector, we can consider the following reduced $4 \times 4$ mass matrix. This matrix is sufficient and convenient to understand most 
of the discussion in the paper. It is given by 
\begin{equation}
\mathcal{M}^2_{\tilde{l}} = \begin{pmatrix}
m_{\tilde{\mu}_L}^2 & 0 & m_{\tilde{\mu}_{LR}}^2 & 0 \\
0 &  m_{\tilde{\tau}_L}^2 & 0 &m^2_{\tilde{\tau}_{LR}} \\
m^2_{\tilde{\mu}_{LR}} & 0 &  m_{\tilde{\mu}_R}^2 &\Delta^{\mu \tau}_{RR} \\
0 & m^2_{\tilde{\tau}_{LR}} & \Delta^{\mu \tau}_{RR} &  m_{\tilde{\tau}_R}^2
\end{pmatrix} \label{eff4by4},
\end{equation}
where, we have taken it to be real for simplicity.  The lightest eigenvalue of the above matrix can be easily estimated. The lower $2\times 2$ block 
can be diagonalized assuming that the flavor violating $\Delta^{\mu\tau}_{RR}$ is much smaller than the flavor diagonal entries.  A second diagonalization
for the stau LR mixing entry can be done in a similar manner. This leads to a  rough estimate
 of the lightest eigenvalue as: 
\begin{equation}
\label{lightesteg}
m_{\lstau}^2\;\simeq\; m_{\tilde{\tau}_R}^2 ( 1 - \delta) - m_\tau \mu \tan \beta,
\end{equation}
where $\delta  =  \frac{\Delta^{\mu \tau}_{RR}} {\sqrt{m_{\tilde{\mu}^{2}_{R}} m_{\tilde{\tau}^{2}_{R}}}}$.
Requiring that the lightest eigenvalue not to be tachyonic, we find an \textit{upper} bound on $\delta$ as follows:
\begin{equation}
\delta\;<\;1 - { m_\tau \mu  \tan \beta \over m_{\tilde{\tau}_R}^2}
\label{deltaeq}
\end{equation}
This condition becomes important  in  regions of the parameter space where  $\mu \gg m_{\tilde{\tau}_R}^2$ and in  regions where  $\tan\beta$ is very large 
such that the second term approaches unity.  For co-annihilations, $\delta$ lowers the lightest eigenvalue of the sleptonic mass matrix. Non-zero $\delta$ shifts the `standard regions' in mSUGRA towards lower values of $M_{1/2}$, for a 
fixed $m_0$. In other words, since the sleptons become lighter, the co-annihilations happen with lighter neutralino masses. 
To illustrate this point let us consider mSUGRA like universal boundary conditions at the GUT scale.
The one exception to the universality of the scalar mass terms particularly slepton mass terms at GUT scale is in terms of the flavor violating mass term $(\Delta_{RR}^{\mu\tau})$.  We will call this model as $\delta$-mSUGRA. Given that the $\Delta^{\mu \tau}_{RR}$ parameter does not run significantly under RG corrections\footnote{This is true as long as we stick to MSSM like particle spectrum and interactions. Additional interactions and particles can modify the flavor structure.}, we can 
use the MSSM RGE with mSUGRA boundary conditions to study the low energy phenomenology. In Appendix [\ref{appA1}], we have presented approximate
solutions for the RGE of soft masses and couplings  in mSUGRA. 
Using approximate formulae, in Fig. (\ref{mom12}) we have plotted, the $\tilde\tau-$co-annihilation condition, $m_{\tilde\chi^0_1} - m_{\lstau} \simeq 0$, with and without flavor mixing. We have chosen $\delta =0.0$, $0.5$ and $\tan\beta =5$. 
As expected from the Eq.(\ref{lightesteg}), the presence of flavor violating $\delta$ shifts the co-annihilation regions more towards the diagonal in the $m_{0}-M_{\half}$ plane. In table \ref{spectrum}, we show the spectrum for two points with $\delta = 0$ and $\delta = 0.5$ which demonstrate that for fixed $m_0$, a lighter neutralino can be degenerate with $m_{\lstau}$ in the presence of $\delta$.

\begin{table}[htdp]
\caption{Spectrum in co-annihilation region with and without $\delta$.}
\begin{center}
\begin{tabular}{c  c  c}
\hline \hline
Parameters & \multicolumn{2}{c}{Mass (GeV)} \\ \hline
$m_{0}$ &  200.0 & 200.0 \\ 
$M_{\half}$& 1031.0 & 458.0  \\ 
$\tan\beta$ & 20 & 20 \\ 
$\delta$ & 0.0 & 0.5 \\ 
$m_{\chi_{1}^{0}}$ & 439.22 & 188.69 \\ 
$m_{\tilde{\tau}_{1}}$ & 439.24  & 188.70 \\ \hline \hline
\end{tabular}
\end{center}
\label{spectrum}
\end{table}

\begin{figure}[ht]
\begin{center}
\includegraphics[width=1.0\textwidth,angle=0]{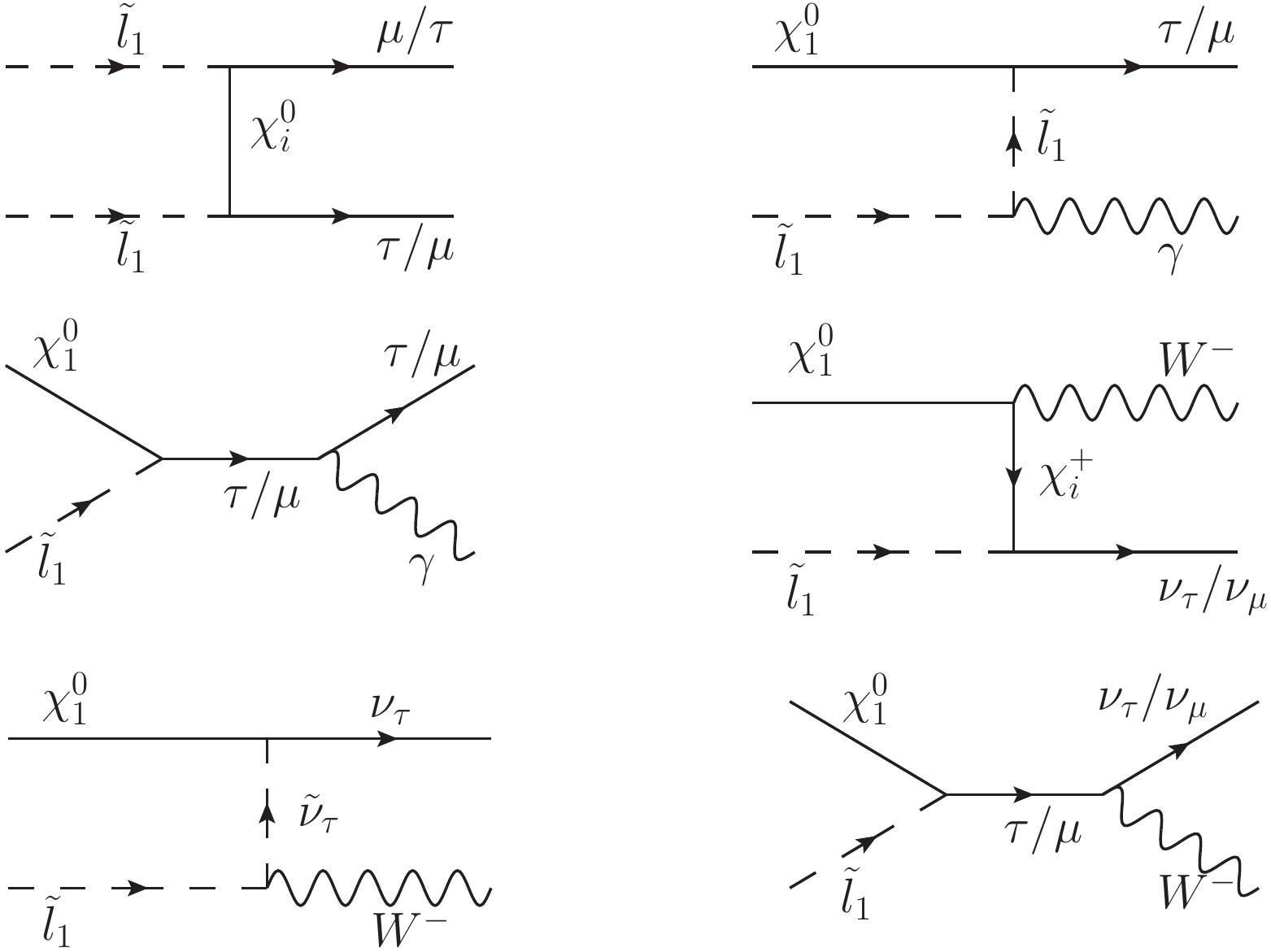}
\end{center}
\caption{Co-annihilation channels appearing in the $\Omega_{DM}$ calculation with $\mu-\tau$ flavor violation in the right handed sector. Notice that 
there are now new final states where either $\mu$ or a $\tau$ could appear.}
\label{coanchannels} 
\end{figure}
Eq.(\ref{lightesteg}) is a rough estimate and not valid for large $\delta$. A more accurate expression is presented in Appendix [ \ref{appB}].  As we will see, this
will not change the conclusions of the present discussion much.  We will revisit  this point again  in the next section.

The presence of $\delta$ also affects the relic density computations in the co-annihilation regions . The thermally averaged cross section on which relic density crucially depends can get significantly modified with $\delta$, where flavor violating scatterings are also now allowed.  The typical $\tilde{\tau}$ co-annihilation processes in the absence of flavor violation  are  $\chi_1^0 \chi_1^0\rightarrow\tau\bar{\tau},\mu \bar{\mu}, e \bar{e}$, $\chiz \stau \rightarrow\tau\gamma$, $\stau \stau \rightarrow\tau\tau$, $\stau \stau^{*} \rightarrow \tau \bar{\tau}$, $\chiz \stau \rightarrow Z \tau$, $\stau \stau^{*} \rightarrow \gamma \gamma$.  In the presence of $\tilde{\mu}_{R} - \tilde{\tau}_{R}$ flavor mixing, the new vertices related to flavor mixing would contribute to the processes with flavor violating final states.  The corresponding Feynman diagrams are shown in Fig.(\ref{coanchannels}), where $\mu/\tau$ would mean that the final state could either  be a $\mu$ or a $\tau$. The relevant Boltzmann equations for the neutralino and the lightest slepton ($\lstau$), continue to remain as in the unflavored co-annihilation case, though the masses and the cross-sections appearing in them change.
 
\begin{figure}[ht]
\begin{center}
\includegraphics[width=1.0\textwidth,angle=0]{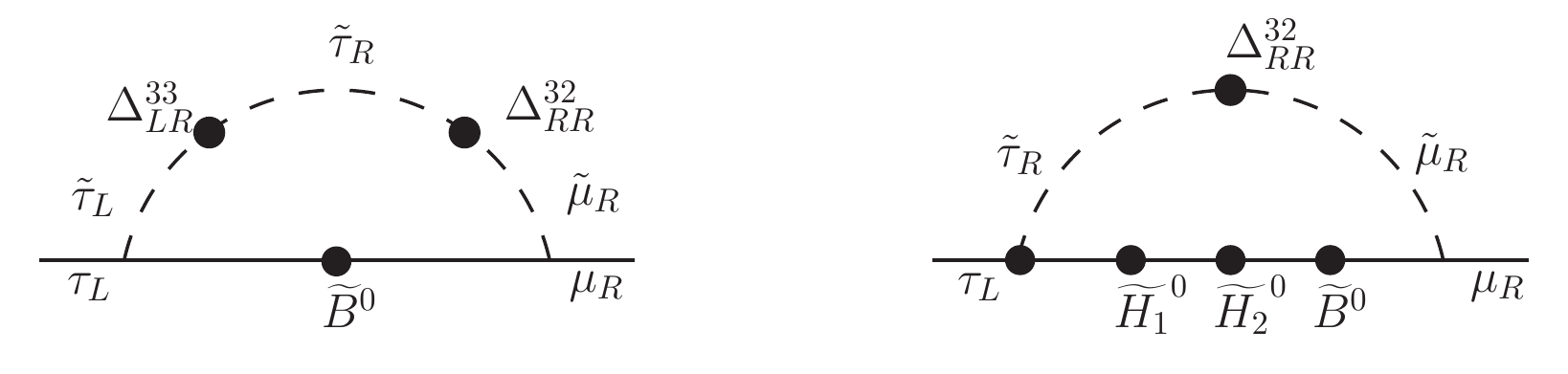}
\end{center}
\caption{{\bf $\tilde{B}^{0}$ and $\tilde{B}^{0}-\tilde{H}^{0}$ contribution in RR-insertion.} The photon can be attached with the charged internal lines.}
\label{feyn_dia}
\end{figure}

We have computed all the possible co-annihilation channels including flavor violation by adding the flavor
violating couplings in the MSSM model file of well known relic density calculator, \texttt{MicrOMEGAs} \cite{Belanger:2010gh}. 
The flavor violating co-annihilations contribute significantly to the total cross section and their relative importance
increases with increasing $\delta$ as expected.  
So far we have not addressed the question whether such large flavor violating entries in the sleptonic mass matrix are  
compatible with the existing flavor violating constraints from rare decay processes like $\tau \to \mu+ \gamma$ or $\tau \to  \mu e e $ etc.  
Constraints from such processes have been discussed in several works.  The constraints on right handed (RR) flavor violating 
sector are different compared to those of left handed (LL)  sector as they only have neutralino contributions and have no chargino contributions. 
Furthermore the two neutralino contributions\footnote{These are the pure $\tilde{B}^0$ and the mixed $\tilde{B}^0-\tilde{H}^0$ diagrams, as depicted in Fig.(\ref{feyn_dia}).}  can have
cancellations amongst each other in certain regions of the parameter space as elaborated in refs. \cite{Hisano:1995nq,Masina:2002mv,Paradisi:2005fk}.
 Following \cite{Masina:2002mv}, the branching ratio for $\tau \to \mu + \gamma $ can be written as in the generalized mass insertion approximation

\begin{align}
\text{BR}(\tau \rightarrow \mu \gamma) =\; &5.78\times 10^{-5}\;
\frac{M_W^4 M_1^2 \tan^2\beta}{|\mu |^2}
\times \left|\delta^{RR}_{23}( I_{B,R} - I_R ) \right|^2,
\label{brlfvs}
\end{align}
where $I_{B,R}$ and $I_R$ are loop functions are given in Appendix [\ref{loopfunc}].

This amplitude is resultant from the two diagrams shown in the mass-insertion approximation in Fig. (\ref{feyn_dia}). The first 
one is a pure Bino ($\tilde{B}^0$) contribution whereas the second one is a  mixed Bino-Higgsino ($\tilde{B}-\tilde{H}_1^0-\tilde{H}_2^0$) contribution.
There is a relative sign difference between these two contributions and thus leads to cancellations  in
some regions of the parameter space. In $\delta$-mSUGRA, these cancellations occur when $m_{\tilde\tau_{R}} \approx 6 M_{1}$ or equivalently $\mu^{2} \simeq m^{2}_{\staur_{R}}$\cite{Masina:2002mv}. In regions
 outside the cancellation region the limit on $\delta_{RR}$ is of $\mathcal{O}( 10^{-1} )$ for $\tan\beta = 10$ and for a slepton mass 
of around 400 GeV \cite{lucanpb} using the present on $\text{BR}(\tau \to \mu + \gamma) \leq 4.4 \times 10^{-8}$ \cite{Nakamura:2010zzi}. In the  cancellation region however the bound on $\delta $ is very weak and $\delta$ could be $\mathcal{O}(1)$. 

\begin{figure}[htp]
\begin{center}
\begin{tabular}{cc}
\includegraphics[width=0.50\textwidth,angle=0]{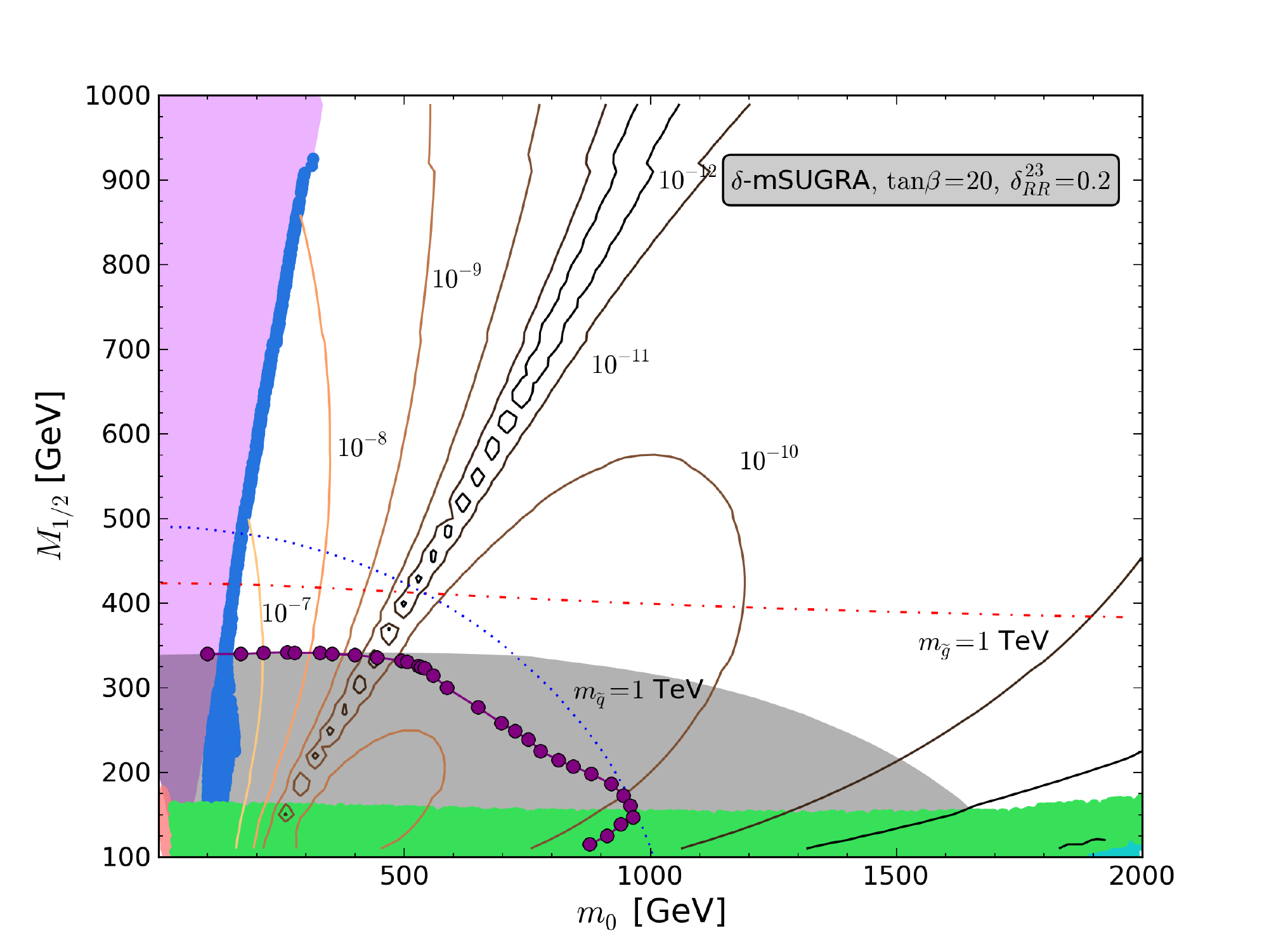} &
\includegraphics[width=0.50\textwidth,angle=0]{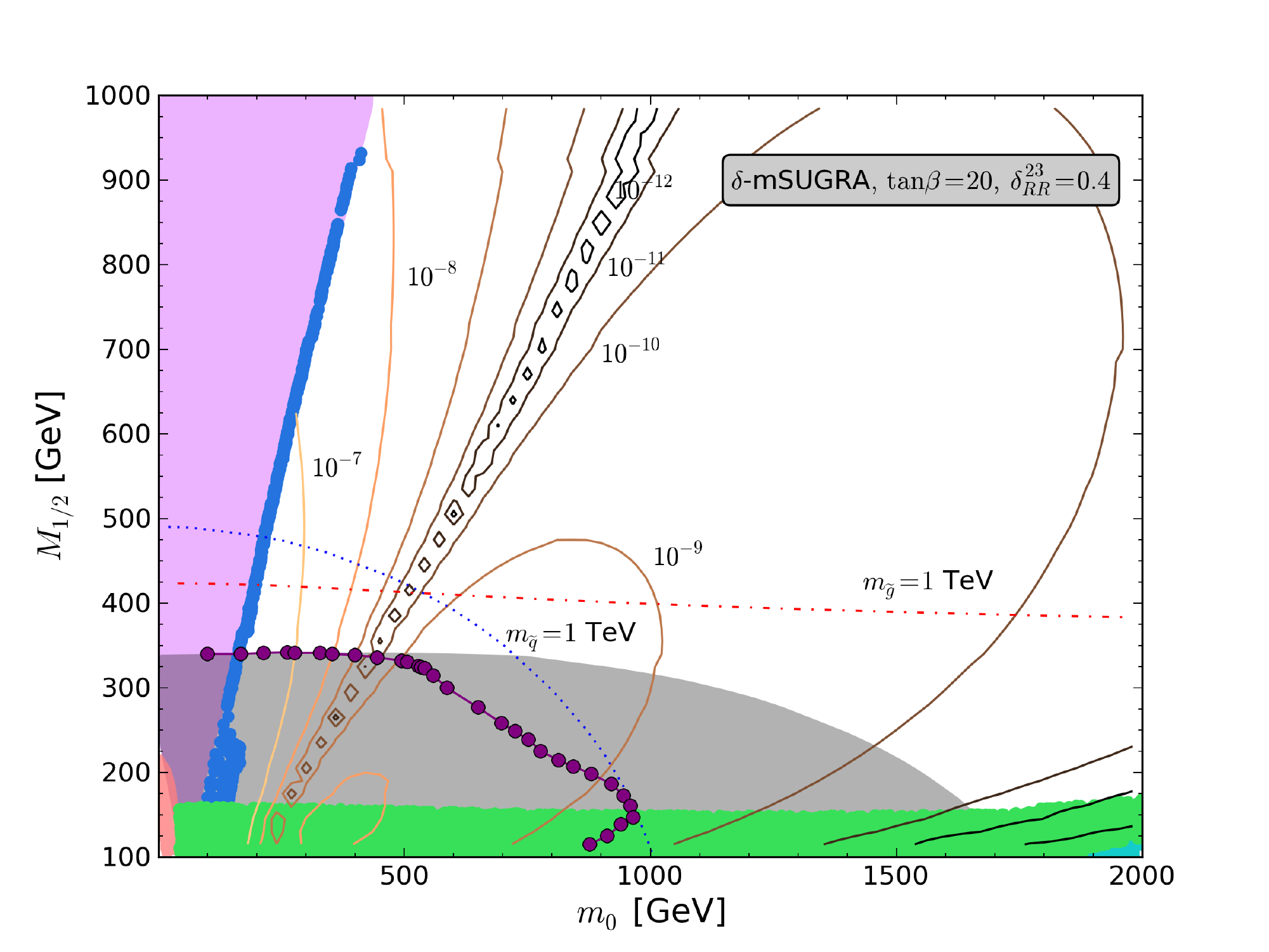} \\
\includegraphics[width=0.50\textwidth,angle=0]{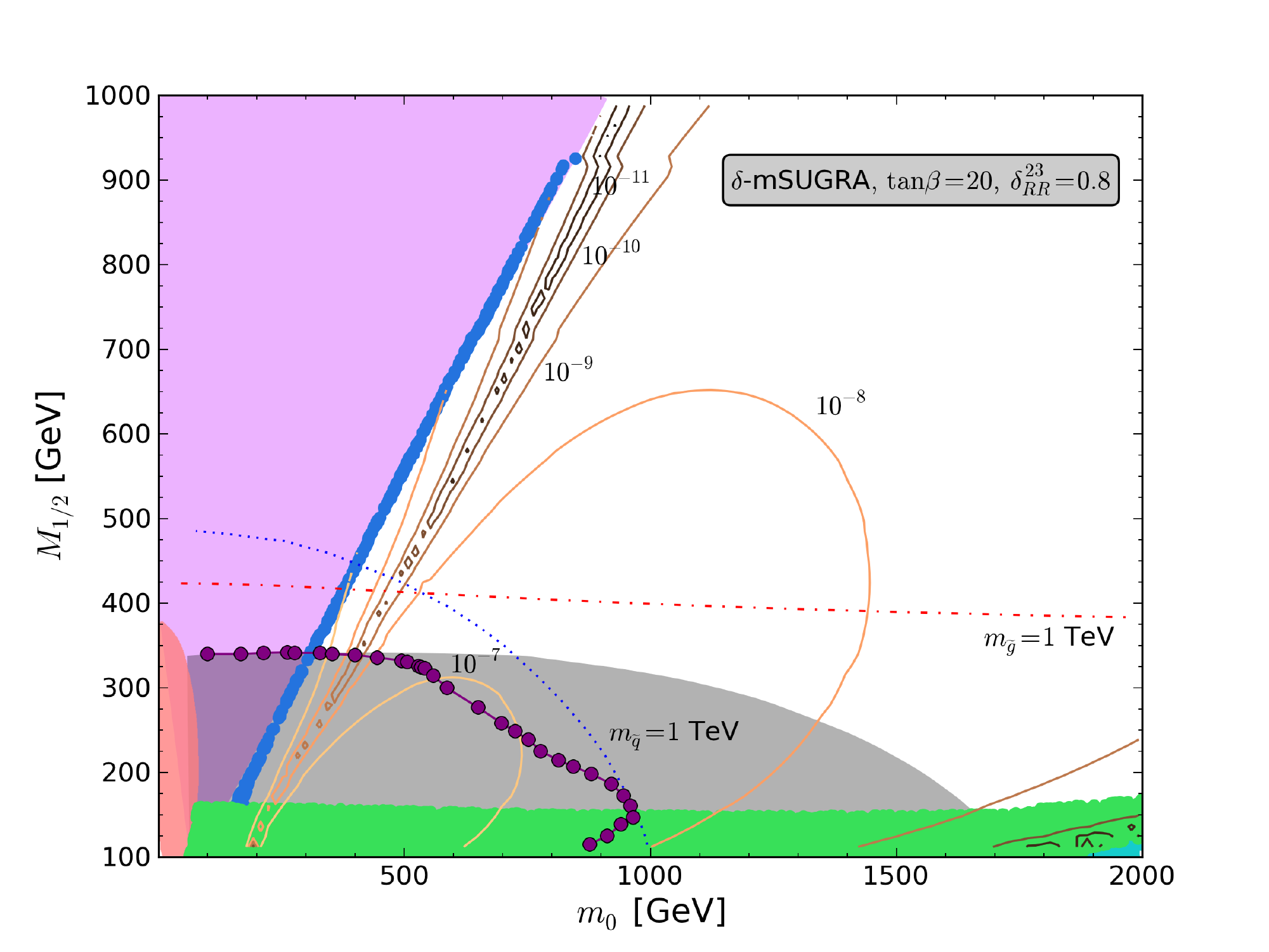} &
\includegraphics[width=0.50\textwidth,angle=0]{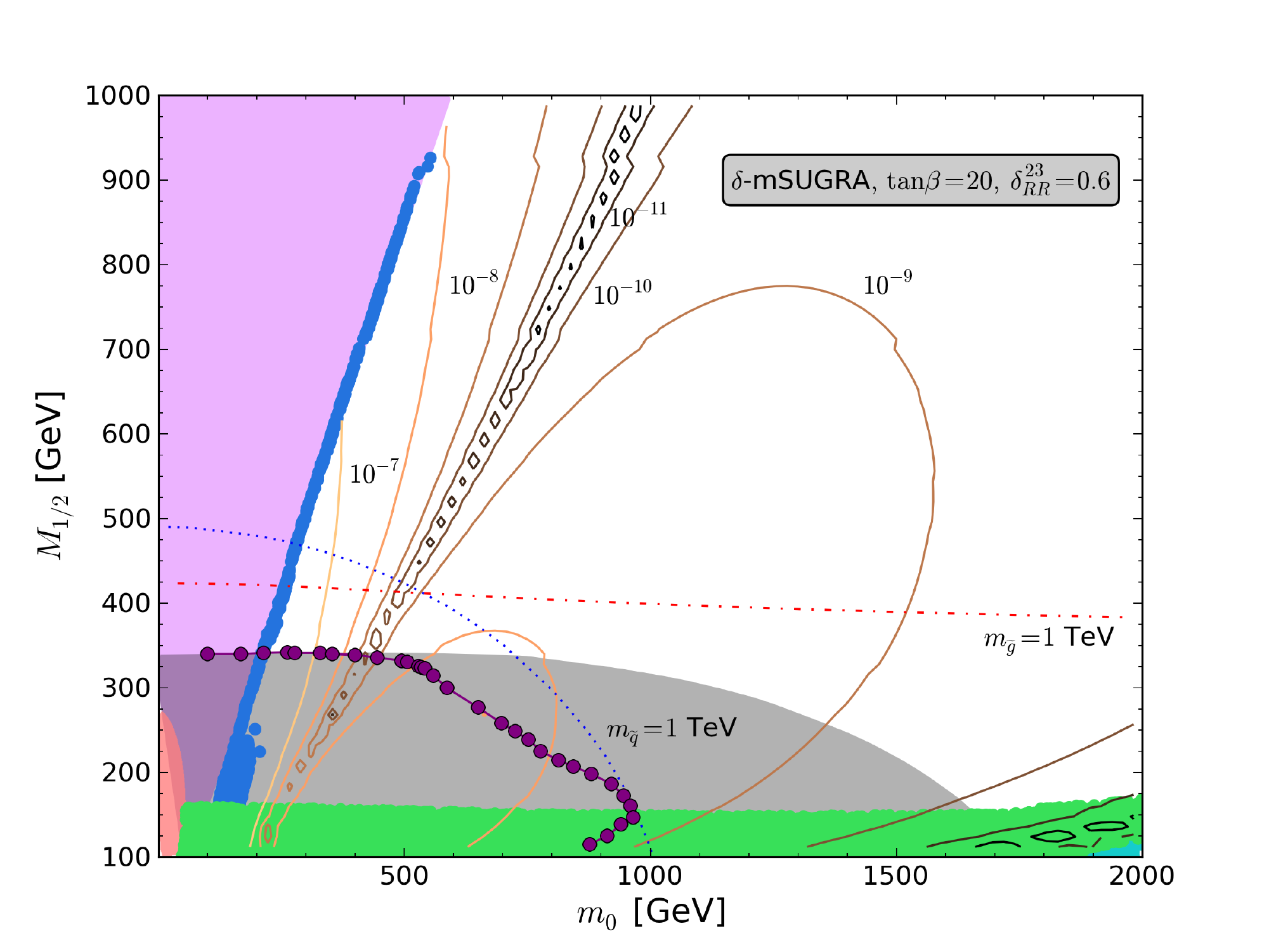} 
\end{tabular}
\caption{{\bf $m_{0}-M_{\half}$ plane in $\delta$-mSUGRA:} The different contour shows branching ratio, BR($\tau\,\rightarrow\,\mu\,\gamma$) for $\delta=0.2,\,0.4,\, 0.6$ and $0.8$ (from top left clockwise) and for $\tan\beta = 20, A_{0} = 0$ and sign$(\mu) > 0$. The blue line indicates WMAP bound satisfied region. The black shaded region is excluded by direct search in LEP for the Higgs boson. The violet dots represent the present limits form LHC \cite{lhcbounds}. The red dot-dashed line indicates 1 TeV contour for gluino and blue dotted line marks the 1 TeV contours for first generation squark mass. The regions where the contours of BR($\tau\,\rightarrow\,\mu\,\gamma$) reaches $\lesssim 10^{-10}$ are the places where cancellations happen. In this region $\delta^{23}_{RR}$ becomes unbounded because of the cancellation between the $\tilde{B}^{0}$ and $\tilde{B}^{0}-\tilde{H}^{0}$ diagrams in Fig.(\ref{feyn_dia}).}
\label{brmsugra}
\end{center}
\end{figure}

A large $\delta \sim \mathcal{O}(1)$\footnote{By definition $\delta$ cannot be larger than 1. Here $\mathcal{O}(1)$ means close to 1.} would increase the flavor violating cross sections in the early universe. The current bounds already push the value of $\delta \sim 10^{-1}$ for reasonable values of slepton mass $\sim 400$ GeV and $\tan\beta \sim 10$. We look for regions where the bound is significantly weakened due to cancellations. 
This would require that there should be significant amount of cancellations among the flavor violating amplitudes to escape
the bound from $\tau \to \mu + \gamma$.   In Fig. [\ref{brmsugra}], 
we have presented the numerical results for mSUGRA with each panel representing a different 
value of $\delta$ ($0.2,\,0.4,\,0.6$ and $0.8$). $\tan\beta$ is fixed to be 20 and sign($\mu$) is positive. The details of the numerical procedures we have
followed are presented in Appendix [\ref{appendix2}]. In all these plots,  we have shown contours of BR$(\tau\rightarrow\mu\gamma)$ and the co-annihilation regions. The other constraints shown on the plot include, 
the  purple region  which is excluded as the LSP is charged, ($m_{\lstau}<m_{\chi_1^0}$);  the translucent black 
shaded region is excluded  by search for a light neutral higgs boson at LEP, $m_h<114.5\ {\rm GeV}$, the light green region where the chargino mass is excluded by Tevatron, $m_{\chi^{\pm}_{i}} < 103.5$ GeV.   The co-annihilation region has been computed including the flavor violating diagrams in the thermally averaged
cross-sections.
 The relic density is fixed by the recent  7-year data of WMAP which
 sets it to be \cite{wmap7},
\begin{equation}
\Omega_{CDM} h^2= 0.1109 \pm 0.0056
\end{equation}
In the blue shaded region the neutralino relic density ($\Omega_{DM}$) is within the $3\sigma$ limit of \cite{wmap7}, i.e., we require it to be 
\begin{equation}
\label{omlim}
0.09\leq \Omega_{DM}h^{2} \leq 0.12\,.
\end{equation}
From the first panel of the figure, for $\delta = 0.2$ we see that there is no overlap in the regions where cancellation in the amplitudes for $\tau\to \mu+\gamma$ happens (around BR$(\tau \rightarrow \mu \gamma) \lesssim 10^{-10}$ ) and the co-annihilation region (blue region). With increasing $\delta$, as can be seen from subsequent panels,  the co-annihilation region moves towards the diagonal of the plane  as the slepton mass becomes lighter, and the cancellation region which  requires $m_{\tilde{\tau}_{R}} \approx 6 M_{1} $  also moves towards the diagonal with increasing $\delta$.  However, within $\delta$-mSUGRA these two regions do not coincide except partially at the top end of the spectrum close to the upper bound of the  the co-annihilation region.

\begin{figure}[htb]
\begin{center}
\begin{tabular}{ll}
\includegraphics[width=.50\textwidth,angle=0]{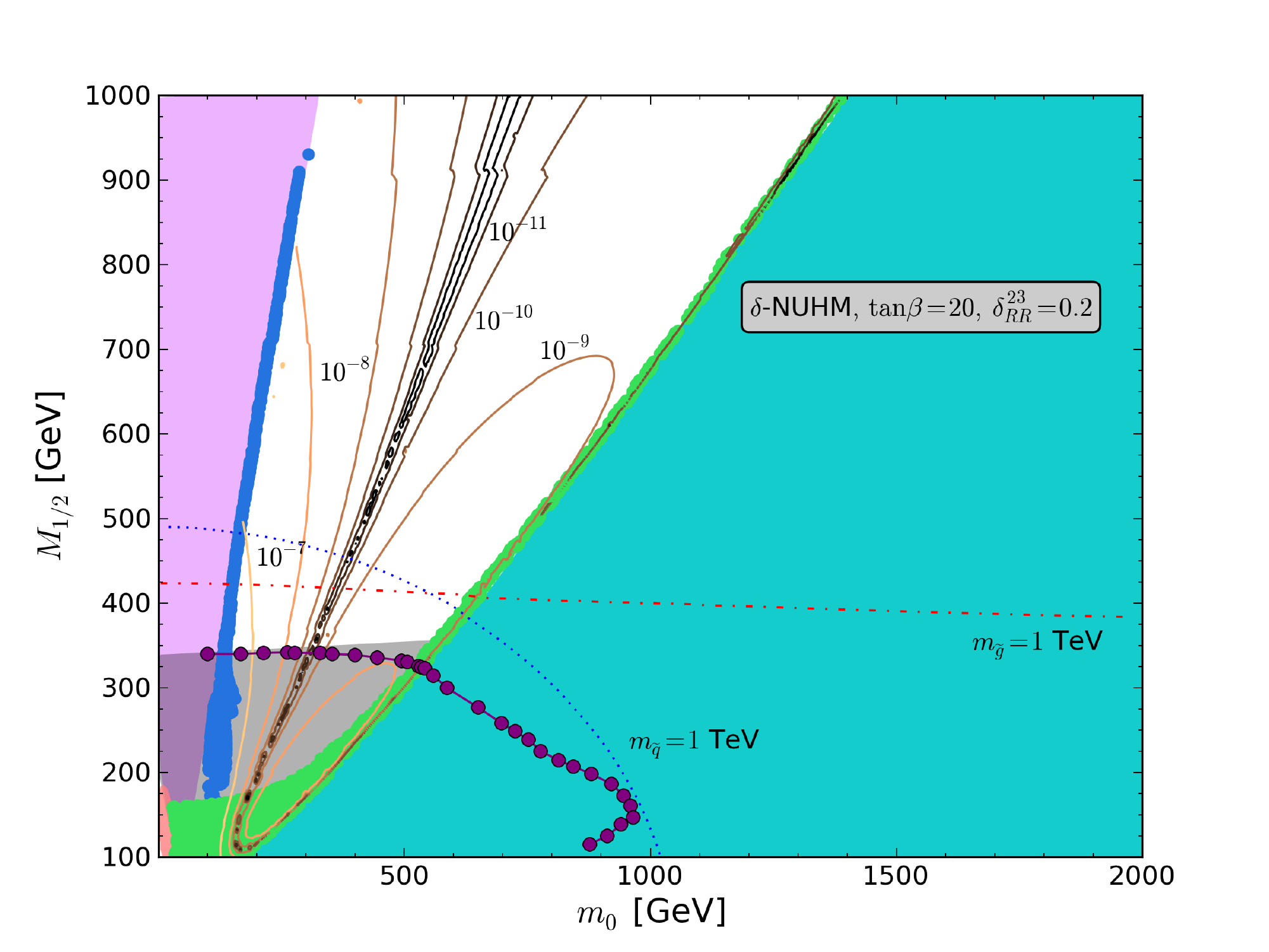} & \includegraphics[width=.50\textwidth,angle=0]{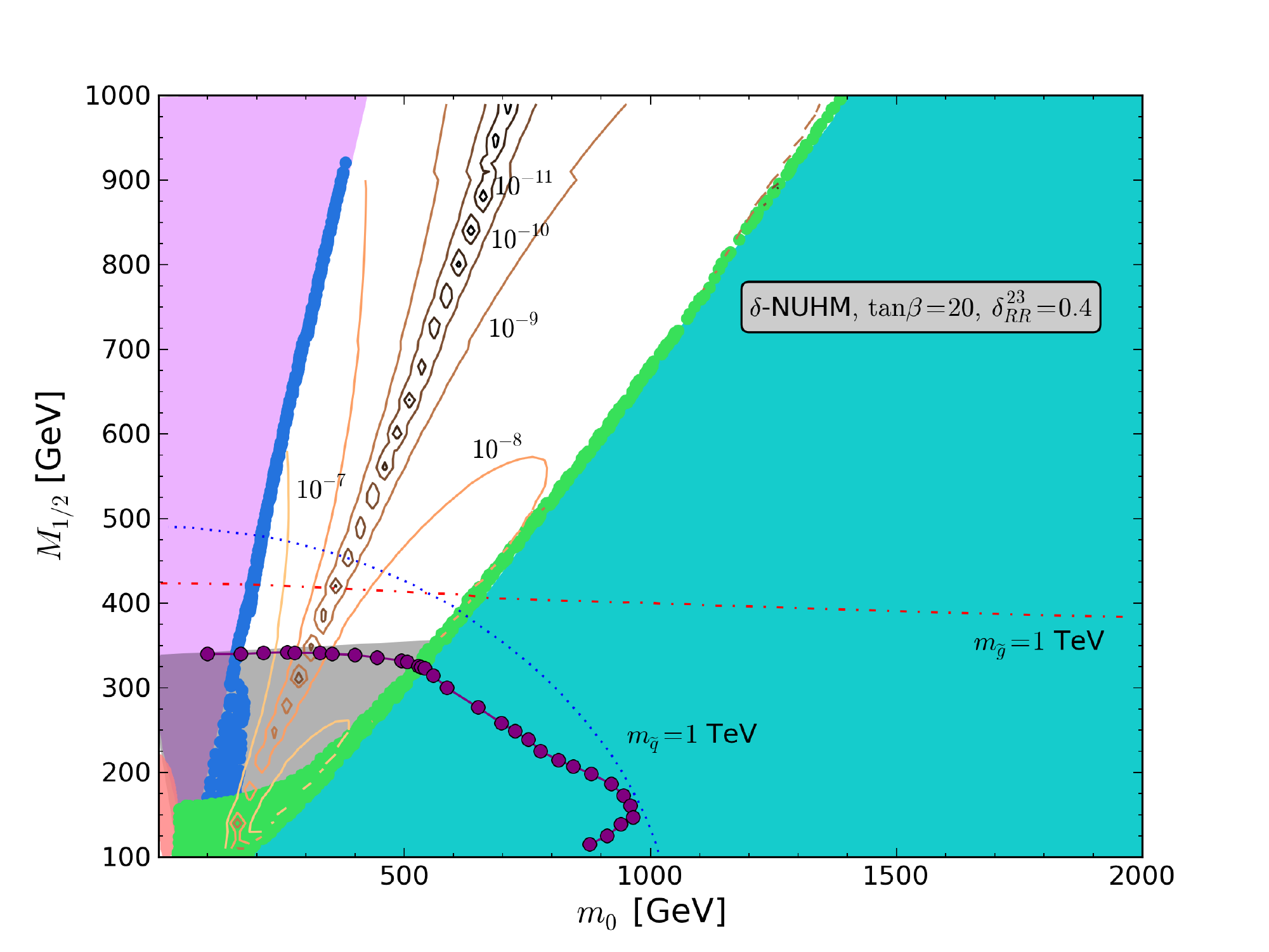} \\
\includegraphics[width=.50\textwidth,angle=0]{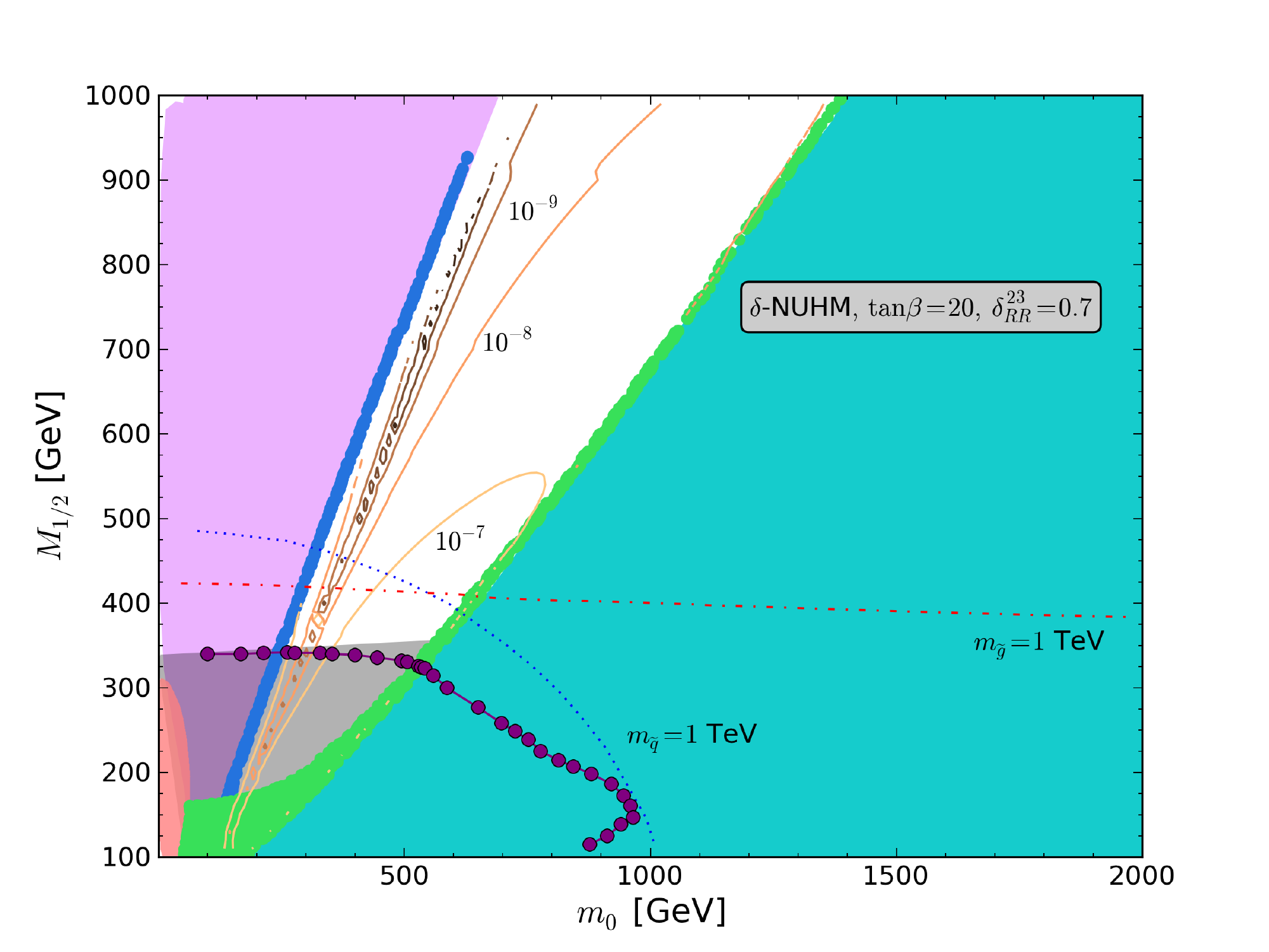} &  \includegraphics[width=.50\textwidth,angle=0]{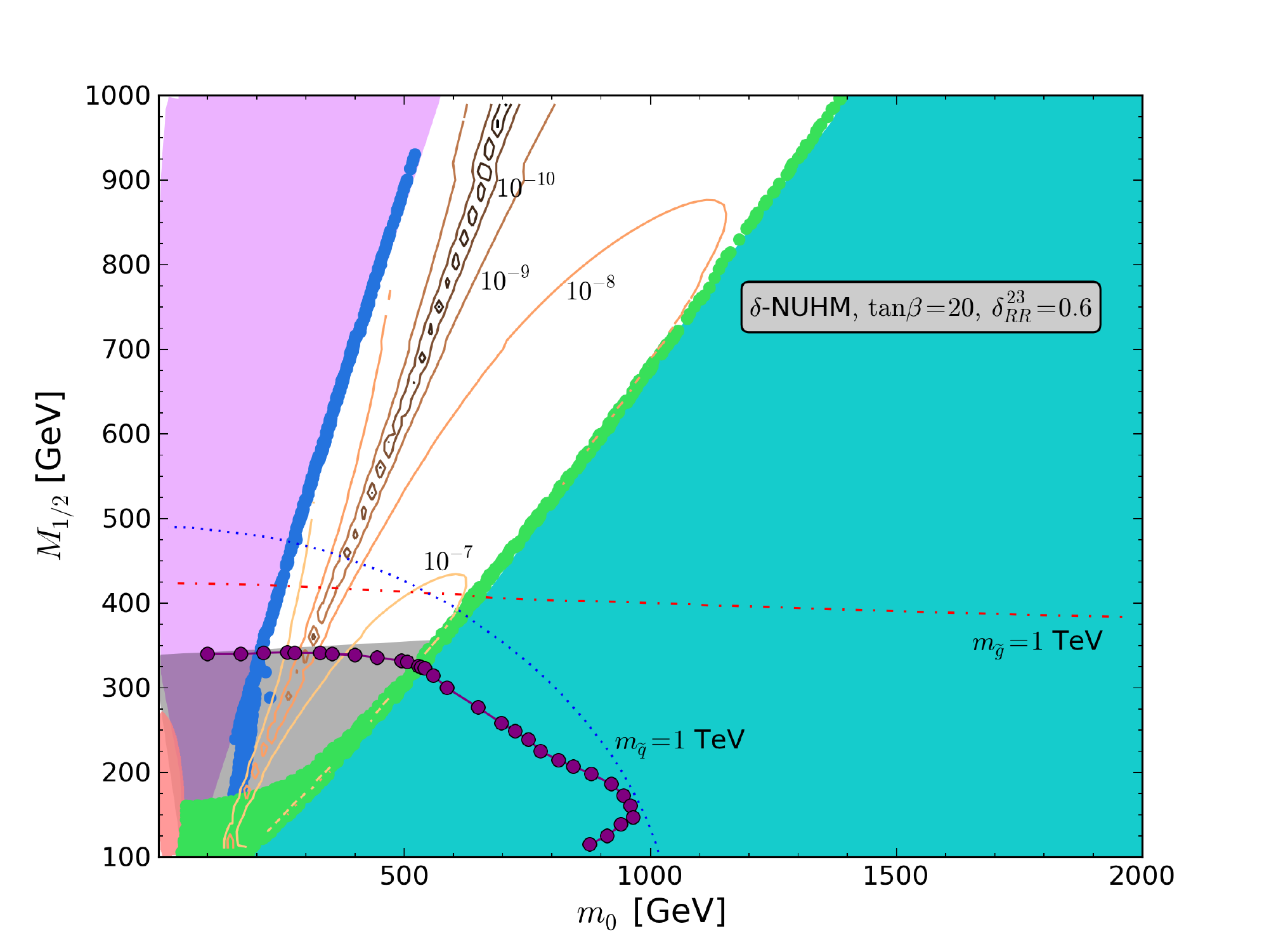} \\
\end{tabular}
\end{center}
\caption{Panels (from top-right in clockwise direction) depict $m_0$-$M_{\half}$ plane with $m_{10}=m_{hd}=0.5\cdot m_{0}$ and $m_{20}=m_{hu}=1.5\cdot m_{0}$ for $\tan\beta=20$, $A_0=0$ and sign$(\mu) > 0$, with $\delta=0.2, 0.4, 0.6,0.7$ respectively. Dark green region indicates inefficient REWSB. Purple region indicates $\lstau$ LSP. Black shade marks the region excluded by unsuccessful search by LEP, $m_h < 114.5\, {\rm GeV}$. The violet dots represent the present limits form LHC \cite{lhcbounds}. The red dot-dashed line indicates 1 TeV contour for gluino and blue dotted line marks the 1 TeV contours for first two generation squark mass. Blue strip bordering $\lstau$ LSP is the co-annihilation region. The different contour marks $BR(\tau\rightarrow\mu\gamma)$. The regions where the contours of BR($\tau\,\rightarrow\,\mu\,\gamma) \lesssim 10^{-10}$ and below are the places where cancellations happen, which can be identified by their `band' like structure.}
\label{nuhmdelta}
\end{figure}

 From Fig. (\ref{brmsugra}) we can see that a  very large ($\delta_{RR}^{23} \gtrsim 0.8$) is required to make the {\it cancellation} region consistent with the co-annihilation region. In $\delta$-mSUGRA having such large $\delta$ is consistent only very specific points of the parameter space (please see Appendix \ref{appB} for more discussion).
Hence, we can infer from the above figures that within the $\delta$-mSUGRA scenario the cancellation and co-annihilation region are disparate and no simultaneous solution exists. While the present discussion
was based on numerical solutions for  a particular $\tan\beta$, one can easily convince oneself that it would be true for any tan$\beta$ by looking at the analytical formulae. In fact, in the 
co-annihilation  region, the  branching fraction can be evaluated in the limit $\left(m_{\tilde\tau_{R}} \rightarrow M_{1}\right)$ and is given as 
\begin{align}
\text{BR}(\tau \rightarrow \mu \gamma) \approx \; &1.134\times10^{-6}\times
\frac{M_{W}^{4}\left|\delta_{23}^{RR}\right|^{2}\tan^2\beta}{M_{1}^{4}}
\label{brlfvcoan}
\end{align}
where, we have used $|\mu|^{2} \approx 0.5 m^{2}_{\tilde{\tau}_{R}} + 20 M^{2}_{1}$ and $ m^{2}_{\tilde{\tau}_{L}} \approx m^{2}_{\tilde{\tau}_{R}} + 2.5 M^{2}_{1}$. It is important to note that, the above expression obviously does not permit any cancellations.  Thus within $\delta$-mSUGRA, flavor violation in the co-annihilation region even if present would be
constrained by the existing leptonic flavor violating constraints. In the following we see that this situation is no longer true in case, when,
one relaxes the strict universality of the $\delta$-mSUGRA and considers simple extensions like non-universal Higgs mass models. 

Before proceeding to $\delta$-NUHM, a couple of observations are important. Firstly, apart from the cancellation regions, the present limits on $BR(\tau \to \mu + \gamma)$ constraint $|\delta| \lesssim 0.11-0.12$ for tan$\beta$ of 20 and slepton mass of around 200 GeV ($M_{{1 \over 2}} \sim 500$ GeV) in the co-annihilation
regions. Since such values  of $\delta$ are allowed by the data, one can consider them to be present in $\delta$-mSUGRA.  A larger value of $\delta$ would be valid for larger slepton masses. As discussed, this would lead to shifts in the parameter 
space of the co-annihilation region corresponding to mSUGRA.  As a result, there is a shift in the spectrum also compared to mSUGRA. The thermally
averaged cross-section are also modified. The shifts would be
largest in the absence of any constraint from lepton flavor violation. For this reason, we look for overlapping regions between the cancellation and co-annihilation regions.  Secondly,
the cancellation region lies within a small narrow band. To the left and right of this band there could be regions of partial cancellations. These are present 
in Figs. (\ref{brmsugra}).  A discussion connected with this issue is present in Appendix [\ref{appB}].

\section{Flavored Co-annihilation in $\delta$-NUHM}
\label{sec3}
As we have seen in the previous section, in $\delta$-mSUGRA, the $\mu$ parameter gets tied up with the neutralino mass in the
co-annihilation region, thus leaving little room for cancellations within the flavor violating amplitudes. In the NUHM models,
which are characterized by non-universal soft masses for the Higgs  alone \cite{Ellis:2002iu}, the $\mu$ remains no longer restricted. 
This can be demonstrated with approximate formulae presented in Appendix [\ref{appA2}]. 
We denote the high scale mass parameters as $m^{2}_{H_{u}}(M_{\text{GUT}}) \equiv m^{2}_{20}$ and $m^{2}_{H_{d}}(M_{\text{GUT}}) \equiv m^{2}_{10}$. For tan$\beta$ = 20, using the approximate expressions in the Appendix [\ref{appA2}], we see that $|\mu|^2$ has the form:
\begin{equation}
|\mu|^2 \approx 0.67 ~m_0^2 + 2.87 ~M_{\half}^2 - 0.027 ~m_{10}^2 - 0.64~ m_{20}^2
\end{equation}
Setting $m_0^2 \approx m_{\tilde{\tau}_R}^2 - 0.15 M_{\half}^2 $ and $M_{1} \approx 0.411 M_{\half}$ and taking the limit $m_{\tilde{\tau}_R} \rightarrow M_{1}$ in the co-annihilation region, we have 
\begin{equation}
|\mu|^2 \approx 17~M_{1}^2 - 0.027 ~m_{10}^2 - 0.64~ m_{20}^2
\end{equation}
thus providing enough freedom\footnote{$|\mu|^2 \approx 20.5 M_{1}^2$ in this limit in mSUGRA as can be seen from the expression below Eq. (\ref{brlfvcoan})}  in terms of $m_{10}$ and $m_{20}$ to 
allow  cancellations in the LFV  amplitudes to co-exist with co-annihilation regions.

The dark matter  phenomenology of NUHM models has been studied by several authors \cite{Ellis:2002iu,nonunivHiggs,Ellis:2008eu,Ellis:2007by,Roszkowski:2009sm,Das:2010kb}. The LSP is a neutralino in large regions of the parameter space and further, it can admit large Higgsino fractions in its composition unlike in mSUGRA. For simplicity, we concentrate on Bino dominated regions in the following. In such a case the lightest neutralino mass, in terms of SUSY parameters is as in mSUGRA:
\begin{equation}
\label{chimass}
m_{\chi^0_1}\,\approx\, 0.411 M_{\half}
\end{equation}

For the lightest slepton mass one can use Eq.(\ref{lightesteg})  where now $m_{\tilde{\tau}_R}^2$ at weak scale will be determined by 
the NUHM boundary conditions at the GUT scale. Similar to the mSUGRA case, approximate solutions can be derived for the NUHM
case also and they are presented in Appendix (\ref{appA2}).  Using the co-annihilation condition $ m_{\lstau}\,\approx\, m_{\chi^0_1} $ 
and the cancellation condition  $m^{2}_{\tilde{\tau}_R} \approx \mu^2$, one can derive expressions for $m_{10}^2$ and $m_{20}^2$
where flavored co-annihilations are of maximal importance. 

The derived expressions for $m^{2}_{10}, m^{2}_{20}$  are however, complicated. We found simpler parameterizations for regions  where the LSP is  Bino dominated and co-annihilations with the $\lstau$ are important. Examples of such  regions are (i)  $m_{20} = 1.5 \cdot m_0$ and $m_{10} = 0.5 \cdot m_{0}$  and  (ii) $m_{20} = 3 \cdot m_0$ and $m_{10} = m_0$. For these values of $m_{10}$ and $m_{20}$, flavored co-annihilations can exist for non-zero $\delta$. In Fig. (\ref{nuhmdelta}), we present in $m_0,M_{\half}$ plane regions consistent with all constraints for $\delta = 0.2, 0.4, 0.6$ and $0.7$, in an analogous fashion as to those
presented in $\delta$-mSUGRA section, Fig. (\ref{brmsugra}). We have chosen $m_{20} = 1.5 \cdot m_0$ and $m_{10} = 0.5 \cdot m_0$ for this plots. The purple region is excluded as the LSP is charged, here $m_{\lstau}<m_{\chi_1^0}$. Dark green region indicates no radiative electroweak symmetry breaking, $|\mu|^2<0$. The  translucent black shaded region is excluded by search for light  neutral higgs boson at LEP, $m_h<114.5\, {\rm GeV}$. As in $\delta$-mSUGRA, we see that with increase in $\delta_{23}^{RR}$, $\lstau-$LSP region increases owing to the  reduction of mass of $\lstau$. The impact of non-universality in the  Higgs sector is negligible for $m_{\lstau}$ in these regions. Analogously, regions excluded by light higgs search ($m_h<114.5\, {\rm GeV}$) are weakly affected in the presence of $\delta$. Moreover, region with $|\mu|^2<0$ is not affected by $\delta$ as it is entirely governed by $m_0,m_{10}$ and $m_{20}$ with maximum contribution from $m_{20}$ and $m_0$. However, as expected the magnitude of $BR(\tau\rightarrow\mu\gamma)$ governed by eq.(\ref{brlfvcoan}), increases with $\delta_{23}^{RR}$.  The last panel of the figure shows regions where cancellation regions overlaps with the co-annihilation regions for $\delta = 0.7$.  For a different set of  values of $m_{10}$ and $m_{20}$, for example, $m_{20} = 3 \cdot m_0$, $m_{10} = m_0$ the  overlap  regions can be found for even smaller values of $\delta$. In these regions flavored co-annihilations play a dominant role.

\section{Channels}
\label{sec4}
\begin{figure}[htp]
\begin{center}
\begin{tabular}{cc}
\includegraphics[width=0.50\textwidth,angle=0]{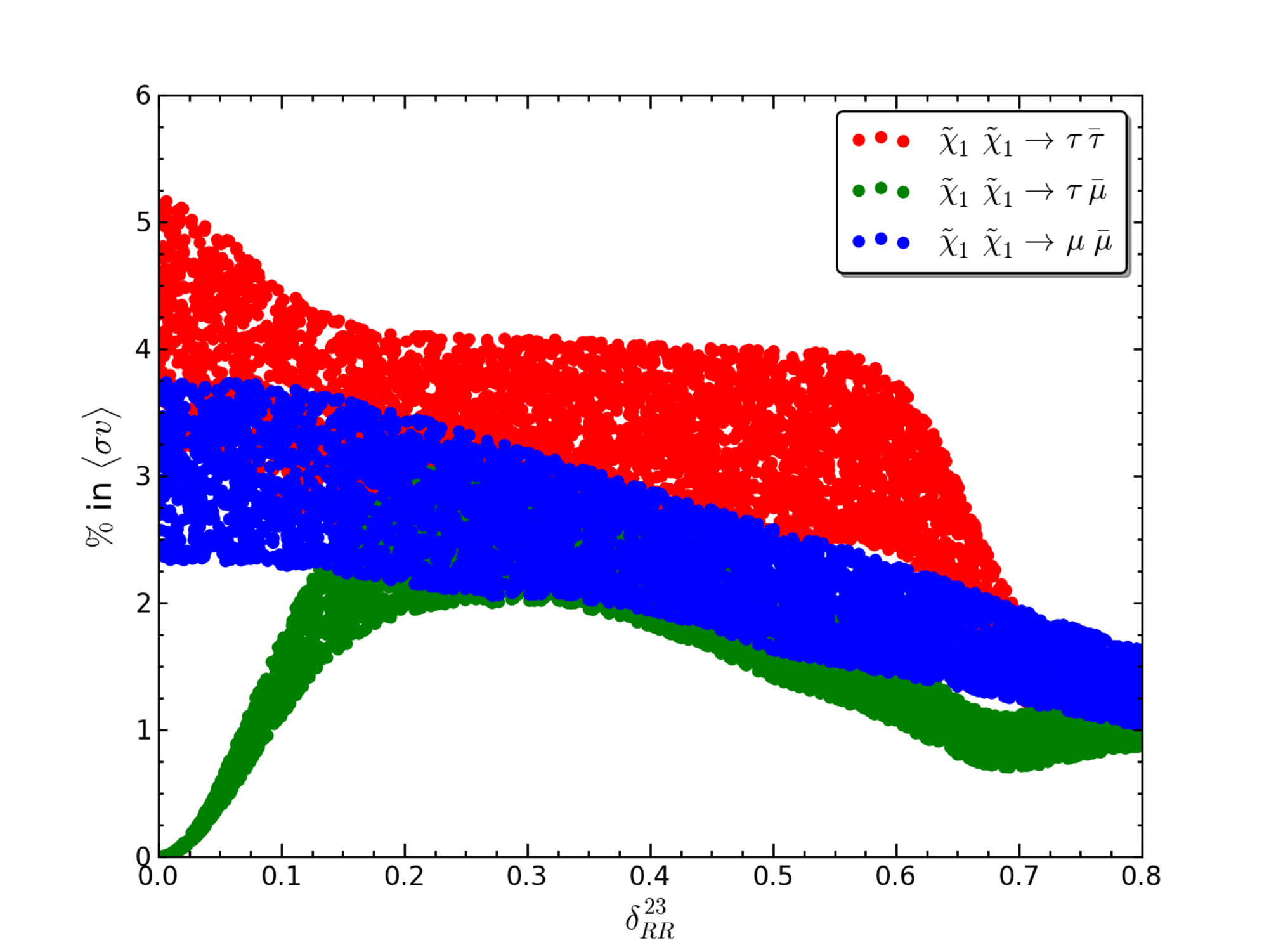} &
\includegraphics[width=0.50\textwidth,angle=0]{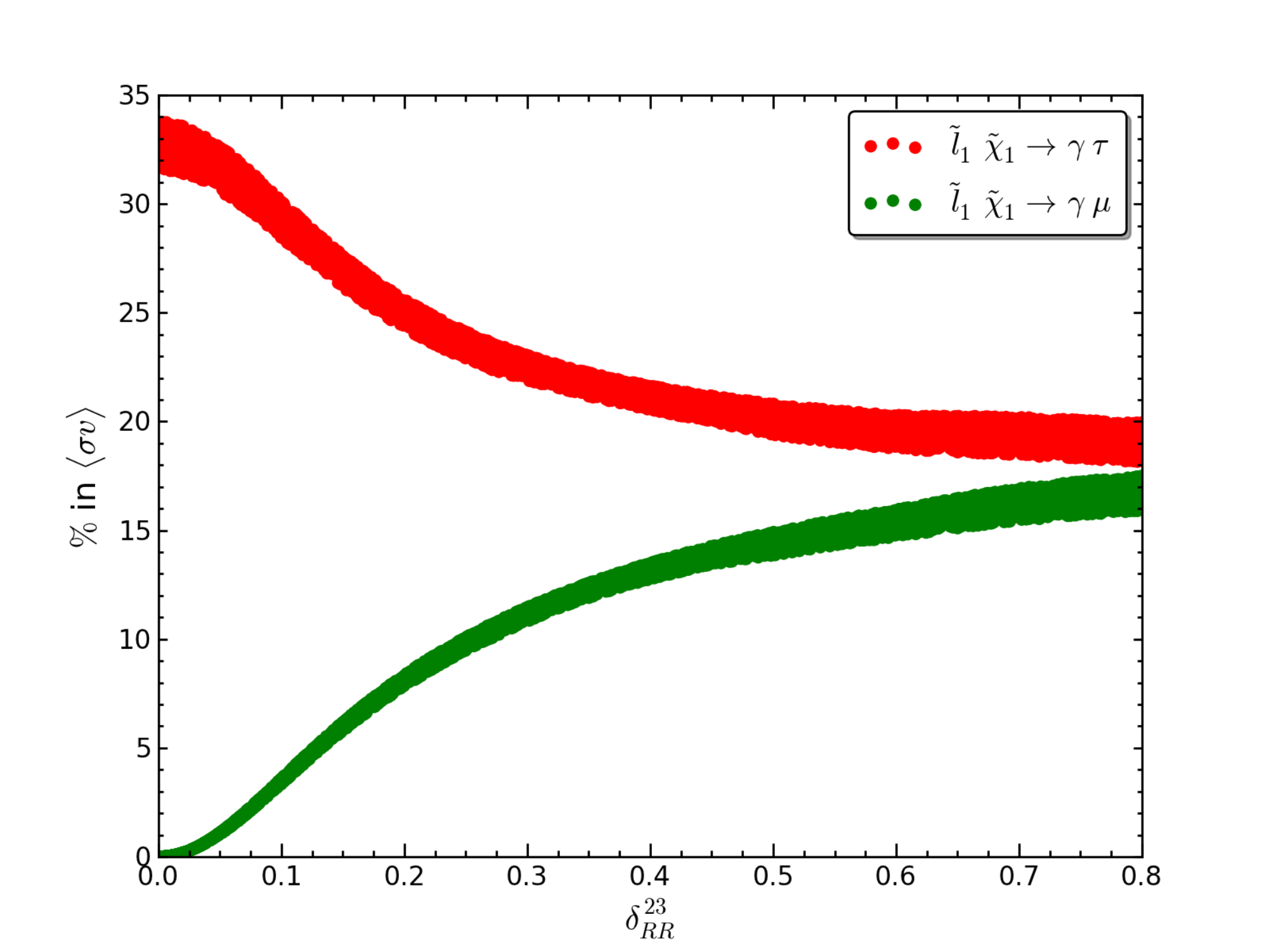} \\
\includegraphics[width=0.50\textwidth,angle=0]{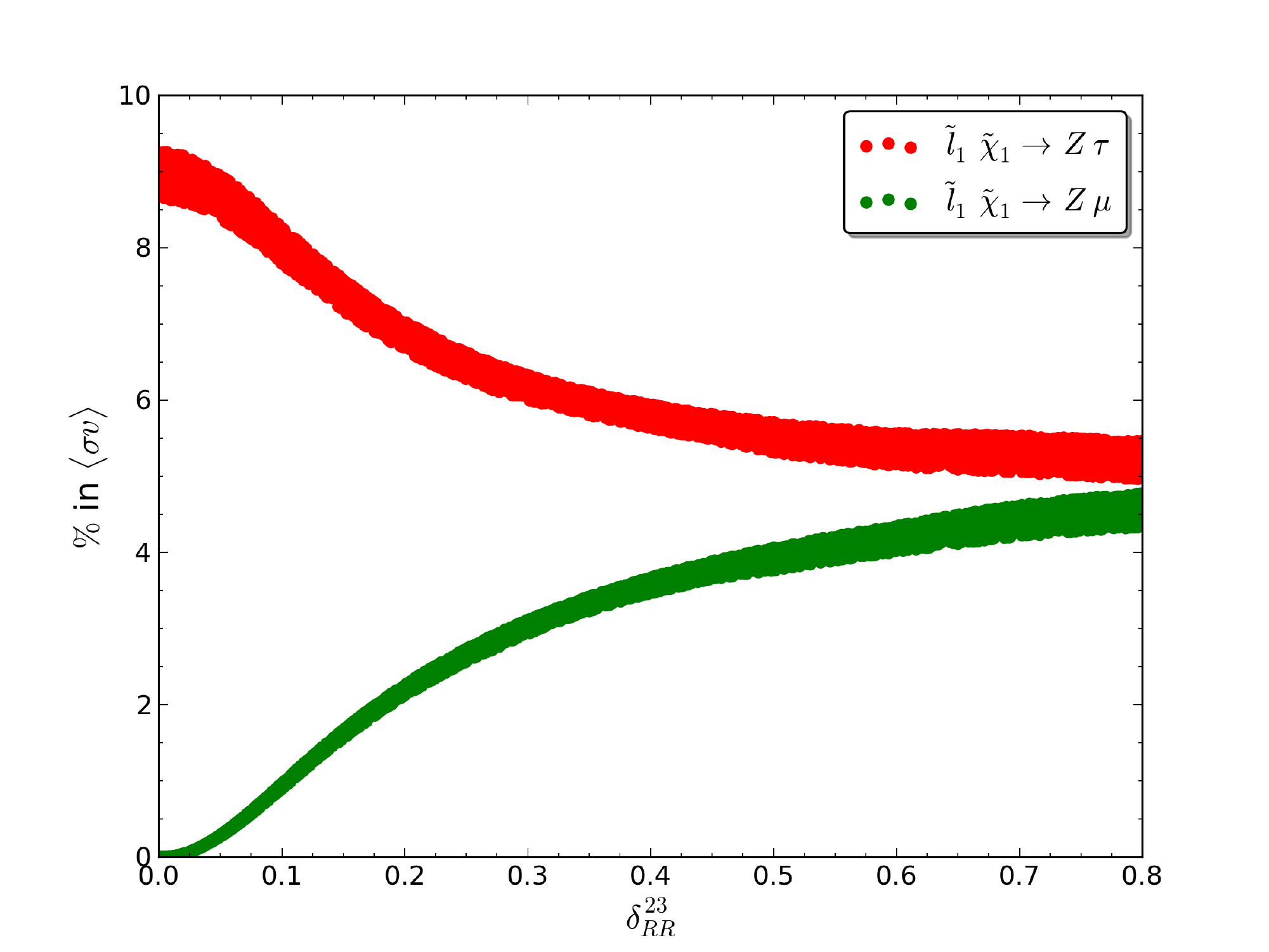} &
\includegraphics[width=0.50\textwidth,angle=0]{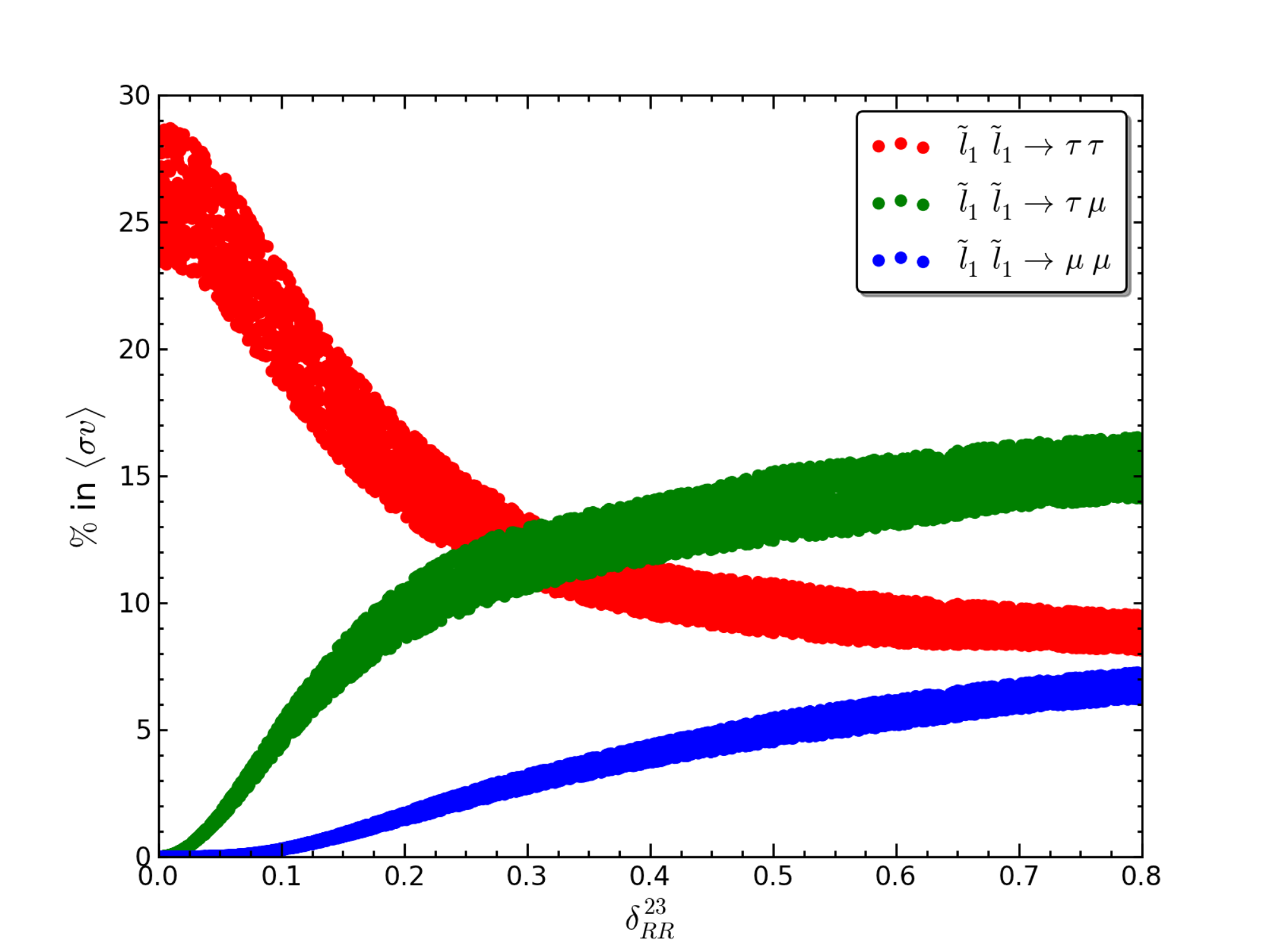} \\
\includegraphics[width=0.50\textwidth,angle=0]{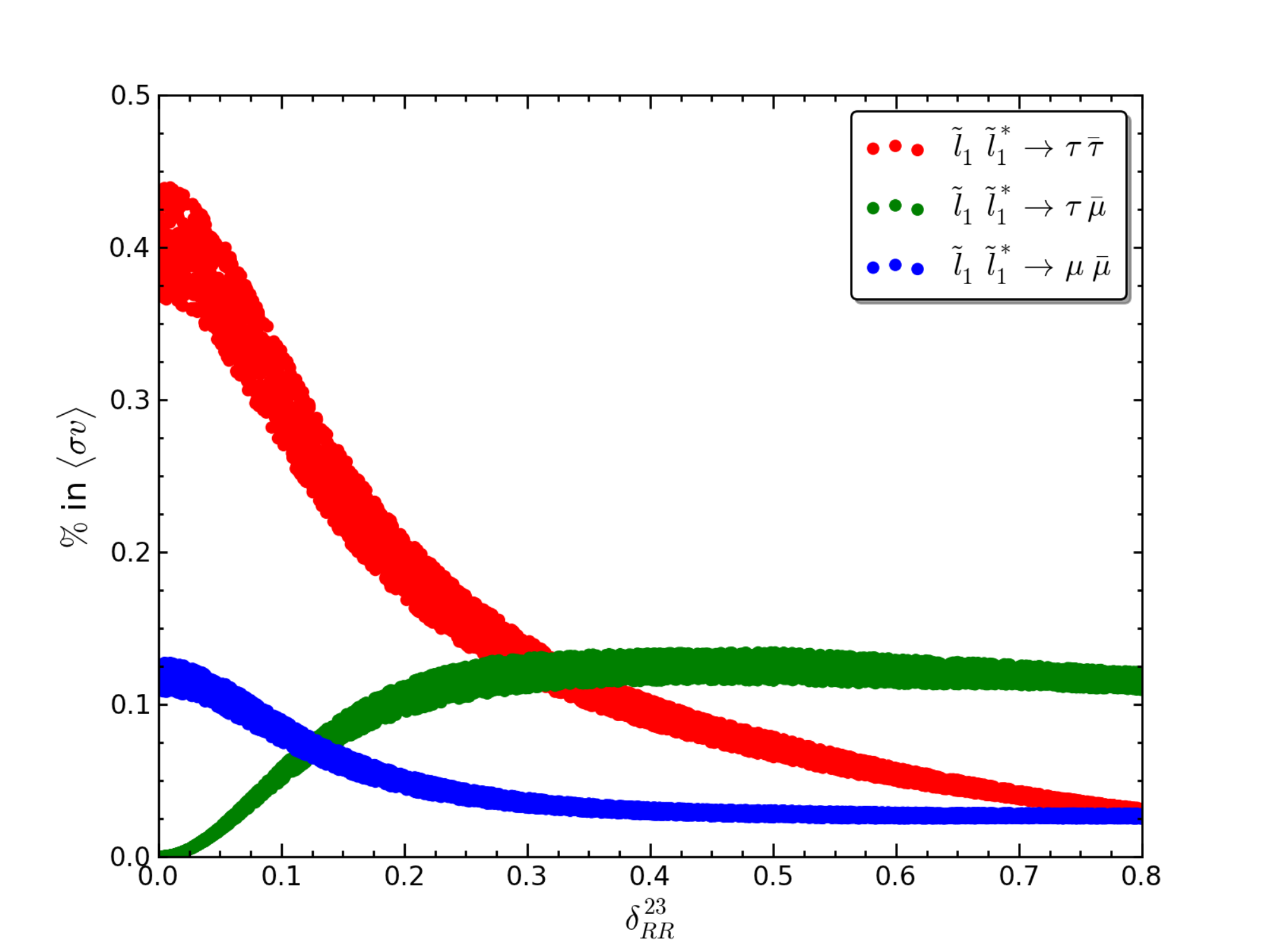} &
\includegraphics[width=0.50\textwidth,angle=0]{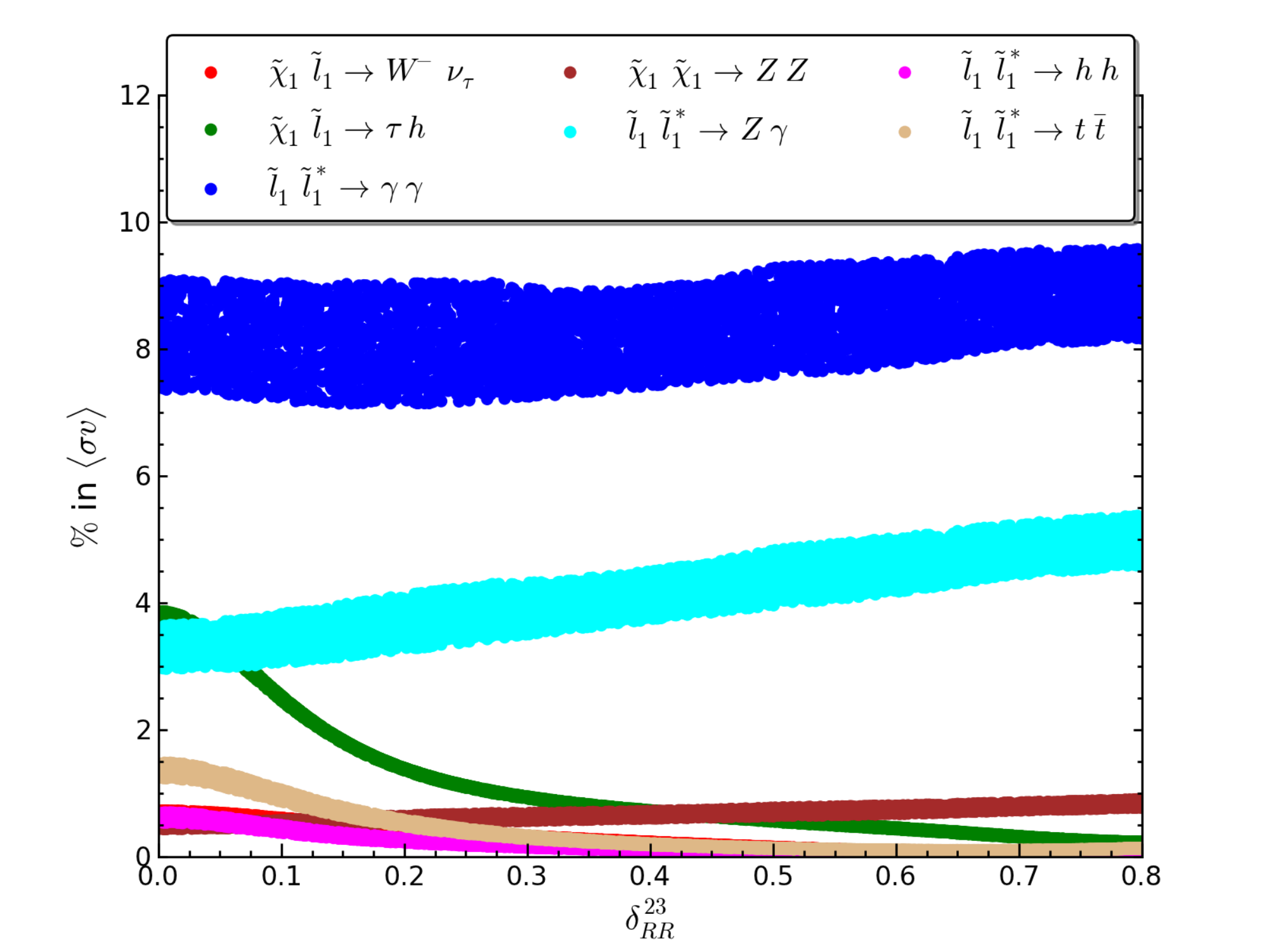} 
\end{tabular}
\end{center}
\caption{{\bf Channels in $\delta$-mSUGRA:} The colored dots show relative contribution of a particular channel to $\sigmav_{tot}$. $M_{1/2} = 500$ GeV and $m_0$ and $\delta$ are varied to fit the co-annihilation condition. Here all the points satisfy WMAP $3\sigma$ bound (\ref{omlim}).  For the above plots $\tan\beta$ is fixed to 20 and sign$(\mu) > 0$. Flavor violating constraints are not imposed here.}
\label{sigvmsug}
\end{figure}

The individual scattering processes involved in the computation of thermally averaged cross-section are called channels. The typical channels which are dominant in the co-annihilation region are $\lstau \lstau \rightarrow l \bar{l}$, $\chiz \lstau \rightarrow \gamma l$, $\chiz \lstau \rightarrow Z l$, $\chiz \chiz \rightarrow l \bar{l}$, $\lstau \lstau^{*} \rightarrow l \bar{l}$ etc. (they are about thirty of them in total). In the presence of flavor violation the number of these processes
would be enlarged to include flavor violating final states. In the present section, we analyze the relative importance of the new flavor violating channels with the
corresponding flavor conserving ones as a function of $\delta$. To, do this we fix $M_{{1 \over 2}}$ and vary $\delta$ and $m_0$. In effect, this corresponds
to the combination of horizontal sections of the co-annihilation regions of all the panels in Fig .(\ref{brmsugra}) (Fig. (\ref{nuhmdelta})) for $\delta$-mSUGRA 
($\delta$-NUHM). 
 In Fig. (\ref{sigvmsug}) we plot the dominant channels as a function of $\delta$ in  $\delta$-mSUGRA. All the points satisfy relic density within WMAP $3\sigma$ bound and lie in the co-annihilation region. Rest of the phenomenological constraints are also imposed. $m_{0}$ is varied from 100 to 600 GeV, whereas $M_{1/2}$ is fixed at 500 GeV, $\tan\beta = 20$ and sign$(\mu) > 0$. The Y-axis is percentage contribution to the thermally averaged cross section, $\langle \sigma v \rangle$ defined by
\begin{align}
\% \; \langle \sigma v \rangle_{ij \rightarrow m n} = \frac{\langle \sigma v \rangle_{ij \rightarrow m n}}{\langle \sigma v \rangle_{total}} \times 100
\end{align}

\begin{table}[htdp]
\caption{$\dfrac{\langle \sigma v \rangle_{channel}}{\langle \sigma v \rangle_{total}}$ for dominant channels for $\delta$-mSUGRA}
\label{sigvnumsug}
\begin{center}
\begin{tabular}{c p{3cm} p{3cm} p{3cm}}
\hline
Parameters & \raggedright{\bf Point I} $M_{\half} = 500.0$ GeV, $\tan\beta = 20$, $m_{0} = 165.6$ GeV & \raggedright{\bf Point II} $M_{\half} = 500.0$ GeV, $\tan\beta = 20$, $m_{0} = 169.6$ GeV & \raggedright{\bf Point III} $M_{\half} = 500.0$ GeV, $\tan\beta = 20$, $m_{0} = 249.0$ GeV  \tabularnewline \hline
$\delta$ & 0.197 & 0.202 & 0.5 \\ 
$\Omega h^{2}$ & 0.0910 & 0.119 & 0.120 \\
$\csttgl$ & 0.206 & 0.227  & 0.181 \\ 
$\csttgm$ & $6.53 \times 10^{-2}$ & $7.47 \times 10^{-2} $ & 0.13 \\ 
$\ststttt$ & 0.211 & 0.181 & 0.116 \\ 
$\ststttm$ & 0.130 & 0.117 & 0.165 \\ 
$\ststtmm$ & $2.10 \times 10^{-2}$ & $1.97 \times 10^{-2}$ & $5.97 \times 10^{-2}$ \\ 
$\ststbtgg$ & 0.110 & $9.65 \times 10^{-2}$ & $9.93 \times 10^{-2}$\\ 
$\csttzl$ & $5.67 \times 10^{-2}$ & $6.23 \times 10^{-2}$  & $4.96 \times 10^{-2}$\\ 
$\csttzm$ & $1.76 \times 10^{-2}$ & $2.02 \times 10^{-2}$  & $3.53 \times 10^{-2}$\\ 
$\ststbtzg$ & $5.00 \times 10^{-2}$  & $4.42 \times 10^{-2}$ & $5.18 \times 10^{-2}$\\ 
$\cctttb$ & $2.02 \times 10^{-2}$  & $2.81 \times 10^{-2}$ & $2.27 \times 10^{-2}$\\ 
$\ccttmb$ & $6.76 \times 10^{-3}$ & $9.50 \times 10^{-3}$ & $8.29 \times 10^{-3}$ \\ 
$\cctmmb$ & $1.73 \times 10^{-2}$  & $2.42 \times 10^{-2}$ & $1.80 \times 10^{-2}$ \\ \hline 
\end{tabular}
\end{center}
\end{table}%


It should be noted that flavor violating constraints are not imposed for $\delta$-mSUGRA in this analysis. The current limits on BR($\tau \to \mu + \gamma$) 
constraint $|\delta| \lesssim  0.11$ in the parameter space presented in the figure.  For those values of $\delta$ we see that the flavor violating channels
contribute up to  $5\%$ of  the dominant channel contribution.  Larger values of $\delta$ are not allowed after the imposition of this constraint as there is
 no overlap between cancellation regions and co-annihilation regions in $\delta$-mSUGRA.  However, to study the features of the channels with 
 respect to $\delta$ it would be useful not to impose the BR($\tau \to \mu + \gamma$)  constraint for the present. 
\begin{figure}[htp]
\begin{center}
\begin{tabular}{cc}
\includegraphics[width=0.50\textwidth,angle=0]{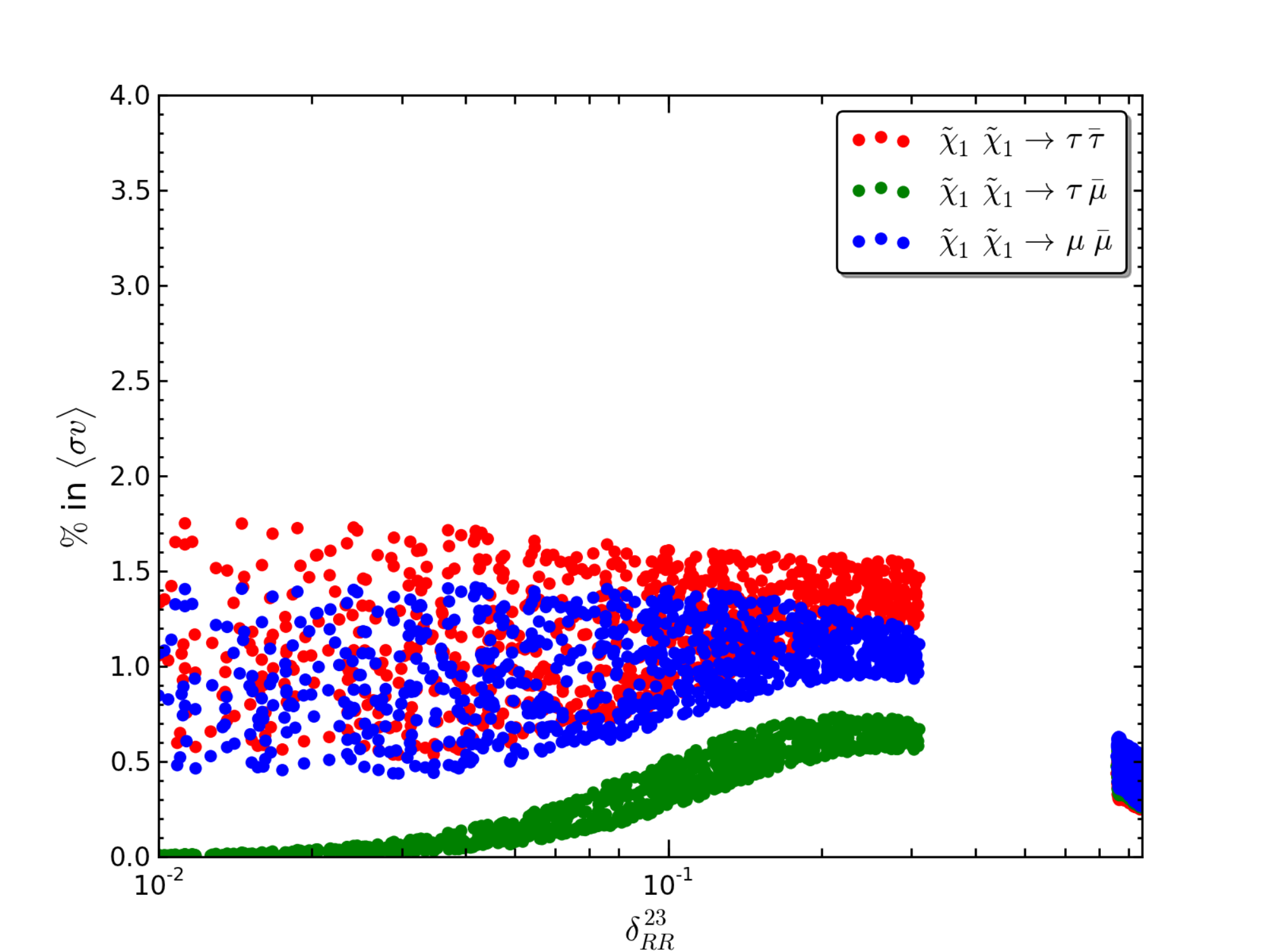} &
\includegraphics[width=0.50\textwidth,angle=0]{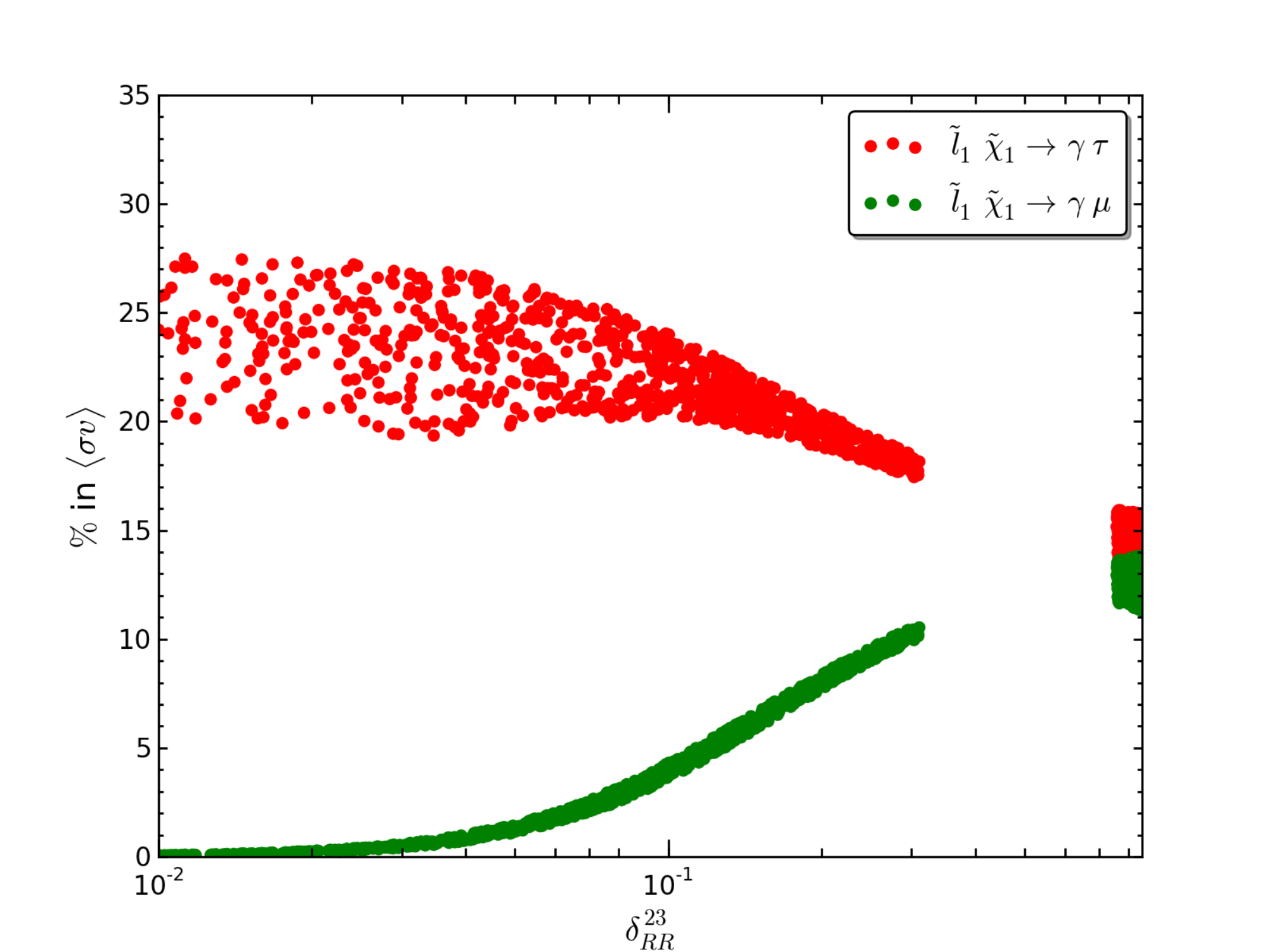} \\
\includegraphics[width=0.50\textwidth,angle=0]{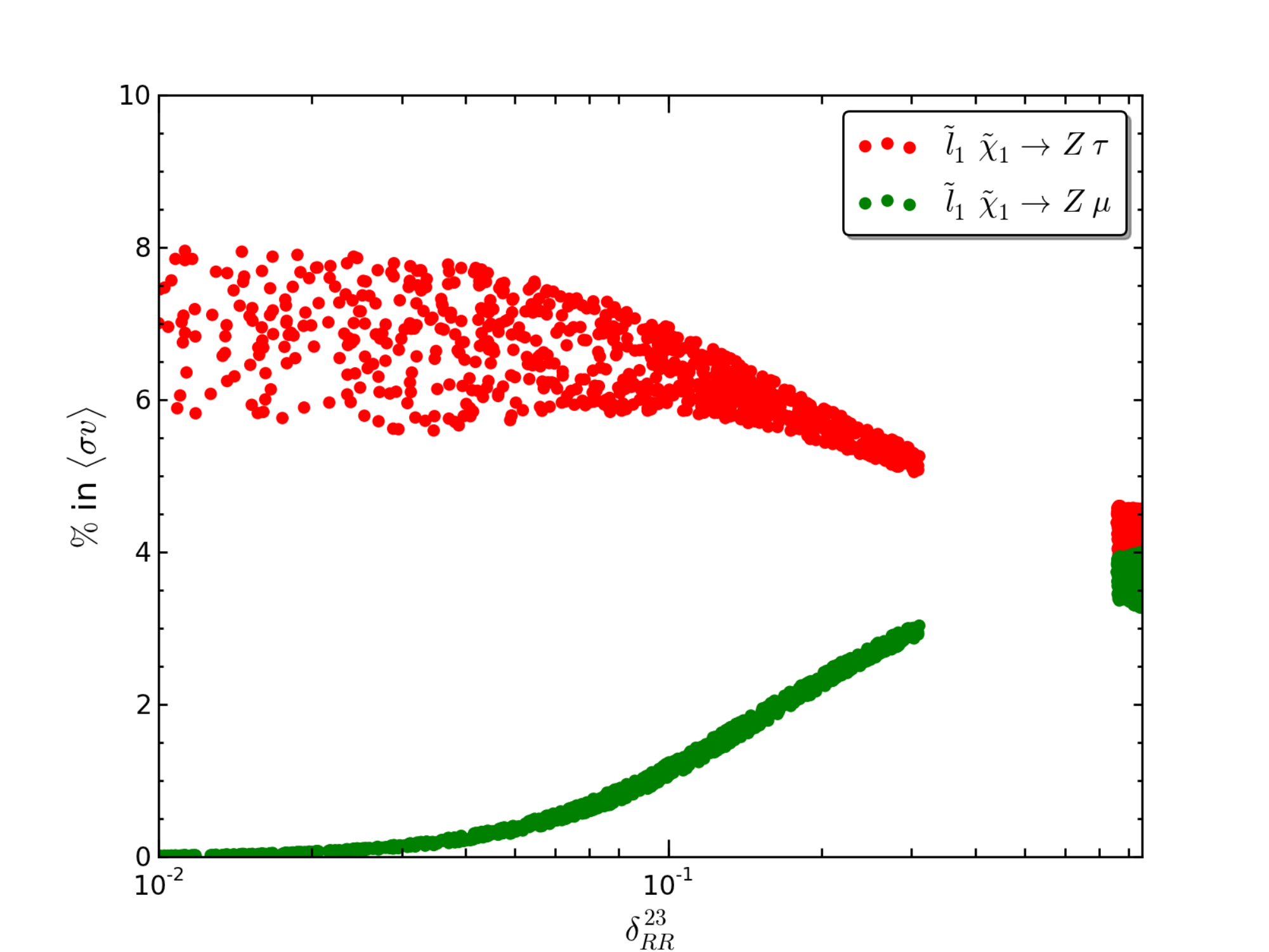} &
\includegraphics[width=0.50\textwidth,angle=0]{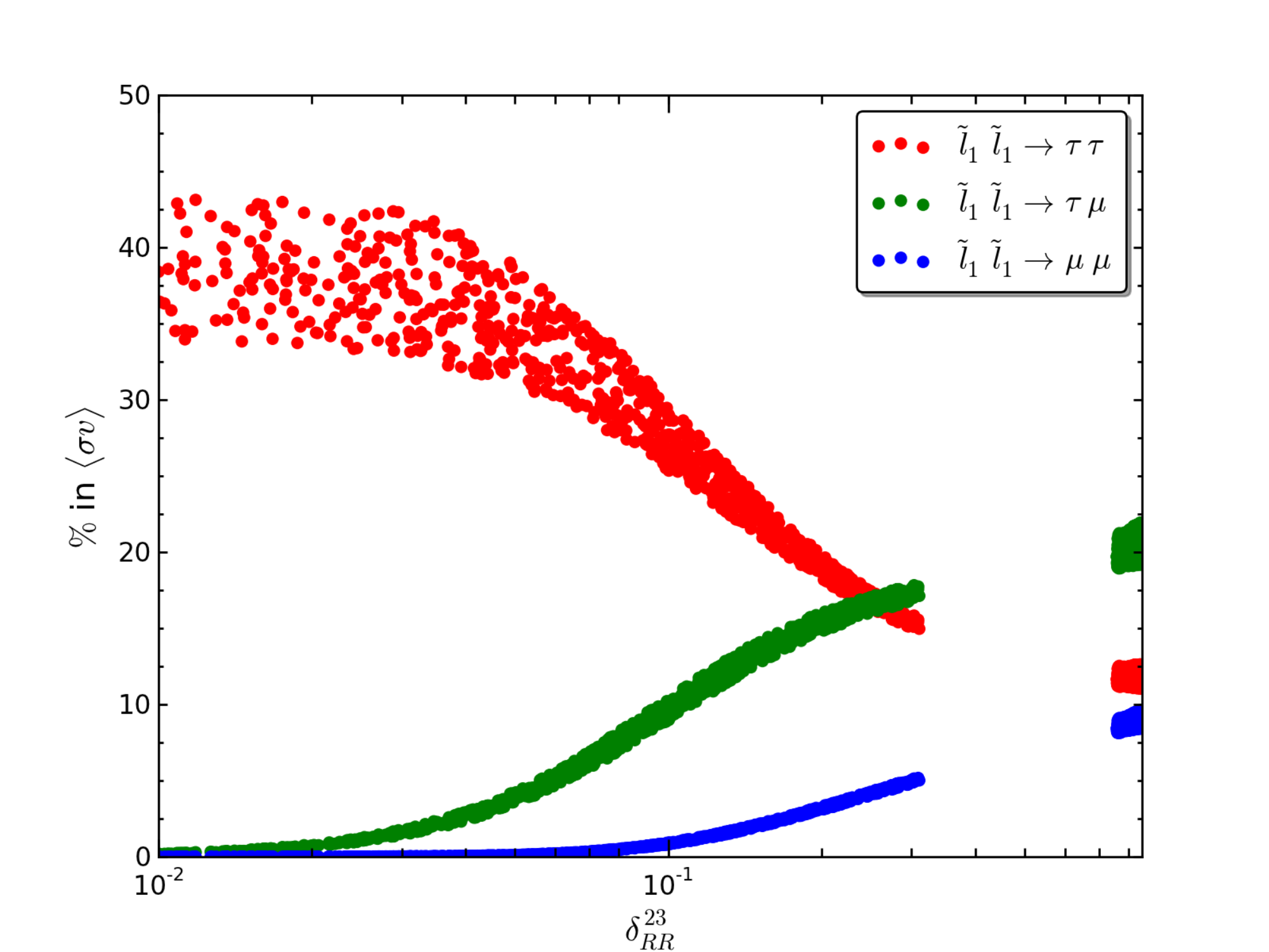} \\
\includegraphics[width=0.50\textwidth,angle=0]{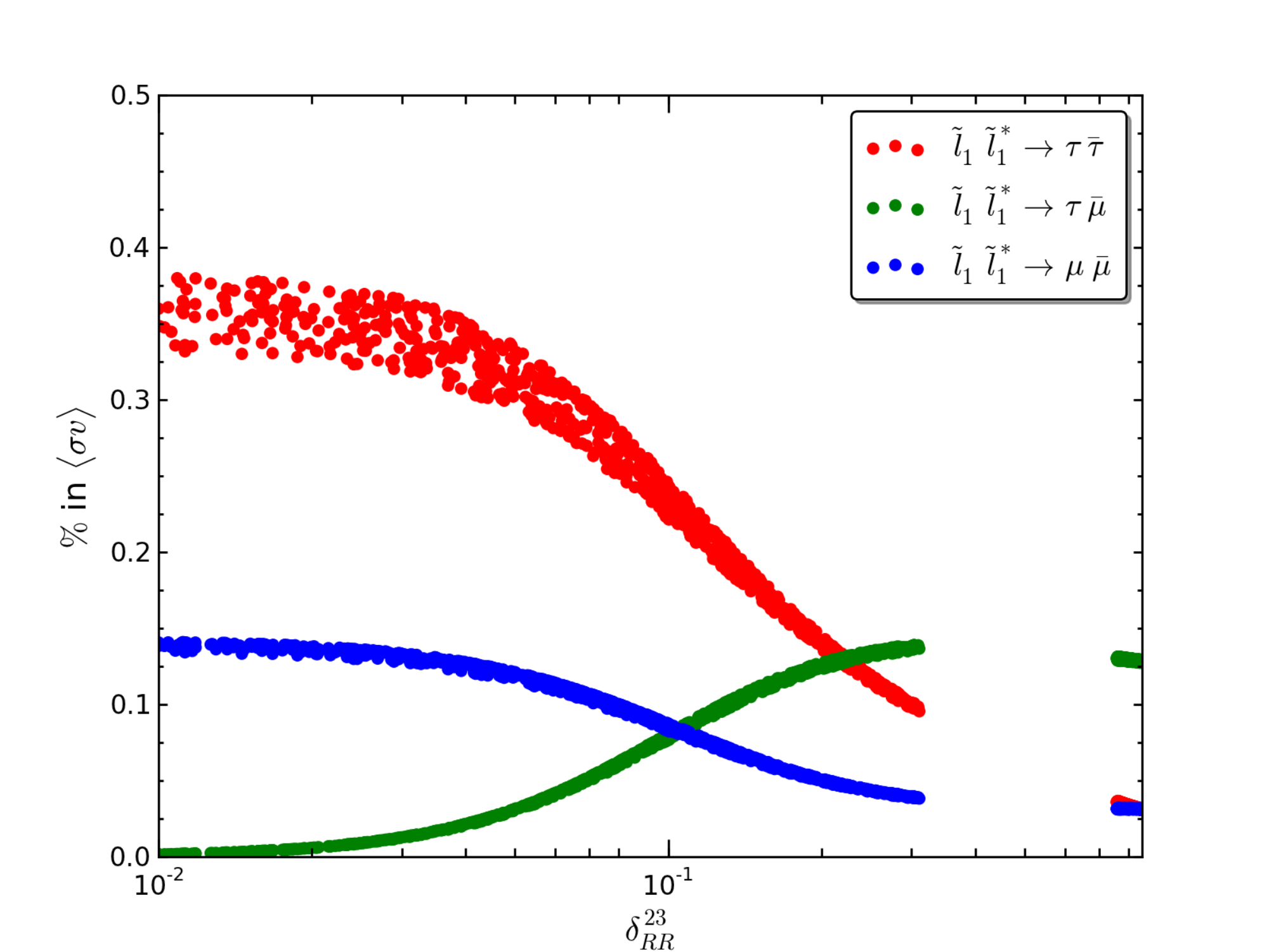} &
\includegraphics[width=0.50\textwidth,angle=0]{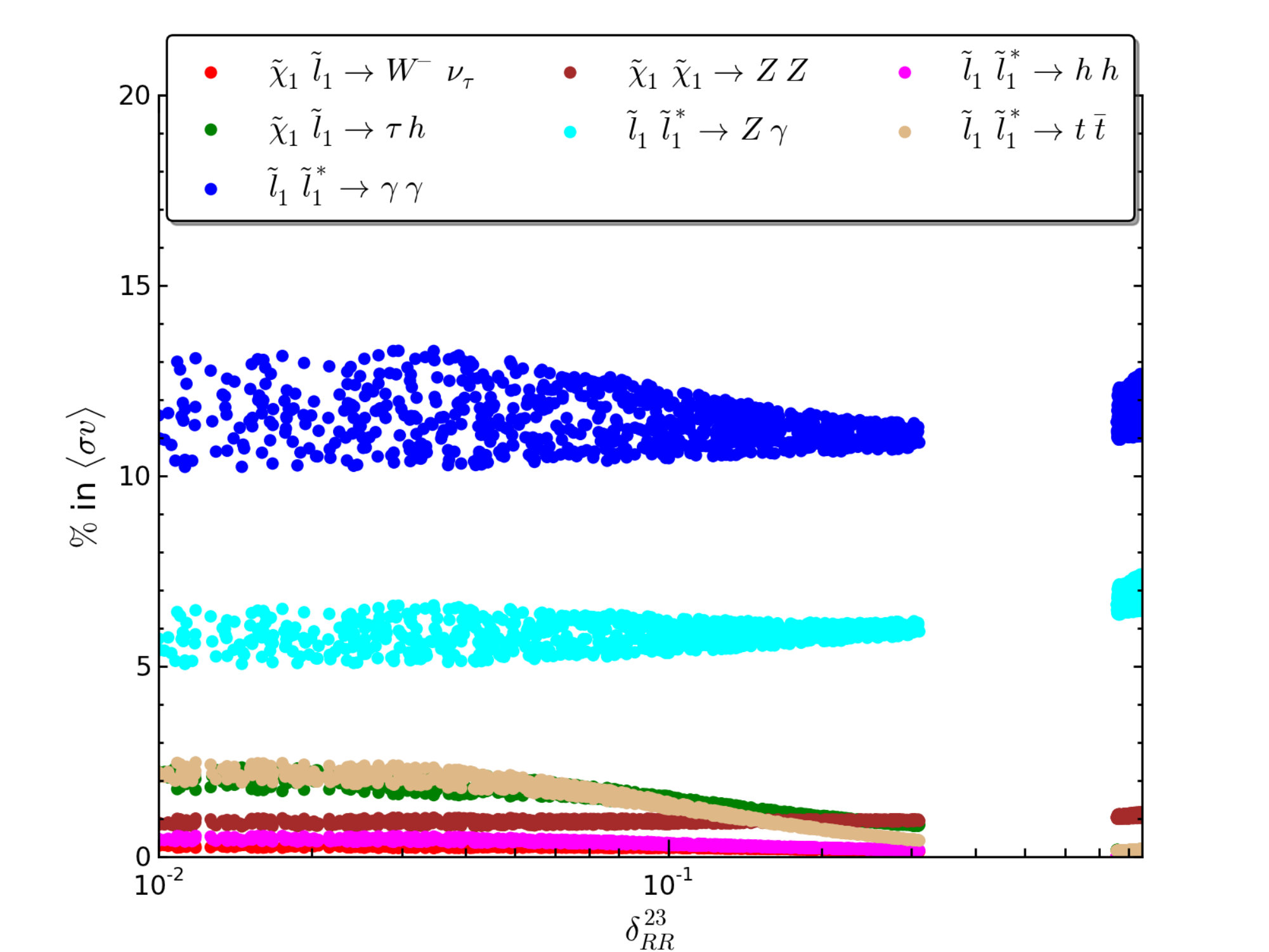} 
\end{tabular}
\end{center}
\caption{{\bf Channels in $\delta$-NUHM:}  The colored dots show relative contribution of a particular channel to $\sigmav_{tot}$. $M_{1/2} = 750 $ GeV and $m_0$ and $\delta$ are varied to fit the co-annihilation condition. Here all the points satisfy WMAP $3\sigma$ bound (\ref{omlim}).  For the above plots $\tan\beta$ is fixed to 20 and sign$(\mu) > 0$. Flavor violating constraints are imposed here, which causes the discontinuous regions in each of the channels.}
\label{sigvnuhm}
\end{figure}

\begin{table}[htdp]
\caption{$\dfrac{\langle \sigma v \rangle_{channel}}{\langle \sigma v \rangle_{total}}$ for dominant channels for $\delta$-NUHM}
\label{sigvnumnuhm}
\begin{center}
\begin{tabular}{c p{3cm} p{3cm} p{3cm}}
\hline
Parameters &  \raggedright{\bf Point IV} $M_{\half} = 750.0$ GeV, $\tan\beta = 20$, $m_{0} = 199.3$ GeV & \raggedright{\bf Point V} $M_{\half} = 750.0$ GeV, $\tan\beta = 20$, $m_{0} = 216.0$ GeV & \raggedright{\bf Point VI} $M_{\half} = 750.0$ GeV, $\tan\beta = 20$, $m_{0} = 592.1$ GeV  \tabularnewline \hline
$\delta$ & 0.01 & 0.12 & 0.767 \\ 
$\Omega h^{2}$ & 0.115 & 0.116 & 0.111 \\
$\csttgl$ & 0.190 & 0.168  & 0.116 \\ 
$\csttgm$ & $4.74 \times 10^{-4} $ & $3.89 \times 10^{-2}$ & $9.89 \times 10^{-2}$ \\ 
$\ststttt$ & 0.388 & 0.280 & 0.134 \\ 
$\ststttm$ & $1.90 \times 10^{-3}$ & 0.127 & 0.227 \\ 
$\ststtmm$ & $2.39 \times 10^{-6}$ & $1.48 \times 10^{-2}$ & $9.37 \times 10^{-2}$ \\ 
$\ststbtgg$ & $0.115$ & 0.123 & $0.129$\\ 
$\csttzl$ & $5.50 \times 10^{-2}$ & $4.88 \times 10^{-2}$ & $3.35 \times 10^{-2}$\\ 
$\csttzm$ & $2.02 \times 10^{-6}$ & $1.11 \times 10^{-2}$ & $2.28 \times 10^{-2}$\\ 
$\ststbtzg$ & $5.67 \times 10^{-2}$ & $6.36 \times 10^{-2}$ & $7.49 \times 10^{-2}$\\ 
$\cctttb$ & $1.14 \times 10^{-2}$ & $1.13 \times 10^{-2}$ & $3.72 \times 10^{-3}$\\         
$\ccttmb$ & $2.80 \times 10^{-5}$ & $1.77 \times 10^{-3}$ & $3.53 \times 10^{-3}$ \\ 
$\cctmmb$& $9.53 \times 10^{-3}$ & $9.87 \times 10^{-3}$ & $4.49 \times 10^{-3}$ \\ \hline 
\end{tabular}
\end{center}
\end{table}%

The upper left panel shows the \% $\sigmav$ for $\chiz \chiz \rightarrow l \bar{l}$, which contributes about $\lesssim 5 \%$ total to $\sigmav$ in this region of parameter space. In this case, the initial state masses are independent of $\delta$ and $m_{0}$, and thus, the only variation comes from the mass of intermediate state particle $(\lstau)$. In Table (\ref{sigvnumsug}), we presented the sample points which are represented in the plot. From the points,  I and II of table \ref{sigvnumsug}, we see that a slight shift of 5 GeV in $m_{0}$ is still allowed by WMAP $3\sigma$ limits, which changes the $\chiz\chiz$ cross-section by about $40\%$. This is the reason why the band of allowed  points is broad in this channel. Other dominant channels are represented in subsequent panels of the figure. From the panel it is obvious that the dominant contribution comes from $\csttgl$ and $\ststttt$ channels. Each of which contribute to about $35\%$ and $25\%$ respectively to $\sigmav$. Most of the flavor violating counterparts of these channels behave as expected, i.e. at large $\delta$, they become comparable to the flavor conserving ones. One exception of this is the $\lstau\lstau$ channel. Here the initial state composition crucially depends on `$\delta$' and also on $\staur_{L}\staur_{R}$ mixing. In such a situation, its clear that the initial state cannot be attributed any flavor quantum number. In fact we find that the $\tilde{\mu}$ (smuon) component of $\tilde{l}_1$ can be large $\sim 50\%$ even for $\delta \approx 0.2$ in some regions of parameter space. We see from the figure that the flavor violating final states dominates over the flavor conserving ones, as $\delta$ grows beyond $\delta \geqslant 0.2$. The exact point of crossing of the flavor violating channels over flavor conserving ones is dependent on the parameter space chosen, crucially on $\tan \beta$ and $\mu$. This is because the effective $\smu_{L}\staur_{R}$ and/or $\smu_{R}\staur_{L}$ coupling generated play an important role in determining the initial state composition. The last two panels shows some of the channels, which contribute negligibly to the $\sigmav$. In Appendix [\ref{appD}] we have given approximate formulae in $m_{\tau}/m_{\mu}\rightarrow 0$ limit for  the dominant cross-sections.  Using these and approximate formulae presented in appendix \ref{appendix1}  features of full numerical analysis can be verified.  More detailed  analysis  of cross-sections in the presence of flavor violation is various dark matter allowed regions will be presented elsewhere \cite{upcoming}. 

\begin{figure}[htp]
\begin{center}
\includegraphics[width=0.9\textwidth,angle=0]{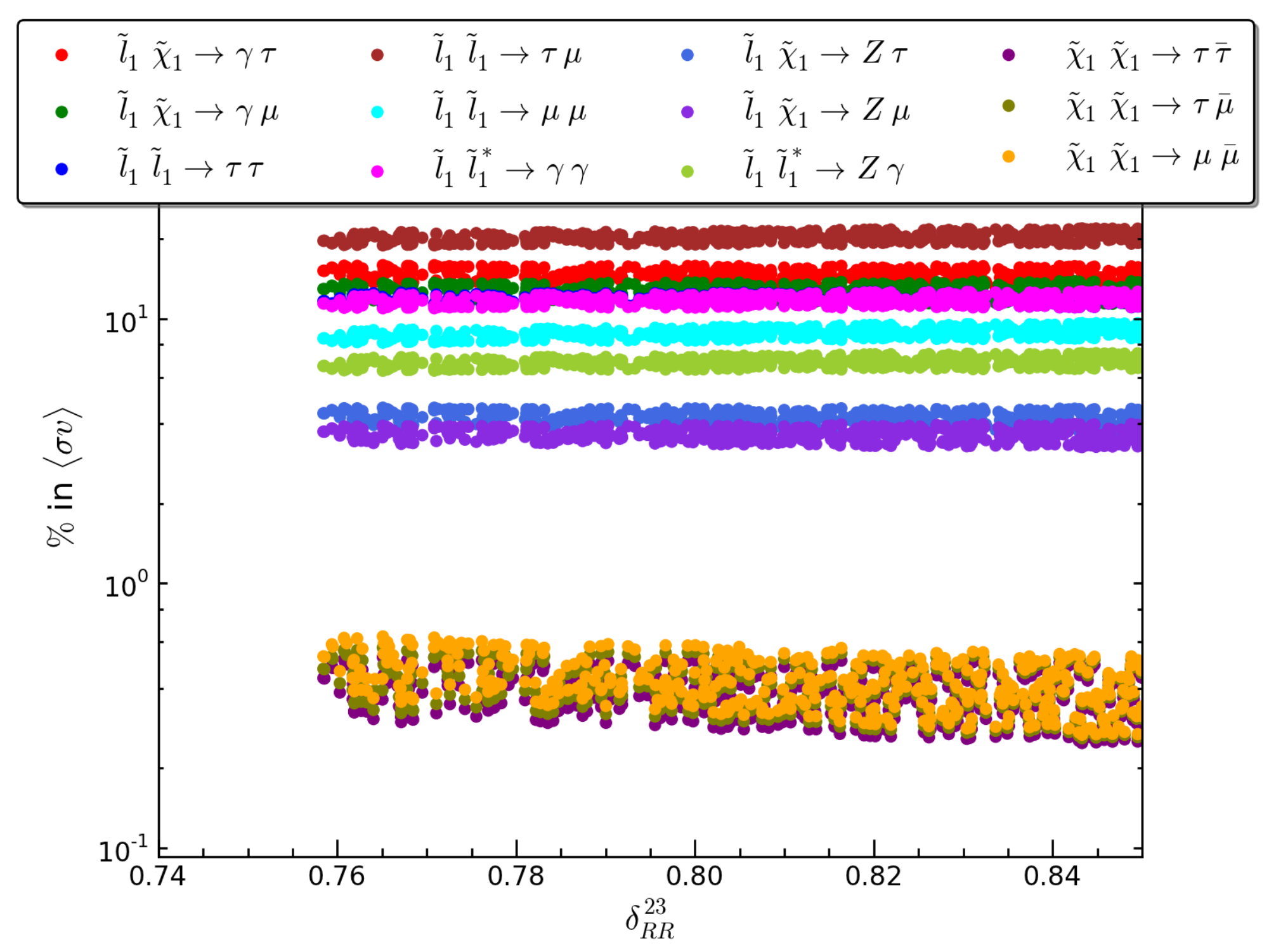}
\end{center}
\caption{Dominant Channels contribution to the $\sigmav_{tot}$. Here $\tan\beta$ is fixed to 20, $M_{1/2} = 750$ GeV and $m_0$ and $\delta$ are varied to fit the co-annihilation condition. Here all the points satisfy WMAP $3\sigma$ bound (\ref{omlim}).}
\label{sigvmag}
\end{figure}

In Fig. \ref{sigvnuhm}, we present similar plots form channels in $\delta$-NUHM case for the parametrization chosen in the previous section. Here we have imposed BR$(\tau \rightarrow \mu \gamma) \leqslant 4.4 \times 10^{-8}$ to be satisfied along with relic density constraints. These channel show a similar pattern here as in $\delta$-mSUGRA. However, as we can see from the panels, there is a gap between $\delta = 0.2$ to $\delta = 0.7$ where the parameter space does not satisfy BR$(\tau \rightarrow \mu \gamma) \leqslant 4.4 \times 10^{-8}$. 
For points below $\delta \leqslant 0.2$, this constraint is satisfied as `$\delta$' is too small to generate appreciable $\tau \to \mu + \gamma$ amplitudes.
 For $\delta \geqslant 0.7$, the constraint is now satisfied because of the overlap between the cancellation regions and co-annihilation regions. The relative contribution in the overlap region is magnified in Fig. \ref{sigvmag} where all the channels contributions are presented between $0.70 \leqslant \delta \leqslant 0.85$. As we can
 see,  flavor violating channels strongly compete with flavor conserving ones. A sample of the points in $\delta$-NUHM is presented in Table \ref{sigvnumnuhm}, where points IV and V represent low $\delta$ values whereas point VI represent the large $\delta$ value signifying overlapping regions. 

Finally a note about relative contribution to relic density. We have 
\begin{align}
\Omega h^{2} &\propto \frac{1}{\sigmav_{tot}} = \frac{1}{\displaystyle \sum_{\text{all channels}} \sigmav_{i}}  \notag \\
& \propto \frac{1}{\sigmav_{tot} \displaystyle \sum_{\text{all channels}} \frac{\sigmav_{i}}{\sigmav_{tot}}}
\end{align}
For small $\delta$ $\left(\sim \mathcal{O}\left(10^{-2}\right)\right)$ where $\sigmav$ contribution to flavor violating channel is small, the estimate of relic density does not modify much from the flavor conserving case. However for large enough $\delta$ $\left(\sim \mathcal{O}\left(10^{-1}\right)\right)$, one tends to overestimate relic density, if one does not consider flavor violating scatterings while computing the thermally averaged cross-section.

\section{Summary and  Outlook}
\label{sec5}

We have generalized the co-annihilation process by including flavor violation in the sleptonic $\mu-\tau$ (RR) sector. The amount
of flavor violation admissible is constrained to be small by the limit on the BR($\tau \to \mu + \gamma$). This constraint is significantly
weakened in regions of the parameter space where cancellations in the amplitudes takes place. We look for regions of  parameter
space where there is a significant overlap between cancellation regions and co-annihilation regions. The search is done in mSUGRA
and NUHM augmented with one single flavor violating parameter in the $\mu -\tau$ (RR) sector. We found that while no significant
overlap is possible in $\delta$-mSUGRA,  $\delta$-NUHM allows for large regions where significant overlap is possible.   

The presence of flavor violation  shifts the lightest slepton co-annihilation regions towards lighter neutralino masses compared to 
mSUGRA. While computing the thermally averaged cross-sections in the overlap regions, we found that  flavor violating processes
could contribute with equal strength and in some cases even dominantly compared to the flavor conserving ones. This is true even for 
$\delta \gtrsim 0.2$ in some regions of the parameter space.  Neglecting the flavor violating
 channels would lead to underestimating the cross section and thus in overestimating  the relic density.  A point to note is that if
 flavor violation is present even within the presently allowed limits, it could still change the dominant channels by about $5\%$ in
 $\delta$-mSUGRA and more in $\delta$-NUHM. 
  Finally, We have probed only  a minor region of the parameter space in the present work demonstrating the existence of such
   regions. A comprehensive  analysis of such regions and the associated phenomenology of their spectrum would be interesting 
   in their own right. 

In this respect, a few comments on flavor violation at the LHC and ILC  are in order. 
Detection of lepton flavor violation at the colliders like LHC is strongly constrained by experimental limits on rare lepton flavor violating decays. 
One standard technique to detect flavor violation at colliders is to study the slepton mass differences using end-point kinematics of cascade decays \cite{Hinchliffe:2000np}. The typical sensitivity being discussed in the literature is $\frac{\Delta m_{\tilde{l}}}{m_{\tilde{l}}}(l_i,l_j) \, = \, \frac{|m_{\tilde{l}_i}- m_{\tilde{l}_j} |}{\sqrt{m_{\tilde{l}_i}m_{\tilde{l}_j}}}\, \simeq \, \mathcal{O}(0.1 )\%$ for $\tilde{e}_L-\tilde{\mu}_L$ and $\mathcal{O}(1)\%$ for  $\tilde{\mu}_L-\tilde{\tau}_L$ \cite{Allanach:2008ib}. In the presence of  $\Delta^{\mu\tau}_{RR}$
splittings are generated in all the three eigenvalues \cite{calibbi2}, $e-\mu\, , \mu-\tau \, , e-\tau$ sectors. 
In the case discussed in this work, the typical splittings are $\mathcal{O}(20) \%$ to $\mathcal{O}(70) \%$ as the constraints from LFV experiments are evaded. Thus, far less sensitivity is required to measure these splittings compared to the regular case.  Further investigations in this direction are however needed. Another interesting aspect of this scenario would be to measure  widths for LFV decay processes like $\tilde{\chi}_2^0 \rightarrow \tilde{\chi}_1^0 l_i^{\pm}l_j^{\mp}$. These widths have been studied for the case of right handed slepton flavor violation in \cite{bartl}. In NUHM, with a comparatively smaller value of $\mu$ one could expect large production cross sections for $\tilde{\chi}_4^0 $ and  $\tilde{\chi}_2^{\pm} $  in the decays of colored particles. In fact, a full Monte Carlo study has been reported by Hisano et al. \cite{Hisano:2008ng} for a particular parameter space point in the model.

At the linear collider, it should be possible to identify the $\tilde{\tau}$ co-annihilation region \cite{Nojiri:1994it,Nojiri:1996fp,Guchait:2002xh,Hamaguchi:2004df,Godbole:2008it} by studying  the polarization of the decay  $\tilde{\tau}_1\rightarrow \tilde{\chi}_1^0 \tau$. In the presence of flavored co-annihilations one should be able to see flavor violating decays of $\tilde{\tau}_1$.  Heavier particles like $\tilde{\tau}_2$ and charginos would also have  flavor violating decays.

Finally lets note that we have considered the cancellations in the dipole operator of the $\tau\,\rightarrow\,\mu$ transitions, it does not guarantee us suppression in amplitudes associated with other operators. For example, in this
region $\tau\,\rightarrow\,\mu\,\eta$ or $\tau\,\rightarrow\,\mu\,\eta^{\prime}$ could be sizable ($\sim 10^{-9}-10^{-10}$) \cite{Brignole:2004ah}, which could be probed in future B-factories. Whereas, $\tau\,\rightarrow\,\mu\,\gamma$ will continue to remain constrained and thus will not be detected.

The focus of the present work has been to introduce new regions of parameter space where flavor effects in the co-annihilation  regions could be important. More generally flavor effects could play a role in any dark matter `regions' of the  SUSY parameter space. Such studies are being explored in \cite{upcoming}.
 
\acknowledgments
We thank Ranjan Laha for participating in this project at the initial stages. We also thank  Yann Mambrini, 
Utpal Chattopadhyay and  Alexander Pukhov for discussions and useful inputs. SKV acknowledges support from 
DST project ``Complementarity between direct and indirect  searches for Supersymmetry" and also  support from 
DST Ramanujan Fellowship SR/S2/RJN-25/2008. RG acknowledges support from 
SR/S2/RJN-25/2008. DC acknowledges partial support from SR/S2/RJN-25/2008. 

\appendix
\section{Approximate Solutions}
\label{appendix1}
\subsection{mSUGRA Case}
\label{appA1}
In the approximation of small Yukawa couplings, we retain only $Y_{t},\,Y_{b},\,Y_{\tau}$ and solve the RGEs semi-analytically. For the first two generations of the particles the dependence on $\tan\beta$  is very weak, so we take them to be valid for all $\tan\beta$. In deriving the approximate expressions we have taken $m_{t}(M_{Z}) = 165 {\rm ~GeV}$, $m_{b}(M_{Z}) = 3 {\rm ~GeV}$ and $m_{\tau}(M_{Z}) = 1.77 {\rm ~GeV}$. For $\tan\beta = 5$,  the first two generation masses at the weak scale are
\begin{align} 
(m^{2}_{Q})_{1,2}(M_{Z}) \;\simeq\quad &\, m_{0}^2 + 6.66\,M_{\half}^2 \\ 
(m^{2}_{D})_{1,2}(M_{Z}) \;\simeq\quad &\, m_{0}^2 + 6.19\,M_{\half}^2 \\
(m^{2}_{U})_{1,2}(M_{Z}) \;\simeq\quad &\, m_{0}^2 +  6.22\,M_{\half}^2 
\end{align}
\begin{align} 
(m^{2}_{L})_{1,2}(M_{Z}) \;\simeq\quad & \, m_{0}^2 + 0.51\,M_{\half}^2 \\
(m^{2}_{E})_{1,2}(M_{Z}) \;\simeq\quad &\,  m_{0}^2 +  0.17\,M_{\half}^2 
\end{align}
Third generation masses strongly depend on $\tan\beta$ than the first two generations. For low $\tan\beta = 5$ their values are as follows
\begin{align}
(m^{2}_{Q})_{3}(M_{Z}) \simeq &- 0.036\, A_{0}^2 + 0.65\,m_{0}^2 + 0.16\, A_{0} M_{\half} + 5.66\,M_{\half}^2 \\
(m^{2}_{U})_{3}(M_{Z}) \simeq &- 0.070\, A_{0}^2 + 0.31\,m_{0}^2 + 0.30\, A_{0} M_{\half} + 4.26\,M_{\half}^2 \\
(m^{2}_{D})_{3}(M_{Z}) \simeq &- 1.70 \times 10^{-3}\, A_{0}^2 + m_{0}^2 + 7.23\times 10^{-3}\, A_{0} M_{\half} + 6.17\,M_{\half}^2 \\
(m^{2}_{L})_{3}(M_{Z}) \simeq &- 7.34 \times 10^{-4}\, A_{0}^2 + m_{0}^2 + 6.29 \times 10^{-4}\, A_{0} M_{\half} + 0.51\,M_{\half}^2 \\
(m^{2}_{E})_{3}(M_{Z}) \simeq &- 1.47 \times 10^{-3}\, A_{0}^2 + m_{0}^2 + 1.26 \times 10^{-3}\, A_{0} M_{\half} + 0.16\,M_{\half}^2\\
m_{H_d}^{2}(M_{Z}) \simeq &- 3.30\times 10^{-3}\, A_{0}^{2} + 0.99\, m_{0}^{2} + 0.01\, A_{0}M_{\half} + 0.48\, M_{\half}^{2} \\ 
m_{H_u}^{2}(M_{Z}) \simeq &- 0.105\, A_{0}^{2} - 0.046\, m_{0}^{2} + 0.46\, A_{0}M_{\half} - 2.95\, M_{\half}^{2} \\ 
|\mu|^{2}(M_{Z}) = &-4158.72 + 0.110\, A_{0}^{2} + 0.084\, m_{0}^{2} - 0.47\, A_{0}M_{\half} + 3.09\, M_{\half}^{2} 
\end{align}
For medium $\tan\beta = 20$ their values are as follows
\begin{align}
(m^{2}_{Q})_{3}(M_{Z}) \simeq &- 0.048\, A_{0}^2 + 0.62\,m_{0}^2 + 0.20\, A_{0} M_{\half} + 5.54\,M_{\half}^2 \\
(m^{2}_{U})_{3}(M_{Z}) \simeq &- 0.070\, A_{0}^2 + 0.33\,m_{0}^2 + 0.30\, A_{0} M_{\half} + 4.32\,M_{\half}^2 \\
(m^{2}_{D})_{3}(M_{Z}) \simeq &- 0.023\, A_{0}^2 + 0.91\, m_{0}^2 + 0.10\, A_{0} M_{\half} + 5.86\,M_{\half}^2 \\
(m^{2}_{L})_{3}(M_{Z}) \simeq &- 0.011\, A_{0}^2 + 0.97\, m_{0}^2 + 8.38 \times 10^{-3}\, A_{0} M_{\half} + 0.50\,M_{\half}^2 \\
(m^{2}_{E})_{3}(M_{Z}) \simeq &- 0.021\, A_{0}^2 + 0.93\, m_{0}^2 + 0.017 \, A_{0} M_{\half} + 0.15\,M_{\half}^2\\
m_{H_d}^{2}(M_{Z}) \simeq &- 0.046\, A_{0}^{2} + 0.83\, m_{0}^{2} + 0.16\, A_{0}M_{\half} + 0.01\, M_{\half}^{2} \\
m_{H_u}^{2}(M_{Z}) \simeq &- 0.105\, A_{0}^{2} - 0.007\, m_{0}^{2} + 0.46\, A_{0}M_{\half} - 2.86\, M_{\half}^{2} \\ 
|\mu|^{2}(M_{Z}) = &-4158.72 + 0.106\, A_{0}^{2} + 0.009\, m_{0}^{2} - 0.46\, A_{0}M_{\half} + 2.87\, M_{\half}^{2}
\end{align}
For high $\tan\beta = 35$ their values are as follows
\begin{align}
(m^{2}_{Q})_{3}(M_{Z}) \simeq &- 0.058\, A_{0}^2 + 0.53\, m_{0}^2 + 0.25\, A_{0} M_{\half} + 5.26\,M_{\half}^2 \\
(m^{2}_{U})_{3}(M_{Z}) \simeq &- 0.064\, A_{0}^2 + 0.33\, m_{0}^2 + 0.27\, A_{0} M_{\half} + 4.35\,M_{\half}^2 \\
(m^{2}_{D})_{3}(M_{Z}) \simeq &- 0.052\, A_{0}^2 + 0.727\, m_{0}^2 + 0.23\, A_{0} M_{\half} + 5.26\,M_{\half}^2 \\
(m^{2}_{L})_{3}(M_{Z}) \simeq &- 0.027\, A_{0}^2 + 0.89\, m_{0}^2 + 0.02\, A_{0} M_{\half} + 0.49\,M_{\half}^2 \\
(m^{2}_{E})_{3}(M_{Z}) \simeq &- 0.055 \, A_{0}^2 + 0.78\, m_{0}^2 + 0.03 \, A_{0} M_{\half} + 0.12\,M_{\half}^2 \\
m_{H_d}^{2}(M_{Z}) \simeq &- 0.105\, A_{0}^{2} + 0.48\, m_{0}^{2} + 0.36\, A_{0}M_{\half} - 0.91\, M_{\half}^{2} 
\end{align}
\begin{align}
m_{H_u}^{2}(M_{Z}) \simeq &- 0.095\, A_{0}^{2} - 0.005\, m_{0}^{2} + 0.41\, A_{0}M_{\half} - 2.81\, M_{\half}^{2} \\ 
|\mu|^{2}(M_{Z}) = &-4158.72 + 0.095\, A_{0}^{2} + 0.005\, m_{0}^{2} - 0.41\, A_{0}M_{\half} + 2.81\, M_{\half}^{2} 
\end{align}
\subsection{NUHM case}
\label{appA2}
In our notation $m_{10}= m_{H_{d}}(M_{GUT})$ and $m_{20} = m_{H_{u}}(M_{GUT})$. For $\tan\beta = 5$, at the weak scale the first two generation masses are
\begin{align} 
(m^{2}_{Q})_{1,2}(M_{Z}) \;\simeq &\quad m_{0}^2 + 6.66\,M_{\half}^2 + 0.009\, (m_{10}^2  - m_{20}^2) \\
(m^{2}_{D})_{1,2}(M_{Z}) \;\simeq &\quad m_{0}^2 + 6.19\,M_{\half}^2 + 0.018\, (m_{10}^2 - m_{20}^2)\\
(m^{2}_{U})_{1,2}(M_{Z}) \;\simeq &\quad m_{0}^2 + 6.22\,M_{\half}^2 - 0.036\, (m_{10}^2 - m_{20}^2)\\
(m^{2}_{L})_{1,2}(M_{Z}) \;\simeq &\quad m_{0}^2 + 0.51\,M_{\half}^2 - 0.027\, (m_{10}^2 - m_{20}^2)\\
(m^{2}_{E})_{1,2}(M_{Z}) \;\simeq &\quad m_{0}^2 + 0.17\,M_{\half}^2 + 0.053\, (m_{10}^2 - m_{20}^2)
\end{align}
Third generation masses strongly depend on $\tan\beta$ than the first two generations. For low $\tan\beta = 5$ their values are as follows
\begin{align}
(m^{2}_{Q})_{3}(M_{Z}) \simeq &- 0.036\, A_{0}^2 + 0.77\,m_{0}^2 + 0.16\, A_{0} M_{\half} + 5.66\,M_{\half}^2 + 7.90 \times 10^{-3}\, m_{10}^2  \nonumber \\ &- 0.125\, m_{20}^2 
\end{align}
\begin{align}
(m^{2}_{U})_{3}(M_{Z}) \simeq &- 0.070\, A_{0}^2 + 0.54\,m_{0}^2 + 0.30\, A_{0} M_{\half} + 4.26\,M_{\half}^2 - 0.035\, m_{10}^2 \nonumber \\ &- 0.196\, m_{20}^2 \\
(m^{2}_{D})_{3}(M_{Z}) \simeq &- 1.70 \times 10^{-3}\, A_{0}^2 + \,m_{0}^2 + 7.23 \times 10^{-3}\, A_{0} M_{\half} + 6.17\,M_{\half}^2 + 0.016\, m_{10}^2 \nonumber \\ &- 0.018\, m_{20}^2\\
(m^{2}_{L})_{3}(M_{Z}) \simeq &- 7.34 \times 10^{-4}\, A_{0}^2 + \,m_{0}^2 + 6.29 \times 10^{-4}\, A_{0} M_{\half} + 0.51\,M_{\half}^2 - 0.027\, m_{10}^2 \nonumber \\ &+ 0.027\, m_{20}^2 \\
(m^{2}_{E})_{3}(M_{Z}) \simeq &- 1.47 \times 10^{-3}\, A_{0}^2 + \,m_{0}^2 + 1.26\times 10^{-3}\, A_{0} M_{\half} + 0.16\,M_{\half}^2 + 0.052\, m_{10}^2 \nonumber \\ &- 0.053\, m_{20}^2 \\
m_{H_{d}}^{2}(M_{Z}) \simeq &- 3.30 \times 10^{-3}\, A_{0}^{2} - 7.32 \times 10^{-3}\, m_{0}^{2} + 0.01\, A_{0} M_{\half} + 0.48\, M_{\half}^{2} + 0.969\, m_{10}^{2} \nonumber \\ &+ 0.027\, m_{20}^{2}\\ 
m_{H_{u}}^{2}(M_{Z}) \simeq &- 0.105\, A_{0}^{2} - 0.70 \, m_{0}^{2} + 0.46\, A_{0}M_{\half} - 2.95\, M_{\half}^{2} + 0.027\, m_{10}^{2} \nonumber \\ &+ 0.625\, m_{20}^{2}\\ 
|\mu|^{2}(M_{Z}) = &-4158.72 + 0.110\, A_{0}^{2} + 0.72 \, m_{0}^{2} - 0.47\, A_{0}M_{\half} + 3.09 \, M_{\half}^{2} + 0.012\, m_{10}^{2} \nonumber \\ &- 0.650\, m_{20}^{2}
\end{align}
For medium $\tan\beta = 20$ their values are as follows
\begin{align}
(m^{2}_{Q})_{3}(M_{Z}) \simeq &- 0.048\, A_{0}^2 + 0.75\,m_{0}^2 + 0.20\, A_{0} M_{\half} + 5.54\,M_{\half}^2 - 6.30 \times 10^{-3}\, m_{10}^2 \nonumber\\ &- 0.120\, m_{20}^2\\
(m^{2}_{U})_{3}(M_{Z}) \simeq &- 0.070\, A_{0}^2 + 0.55\,m_{0}^2 + 0.30\, A_{0} M_{\half} + 4.32\,M_{\half}^2 - 0.034\, m_{10}^2 \nonumber\\ &- 0.190\, m_{20}^2\\
(m^{2}_{D})_{3}(M_{Z}) \simeq &- 0.023\, A_{0}^2 + 0.94\,m_{0}^2 + 0.10\, A_{0} M_{\half} + 5.86\,M_{\half}^2 - 0.015\, m_{10}^2 \nonumber\\ &- 0.015\, m_{20}^2 
\end{align}
\begin{align}
(m^{2}_{L})_{3}(M_{Z}) \simeq &- 0.011\, A_{0}^2 + 0.98\,m_{0}^2 + 8.38 \times 10^{-3}\, A_{0} M_{\half} + 0.50\,M_{\half}^2 - 0.038\, m_{10}^2 \nonumber\\ &+ 0.027\, m_{20}^2 \\
(m^{2}_{E})_{3}(M_{Z}) \simeq &- 0.021\, A_{0}^2 + 0.95\,m_{0}^2 + 0.017\, A_{0} M_{\half} + 0.15\,M_{\half}^2 + 0.030\, m_{10}^2 \nonumber\\ &- 0.053\, m_{20}^2 \\
m_{H_{d}}^{2}(M_{Z}) \simeq &- 0.046\, A_{0}^{2} - 0.11\, m_{0}^{2} + 0.16\, A_{0}M_{\half} + 0.01\, M_{\half}^{2} + 0.913\, m_{10}^{2} \nonumber\\ &+ 0.030\, m_{20}^{2}\\ 
m_{H_{u}}^{2}(M_{Z}) \simeq &- 0.105\, A_{0}^{2} - 0.67\, m_{0}^{2} + 0.46\, A_{0}M_{\half} - 2.86\, M_{\half}^{2} + 0.030\, m_{10}^{2} \nonumber\\ &+ 0.634\, m_{20}^{2}\\ 
|\mu|^{2}(M_{Z}) = &-4158.72 + 0.106\, A_{0}^{2} + 0.67\, m_{0}^{2} - 0.46\, A_{0}M_{\half} + 2.87\, M_{\half}^{2} - 0.027\, m_{10}^{2} \nonumber \\ &- 0.636\, m_{20}^{2}
\end{align}
For high $\tan\beta = 35$ their values are as follows
\begin{align}
(m^{2}_{Q})_{3}(M_{Z}) \simeq &- 0.058\, A_{0}^2 + 0.69\, m_{0}^2 + 0.25\, A_{0} M_{\half} + 5.26\,M_{\half}^2 - 0.037\, m_{10}^2 \nonumber\\ &- 0.120\, m_{20}^2\\
(m^{2}_{U})_{3}(M_{Z}) \simeq &- 0.064\, A_{0}^2 + 0.55\, m_{0}^2 + 0.27\, A_{0} M_{\half} + 4.35\,M_{\half}^2 - 0.029 \, m_{10}^2 \nonumber\\ &- 0.194\, m_{20}^2\\
(m^{2}_{D})_{3}(M_{Z}) \simeq &- 0.052\, A_{0}^2 + 0.82\, m_{0}^2 + 0.23\, A_{0} M_{\half} + 5.26\,M_{\half}^2 - 0.081\, m_{10}^2 \nonumber\\ &- 0.010\, m_{20}^2 \\
(m^{2}_{L})_{3}(M_{Z}) \simeq &- 0.027\, A_{0}^2 + 0.93\, m_{0}^2 + 0.02\, A_{0} M_{\half} + 0.49\,M_{\half}^2 - 0.063\, m_{10}^2 \nonumber\\ &+ 0.027\, m_{20}^2 
\\
(m^{2}_{E})_{3}(M_{Z}) \simeq &- 0.055\, A_{0}^2 + 0.85\, m_{0}^2 + 0.03\, A_{0} M_{\half} + 0.12\,M_{\half}^2 - 0.019\, m_{10}^2 \nonumber\\ &- 0.054\, m_{20}^2 
\end{align}
\begin{align}
m_{H_{d}}^{2}(M_{Z}) \simeq &- 0.105\, A_{0}^{2} - 0.35\, m_{0}^{2} + 0.36\, A_{0}M_{\half} - 0.91\, M_{\half}^{2} + 0.789\, m_{10}^{2} \nonumber\\ &+ 0.038\, m_{20}^{2}\\ 
m_{H_{u}}^{2}(M_{Z}) \simeq &- 0.095\, A_{0}^{2} - 0.67\, m_{0}^{2} + 0.41\, A_{0}M_{\half} - 2.81\, M_{\half}^{2} + 0.036\, m_{10}^{2} \nonumber\\ &+ 0.629\, m_{20}^{2}\\ 
|\mu|^{2}(M_{Z}) = &-4158.72 + 0.095\, A_{0}^{2} + 0.67\, m_{0}^{2} - 0.41\, A_{0}M_{\half} + 2.81\, M_{\half}^{2} - 0.036\, m_{10}^{2} \nonumber \\ &- 0.629\, m_{20}^{2}
\end{align}


\section{Lightest Slepton Mass in $\delta$-mSUGRA at Large $\delta$}
\label{appB}
From the plots presented in section \ref{sec2}, Fig.( \ref{brmsugra}), for the case of $\delta$-mSUGRA, the following two things can be inferred: (a) the co-annihilation condition increasingly moves towards the diagonal in $\left(m_{0}, M_{1/2}\right)$ plane with increasing $\delta$ and (b) the cancellation region are almost independent of the value of $\delta$ in $\left(m_{0}, M_{1/2}\right)$ plane. The question then arises if there is some region at large `$\delta$' where the two regions coincide. In the present appendix, we explore this question. The analysis presented here is based on the approximate solutions of Appendix [\ref{appA1}] and we will comment on the full numerical solutions at the end of the section.

The effective $4\times4$ matrix of eq. (\ref{slep6}) can be diagonalized as follows. First the lower $2\times2$ block is rotated by an angle $\theta$, given by,
\begin{align}
\tan\,2\theta_{\mu\tau} = \frac{2\, \Delta_{RR}}{\msmu-\mstau}.
\end{align}
The eigenvlaues of this lower block can be easily read off from the mass matrix. They are

\begin{figure}[htbp]
\label{appBf1}
\begin{center}
\begin{tabular}{cc}
\includegraphics[width=0.5\textwidth]{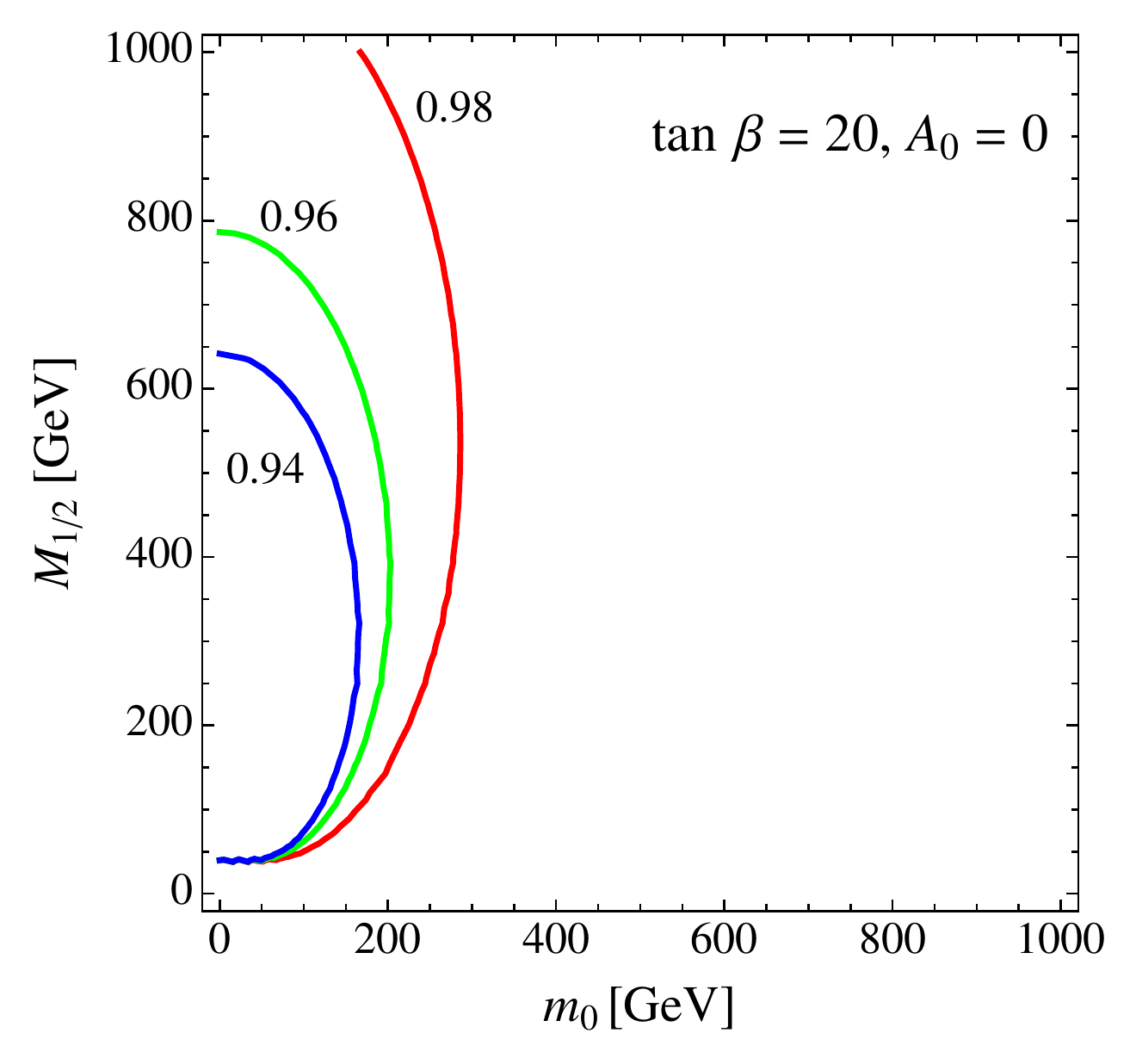} & \includegraphics[width=0.5\textwidth]{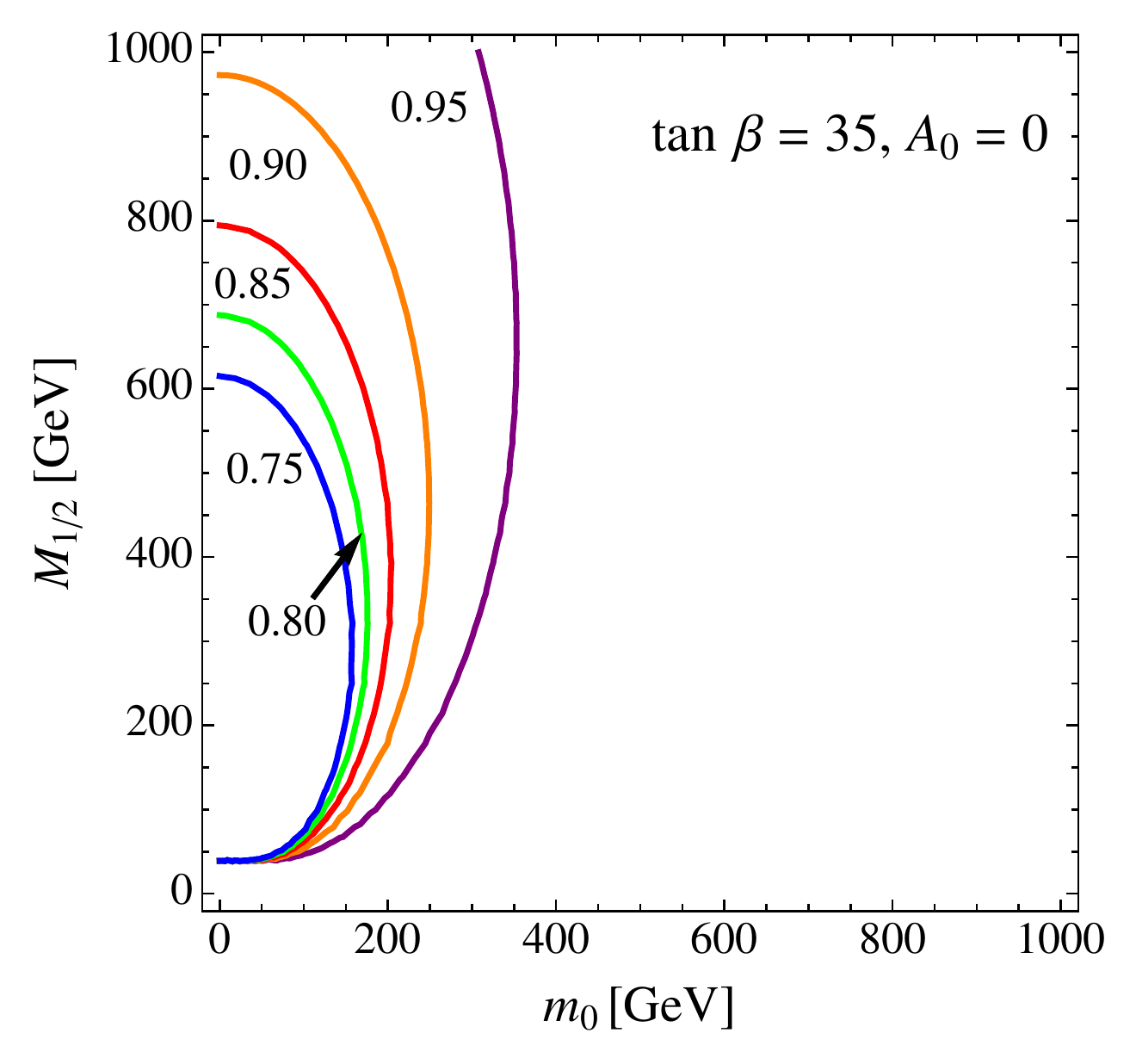}
\end{tabular}
\caption{{\bf $\delta$ Contours:} Upper bounds on $\delta$ in various of the parameter space using the non-tachyonic condition.}
\end{center}
\end{figure}

\begin{equation}
\lambda^{2}_{\pm} = \half \Bigg[(m^{2}_{\smu_{R}}+m^{2}_{\staur_{R}}) \pm \sqrt{(m^{2}_{\smu_{R}}-m^{2}_{\staur_{R}})^{2} + 4\, \Drr^{2}} \Bigg]
\end{equation}
For $\msmu \simeq \mstau$ (which is true for low $\tan\beta$ regions), the eigenvalues have the following form:
\begin{align}
\lambda^{2}_{\pm} &\simeq \bar{m}^{2} \pm \Drr \\
&\simeq \bar{m}^{2} (1\pm \drr)
\end{align}
where $\bar{m}^{2} \equiv \displaystyle \half \left(\msmu+\mstau \right)$. Next we have to diagonalize the $\staur_{LR}$ entry. The eigenvalues after this rotation are approximately given as
\begin{align}
\label{egs1}
\Gamma^{2}_{\pm} &\simeq \half \Bigg[(m^{2}_{\staur_{L}} + \lambda^{2}_{-}) \pm \sqrt{(m^{2}_{\staur_{L}} - \lambda^{2}_{-})^{2} + 4\, \cos^{2}\theta_{\mu\tau} \,\Delta^{2}_{{\staur}_{LR}}} \Bigg] \\
\intertext{In the limit $\left(m^{2}_{\staur_{L}} - \lambda^{2}_{-} \right) \gg \Delta_{\staur_{LR}}$ (the corresponding angle is very small in this limit)\footnote{We will consider the opposite limit at the end of this section.}, which is the case for large $\delta$, we can write the above eigenvalues as}
\Gamma^{2}_{\pm} &\simeq \half \Bigg[(m^{2}_{\staur_{L}} + \lambda^{2}_{-}) \pm (m^{2}_{\staur_{L}} - \lambda^{2}_{-}) \left\{ 1 + \frac{2\, \cos^{2}\theta_{\mu\tau} \, \Delta^{2}_{{\staur}_{LR}}}{(m^{2}_{\staur_{L}} - \lambda^{2}_{-})^{2}} \right\} \Bigg] \label{lambdaplus}
\end{align}

\begin{figure}[htbp]
\label{appBf2}
\begin{center}
\begin{tabular}{cc}
\includegraphics[width=0.5\textwidth]{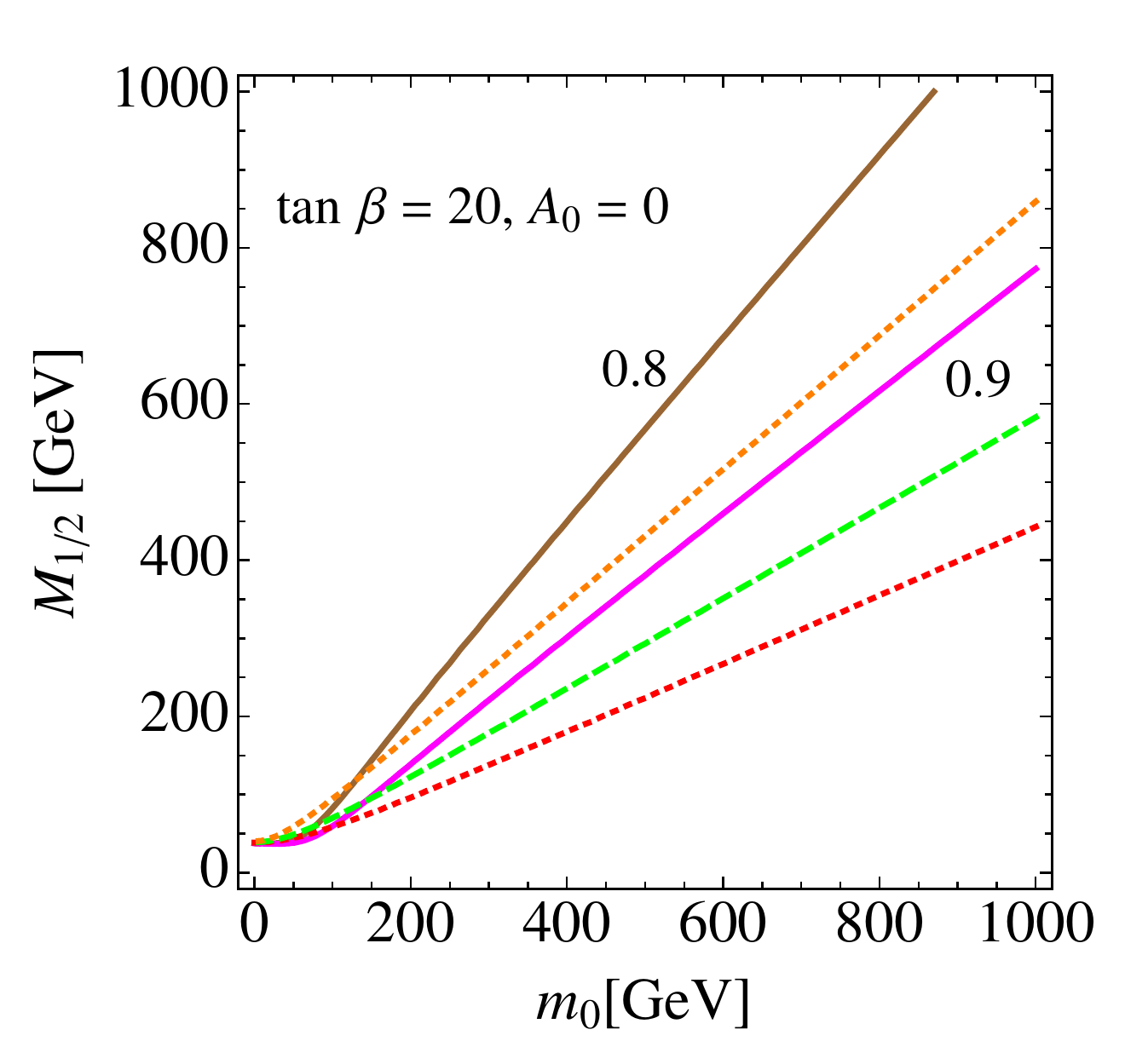} & \includegraphics[width=0.5\textwidth]{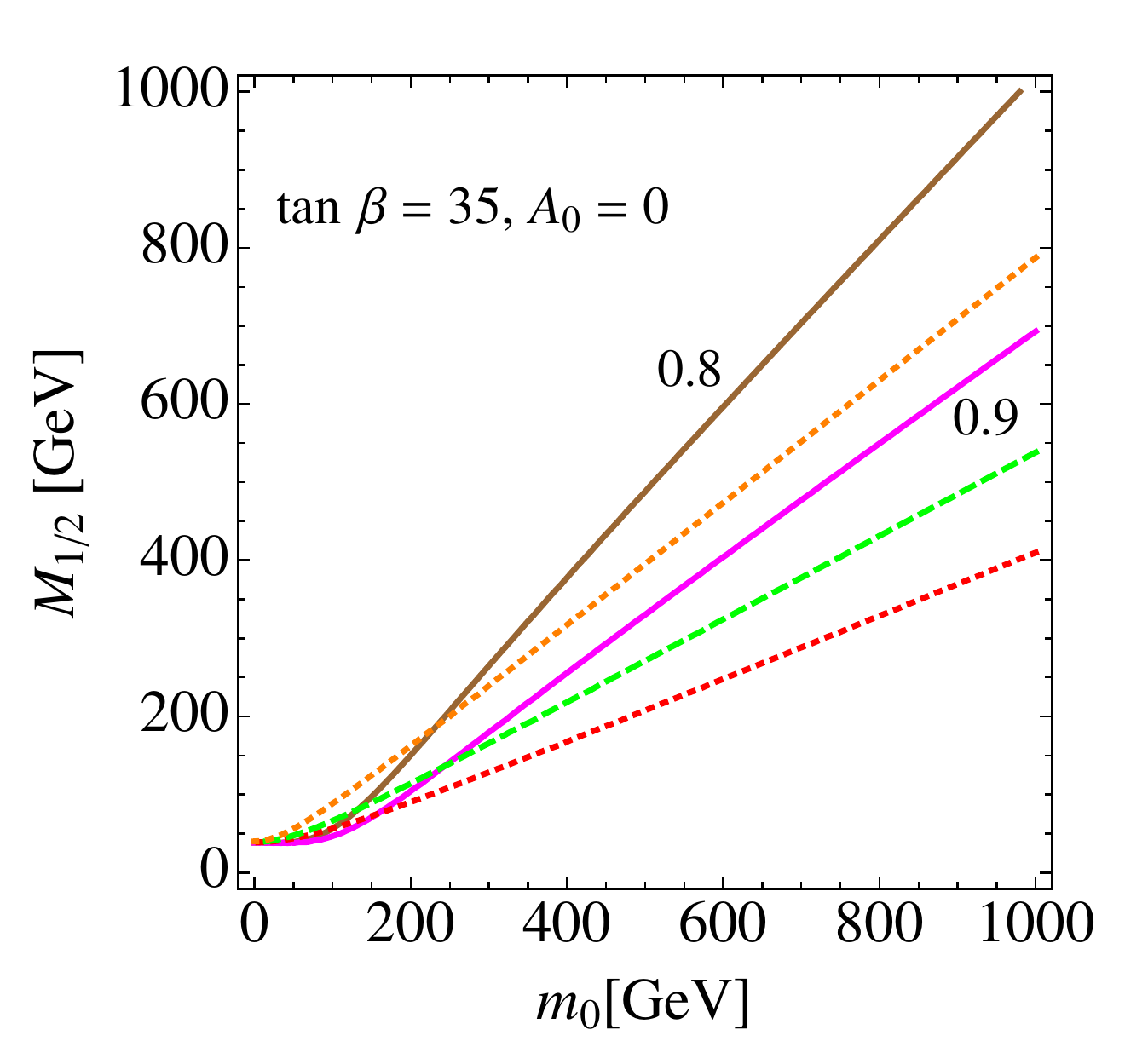}
\end{tabular}
\caption{Cancellation and Co-annihilation region in $\delta$-mSUGRA}
\end{center}
\end{figure}

So, the lightest eigenvalue of the effective $4\times4$ mass matrix of eq.(\ref{slep6}) is given as 
\begin{align}
\Gamma^{2}_{-} &\simeq \lambda^{2}_{-} - \frac{ \cos^{2}\theta_{\mu\tau} \,  \Delta^{2}_{{\staur}_{LR}}}{m^{2}_{\staur_{L}} - \lambda^{2}_{-}} \\
&\simeq \bar{m}^{2}( 1 - \drr) - \frac{\cos^{2}\theta_{\mu\tau} \, \Delta^{2}_{{\staur}_{LR}}}{m^{2}_{\staur_{L}}- \bar{m}^{2}( 1 - \drr)} \label{del1}
\end{align}
Which essentially suppresses the left-right mixing term compared to eq. (\ref{lightesteg}). 
And demanding the lightest eigenvalue to be non-tachyonic we get an upper bound on $\drr$ as below 
\begin{align}
\label{delb}
\drr	 \leq 1 - \frac{\cos^{2}\theta_{\mu\tau}  \, \Delta^{2}_{{\staur}_{LR}}}{m^{2}_{\staur_{L}} \mstau}
\end{align}
Which matches with eq. (\ref{deltaeq}) in the limit $\cos \theta_{\mu\tau} \rightarrow 1$. 

In Fig.(10) we have plotted the tachyonic condition (R.H.S of eq. (\ref{delb})) using the approximate results of Appendix [\ref{appendix1}]. It has been plotted for two values of $\tan\beta$ 20 and 35. The contours represents the upper bounds on $\delta$ in those regions of the parameter space to avoid tachyonic leptons. As we can see, increasing $\tan\beta$, tightens the bound a bit. In Fig.(11) we have shown the cancellation condition $\mu^{2} \simeq m^{2}_{\staur_{R}}$ and the co-annihilation condition $m_{\lstau} \simeq m_{\chiz}$ for two values of $\delta = 0.8$ and 0.9. In both the pannels, the brown and magenta solid lines indicate co-annihilation condition for $\delta = 0.8$ and $0.9$ respectively. The green dashed line satisfy the cancellation condition, whereas the orange and red dashed lines satisfy the cancellation condition with the $\mu$ parameter being 30\% corrected than its tree level value. Comparing the Figs.(10) and (11)  we  can see that there could be some points which could evade both the tachyonic condition as well as have cancellations amongst the LFV amplitudes and still satisfy the co-annihilation condition. However in practice  in full numerical calculation, we could not find any points consistent with both these conditions as other phenomenological constraints rule them out. As can be seen from the figure,  a 30\% correction to the $\mu$ parameter could shift the overlapping region to very small values of $\left( m_{0}, M_{1/2}\right)$ or no overlap at all for $\delta \simeq 0.9$. This approximates  the implications of adding the full 1-loop effective corrections to the SUSY scalar potential.  However, the co-annihilation region  could allow for partial calculations in LFV amplitudes. Such regions are difficult to distinguish in a numerical analysis.

We will now return to Eq. (\ref{egs1}) and consider the limit $\left(m^{2}_{\staur_{L}} - \lambda^{2}_{-} \right) \ll \Delta_{\staur_{LR}}$ , which is an
interesting limit  as it is relevant for the regions which appear in channels plots discussed in section \ref{sec4}.  From eq. (\ref{lambdaplus}), there could be a value of $\delta$ as well as parameter space in $\left( m_{0}, M_{1/2}, 
\tan\beta \right)$ where $m^{2}_{\staur_{L}} \simeq \lambda^{2}_{-}$. In these regions, the corresponding mixing angle is very large and the subsequent diagonalization is very different. It turns out that at least three mixing angles in the slepton mass matrix are large in this parameter space. The plots presented in Figs.(\ref{sigvmsug})
and (\ref{sigvnuhm}) contain these regions.  More details of these regions will be discussed in \cite{upcoming}.

\section{Numerical Procedures}
\label{appendix2}
\subsection{\texttt{SuSeFLAV} and \texttt{MicrOMEGAs}}
The numerical analysis is done using publicly available package \texttt{MicrOMEGAs} \cite{Belanger:2010gh} and \texttt {SuSeFLAV} \cite {suseflav_docu}.  \texttt {SuSeFLAV} is a fortran package which computes the supersymmetric spectrum by considering lepton flavor violation. The program solves complete MSSM RGEs with complete $3\times3$ flavor mixing at 2-loop level and full one loop threshold corrections \cite{Pierce:1996zz} to all MSSM parameters and relevant SM parameters, with conserved R-parity. Also, the program computes branching ratios and decay rates for rare flavor violating processes such as $\mu$ $\rightarrow$ e$\gamma$, $\tau$ $\rightarrow$ e$\gamma$, $\tau$ $\rightarrow$ $\mu$ $\gamma$, $\mu$ $\rightarrow$ e$\gamma$, $\mu^-$ $\rightarrow$ $e^+$ $e^-$ $e^-$, $\tau^-$ $\rightarrow$ $\mu^+$ $\mu^-$ $\mu^-$, $\tau^-$ $\rightarrow$ $e^+$ $e^-$ $e^-$, $B\,\rightarrow\, s\,\gamma$  and $(g-2)_{\mu}$.
 
 In the present analysis we use $M_t^{pole}=173.2\,{\rm GeV}$, $M_b^{pole}=4.23\, {\rm GeV}$ and $M_{\tau}^{pole}=1.77\, {\rm GeV}$. In determining the lightest higgs mass ($m_h$) we use approximations for one loop correction which are mostly  top-stop enhanced \cite{Heinemeyer:1999be}. We use complete $6\times6$ slepton mass matrix to correctly evaluate the inter-generational mixings and masses in the presence of flavor violation. 
 
Moreover we consider flavor violating couplings stemming from lepton flavor violation in the RR sector of $\tilde{\tau}-\tilde{\mu}$.  

\begin{figure}[htb]
\begin{center}
\begin{tabular}{ll}
 (a) & \hspace{1cm}(b) \\
 \includegraphics[width=.40\textwidth,angle=0]{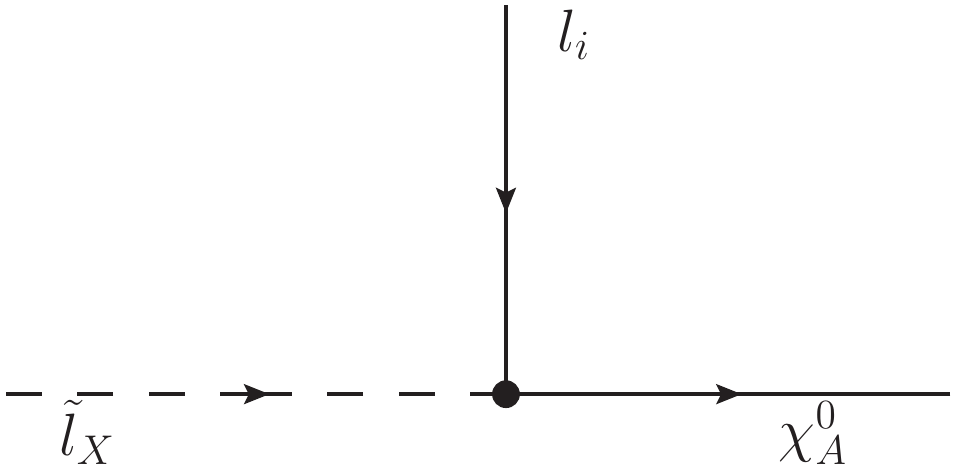} 
 & \hspace{1cm}\includegraphics[width=.40\textwidth,angle=0]{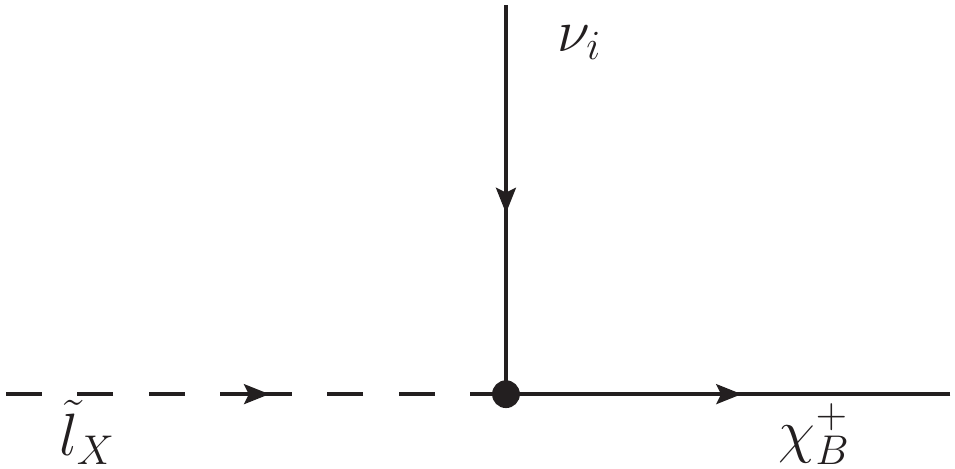} 
 \end{tabular}
\end{center}
 \caption{(a) Neutralino-slepton-lepton vertex and (b) Slepton-lepton-chargino vertex.}
\label{vertex}
\end{figure}

 \begin{itemize}
 \item {\bf Neutralino-slepton-lepton:}  
 
The interaction Lagrangian for neutralino-slepton-lepton is written as
\begin{align}
\mathcal{L}\ =\ \bar{l}_{i}\left(\Sigma^{L}_{iAX}\, P_{L} + \Sigma^{R}_{iAX}\, P_{R}\right)\, \chi^{0}_{A}\, \tilde{l}_{X} + h.c.
\end{align}

Where the coefficients are defined as
\begin{align}
\Sigma^{R}_{iAX}\ &=\  K_{1} \left[\cw (O_{N})_{A2} + \sw (O_{N})_{A1} \right] U_{X,i} \mw \cb - m_{l_{i}} \cw (O_{N})_{A3} U_{X,i+3}
\intertext{and}
\Sigma^{L}_{iAX}\ &= - K_{1}\left[ 2 \sw\ \mw\ \cb\ U_{X,i+3}\ (O_{N})_{A1} + m_{l_{i}} \cw\ U_{X,i}\ (O_{N})_{A3} \right]
\intertext{where}
&\quad K_{1} = \frac{e}{\sqrt{2}\sw} \frac{1}{\mw \cb \cw} \label{k1}
\end{align}

The Interaction Lagrangian for chargino-slepton-neutrino is

\begin{align}
\mathcal{L}\ =\ \bar{\nu}_{i}\left(\Pi^{L}_{iBX}\, P_{L} + \Pi^{R}_{iBX}\, P_{R}\right)\, \chi^{+}_{B}\, \tilde{l}_{X} + h.c.
\end{align}

Where the coefficients are
\begin{align}
\Pi^{R}_{iBX} &= - \frac{e}{\sw}\, (O_{L})_{B1} U_{X,i} \\
\Pi^{R}_{iBX} &= \frac{e}{\sw}\, \frac{m_{l_{i}}}{\sqrt{2} \mw \cb }\, (O_{L})_{B2} U_{X,i+3}
\end{align}

Where, $U_{X,i}$ is the $6\times6$ matrix which diagonalizes the sleptonic mass matrix, here the indices $i = 1$ to $3$ and  $X = 1$ to $6$. $(O_{N})_{Am}$ is the $4\times4$ neutralino mixing matrix, where $A,m = 1$ to $4$ and $(O_{L})_{Bn}$ is the $2\times2$ chargino left eigenvector matrix, where $B, n = 1, 2$. $m_{l_{i}}$ is the mass of the lepton $l_{i}$. In our notation $P_{L} = \frac{1-\gamma_{5}}{2}$ and $P_{R} = \frac{1+\gamma_{5}}{2}$.
\end{itemize}
These couplings are programmed into \texttt{MicrOMEGAs} through \texttt{CalcHEP} \cite{Pukhov:1999gg} package. 

\subsection{Constraints Imposed}
\begin{itemize}
\item We check for efficient radiative electroweak symmetry breaking, requiring $|\mu|^2>0$ for valid points.
\item We require $m_{\tilde{\tau}}>m_{\chi^0}$ as LSP is neutral. Regions for which this condition is not true is excluded as $\tilde{\tau}$ LSP regions. 
\item We impose lower bounds on various sparticle masses that results from collider experiments. $m_h>114.1(GeV)$,  $m_{\chi^{\pm}} > 103.5(GeV)$ and $m_{\tilde{\tau}}>90 (GeV)$ \cite{Barate:2003sz}.
\item $2.0\times10^{-4}\leq\,BR(b\,\rightarrow\,s\,\gamma)\,\leq\, 4.5\times 10^{-4}$ \cite{Nakamura:2010zzi}.
\item We also check for the D-flat directions, while checking for the EWSB condition and charge and color breaking minima \cite{Frere:1983ag,AlvarezGaume:1983gj,Claudson:1983et}.
\end{itemize}

\section{Loop Functions}
\label{loopfunc}

 In this appendix we define the relevant loop functions that contribute to the amplitudes of flavor violating leptonic process $BR(\tau\rightarrow\mu \gamma)$  as presented in the appendix of \cite{Masina:2002mv}
\begin{equation}
x_L = \frac{M_1^2}{m_L^2}, \,\,\,\,\,x_R = \frac{M_1^2}{m_R^2}, \,\,\,\,\,y_L = \frac{|\mu^2|}{m_L^2},\,\,\,\,\, y_R = \frac{|\mu^2|}{m_R^2}
\end{equation}

$I_{B,R}$ and $I_R$ are defined as follows, 
\begin{eqnarray}
I_{B,R}(M_1^2,m_L^2,m_R^2) = -\frac{1}{m_R^2 - m_L^2} \left[y_R \,h_1(x_R) - \frac{y_L\, g_1(x_L) - y_R \,g_1 (x_R)}{1- \frac{m_L^2}{m_R^2}}\right]
\end{eqnarray}

\begin{equation}
I_{R}(m_R^2,M_1^2,\mu^2) = \frac{1}{m_R^2}\frac{y_R}{y_R - x_R} [h_1 (x_R) - h_1(y_R)]
\end{equation}

The functions $g_1$ and $h_1$ are defined as follows, 
\begin{equation}
g_1(x) = \frac{1 - x^2 + 2x \ln(x)}{(1 -x)^3},\,  \, \, \, \,  h_1(x) = \frac{1 + 4x - 5x^2 + (2x^2 + 4 x) \ln(x)}{(1 -x)^4}
\end{equation}

\section{Cross-Sections}
\label{appD}
In this appendix we present the approximate formulae for the relevant cross sections. We do not attempt to discuss a complete comparison of the analytical  expressions and full numerical results in the present paper, that is left for an upcoming publication. 
These expressions generalize the existing expressions \cite{Nihei:2002sc} in the literature to include full flavor violation in the sleptonic sector. The expressions are presented only for the dominant channels and in the limit $m_{\tau}, m_{\mu} \rightarrow 0$.  More detailed expressions and their simplifications will be discussed elsewhere \cite{upcoming}.
 
\begin{align}
\sigma_{channel} = \frac{\rm Numerator}{\rm Denominator}
\end{align}
\subsection{$\lstau \lstau \rightarrow \tau \tau $}
The cross-section of $\lstau\; \lstau \rightarrow \tau\; \tau$  process is as follows. This process involves $t$- and $u$-channel $\chiz$ exchange. In the following and in rest of the cross-sections $e$ is the electric charge, $\theta_{\rm W}$ is the weak mixing angle and $\mw$ is the mass of the W-boson. We get the simplified form of the above cross-section in the limit of $m_{\tau} \rightarrow 0$ as below, where the numerator is 
\begin{align}
e^4 &\left[ - \SP^2 \SM^2 \sqrt{s \left(s-4 m_{\lstau}^2\right)} - \SP^2 \SM^2 \left(s+2 m_{\chiz }^2-2 m_{\lstau}^2\right) \right. \nonumber \\ 
&\times \log \left|\frac{s + 2 m_{\chiz }^2-2 m_{\lstau}^2 - \sqrt{s \left(s-4 m_{\lstau}^2\right)}}{s + 2 m_{\chiz }^2 - 2 m_{\lstau}^2 + \sqrt{s \left(s-4 m_{\lstau}^2\right)}}\right| \notag \\ 
&-\frac{1}{s+ 2 m_{\chiz }^2-2 m_{\lstau}^2}\log \left|\frac{s + 2 m_{\chiz }^2 - 2 m_{\lstau}^2 - \sqrt{s \left(s-4 m_{\lstau}^2\right)}}{s + 2 m_{\chiz }^2 - 2 m_{\lstau}^2 + \sqrt{s \left(s-4 m_{\lstau}^2\right)}}\right| \notag \\ 
&\times \bigg\{2 \SP^2 \SM^2 m_{\chiz }^4+2 \SP^2 \SM^2 m_{\lstau}^{4} \notag \\  
&+  m_{\chiz }^2 \left(\left(\SP^2+\SM^2\right)^2 s-4 \SP^2 \SM^2 m_{\lstau}^2\right)\bigg\} \notag \\ 
&- \frac{\sqrt{s \left(s-4 m_{\lstau}^2\right)}}{2 \left(m_{\chiz }^4+m_{\lstau}^4+m_{\chiz }^2 \left(s-2 m_{\lstau}^2\right)\right)} \left. \bigg\{\left. 4 \SP^2 \SM^2 m_{\chiz}^4 \right. \right. \notag \\ 
&+ \left. \left. 4 \SP^2 \SM^2 m_{\lstau}^4 -  \left.\left(\SP^2 - \SM^2\right)^{2} m_{\chiz}^{2}\, s \right. \right. \right. \notag \\ 
&+ \left. \left. \left. 2 \SP^2 \SM^2 m_{\chiz}^{2}\, s - 8 \SP^2 \SM^2 m_{\chiz}^{2}\, m_{\lstau}^2\right.\right.\right.\bigg\}\bigg]
\end{align}
And the denominator is 
\begin{align}
32\pi s\;  M^{4}_{\rm W} \cos^4\beta \cos^4\theta_{\rm W} \sin^4\theta_{\rm W} \left(s-4 m_{\lstau}^2\right)
\end{align}

Following appendix \ref{appendix2} the coupling structure is:
\begin{itemize}
\item \underline{$\bar{\tau} - \chiz - \lstau$:}
\begin{align}
&K_{1} \left( \SP P_{R} + \SM P_{L} \right)
\intertext{where}
 \SP &= \left[\cw\ ON(1,2) + \sw\ ON(1,1) \right] \cb\ \mw\ U(1,3)  \\ 
\intertext{and}
\SM &= - \, 2 \sw\ \mw\ \cb\ U(1,6)\ ON(1,1) 
\end{align}
Where $K_{1}$ is already defined in eq. (\ref{k1}).
\end{itemize}

\subsection{$\lstau \lstau \rightarrow \mu \tau$}

The simplified form of the $\lstau\; \lstau \rightarrow \mu\; \tau$ cross-section in the limit of $m_{\tau},m_{\mu} \rightarrow 0$ is calculated below. This process involves $t$- and $u$-channel $\chiz$ exchange. The numerator of the cross-section is

\begin{align}
e^4 &\ \left[ -  \left(\SP^2 \LM^2+\SM^2 \LP^2\right) \sqrt{s \left(s-4 m_{\lstau}^2\right)} - \left(\SP^2 \LM^2+\SM^2 \LP^2\right) \right. \nonumber \\ 
&\ \quad \times \left(s+2 m_{\chiz }^2-2 m_{\lstau}^2\right) \log \left|\frac{s + 2 m_{\chiz }^2-2 m_{\lstau}^2 - \sqrt{s \left(s-4 m_{\lstau}^2\right)}}{s + 2 m_{\chiz }^2 - 2 m_{\lstau}^2 + \sqrt{s \left(s-4 m_{\lstau}^2\right)}}\right| \notag \\ 
&\ \quad -\frac{2}{s+ 2 m_{\chiz }^2-2 m_{\lstau}^2}\log \left|\frac{s + 2 m_{\chiz }^2 - 2 m_{\lstau}^2 - \sqrt{s \left(s-4 m_{\lstau}^2\right)}}{s + 2 m_{\chiz }^2 - 2 m_{\lstau}^2 + \sqrt{s \left(s-4 m_{\lstau}^2\right)}}\right| \notag \\ 
&\ \quad \times \bigg\{ \left(\SP^2 \LM^2+\SM^2 \LP^2\right) m_{\chiz }^4 + \left(\SP^2 \LM^2+\SM^2 \LP^2\right) m_{\lstau}^{4} \notag \\  
&\ \quad +  m_{\chiz }^2 \left( \left(\SP^2+\SM^2\right) \left(\LM^2+\LP^2\right) s - 2 \left(\SP^2 \LM^2+\SM^2 \LP^2\right) m_{\lstau}^2\right)\bigg\} \notag \\ 
&\ \quad - \frac{\sqrt{s \left(s-4 m_{\lstau}^2\right)}}{\left(m_{\chiz }^4+m_{\lstau}^4+m_{\chiz }^2 \left(s-2 m_{\lstau}^2\right)\right)} \left. \bigg\{\left. 2 \left(\SP^2 \LM^2+\SM^2 \LP^2\right) m_{\chiz}^4 \right. \right. \notag \\ 
&\ \quad + \left. \left. 2 \left(\SP^2 \LM^2+\SM^2 \LP^2\right) m_{\lstau}^4 +  \left.  \left(\SP^2 \left(2 \LM^2-\LP^2\right) - \SM^2 \left(\LM^2-2 \LP^2\right) \right)\, m_{\chiz}^{2}\, s \right. \right. \right. \notag \\ 
&\ \quad - \left. \left. \left. 4 \left(\SP^2 \LM^2+\SM^2 \LP^2\right) m_{\chiz}^{2}\, m_{\lstau}^2\right.\right.\right.\bigg\}\bigg]
\end{align}

And the denominator is 
\begin{align}
32 \pi s \; M^{4}_{\rm W} \cos^4\beta \cos^4\theta_{\rm W} \sin^4\theta_{\rm W} \left(s - 4 m_{\lstau}^2\right)
\end{align}

Here the coupling structure is:
\begin{itemize}
\item \underline{$\bar{\mu} - \chiz - \lstau$:}
\begin{align}
&K_{1} \left( \LP P_{R} + \LM P_{L} \right)
\intertext{where}
 \LP &= \left[\cw\ ON(1,2) + \sw\ ON(1,1) \right] \cb\ \mw\ U(1,2)  \\ 
\intertext{and}
\LM &= - \, 2 \sw\ \mw\ \cb\ U(1,5)\ ON(1,1) 
\end{align}
\end{itemize}

\subsection{$\chiz \chiz \rightarrow \bar{\tau}/\bar{\mu} \tau$}

In the limit $m_{\tau} \rightarrow 0$ the cross-section for $\chiz\ \chiz \rightarrow \bar{\tau}\ \tau$ is calculated. This process involves $t$- and $u$-channel $\lstau$ exchange. The numerator is 
\begin{align}
 e^4 &\ \left\{\frac{ \sqrt{s \left(s-4 m_{\chiz }^2\right)}}{s
m_{\lstau}^2+\left(m_{\chiz }^2-m_{\lstau}^2\right)^2} \Big\{\left(\SP^4+4 \SP^2 \SM^2+\SM^4\right)
s\, m_{\lstau}^2+2 \left(\SP^4+3 \SP^2 \SM^2+\SM^4\right) \right. \notag \\ 
&\ \quad \left. \times \left(m_{\chiz }^2-m_{\lstau}^2\right)^2\right\} -
\frac{2}{-s + 2 m_{\chiz}^2 - 2 m_{\lstau}^2} \log \left[\frac{- \sqrt{s \left(s-4 m_{\chiz }^2\right)} + \left(s-2 m_{\chiz }^2+2 m_{\lstau}^2\right)}{\sqrt{s \left(s-4 m_{\chiz }^2\right)} +  \left(s-2 m_{\chiz }^2 + 2 m_{\lstau}^2\right)}\right] \notag \\
&\ \quad \times \Big\{s \left(-2 \SP^2 \SM^2 m_{\chiz }^2+\left(\SP^4+4 \SP^2 \SM^2+\SM^4\right) m_{\lstau}^2\right) + 2 \left(\SP^4+3 \SP^2 \SM^2+\SM^4\right)   \notag \\ 
&\ \quad \times \left(m_{\chiz }^2-m_{\lstau}^2\right)^2 \Big\}\Bigg\}
\intertext{And the denominator is}
&\ \qquad \qquad \qquad 128 \pi \, s\, \mwfr \cbfr \swfr \cwfr \left(s-4 m_{\chiz }^2\right)
\end{align}

Whereas in the limit $m_{\tau}, m_{\mu} \rightarrow 0$ the cross-section for $\chiz\ \chiz \rightarrow \bar{\mu}\ \tau$ is calculated. This process involves $t$- and $u$-channel $\lstau$ exchange. The numerator is 
\begin{align}
e^4 &\ \left\{\frac{ \sqrt{s \left(s-4 m_{\chiz }^2\right)}}{s m_{\lstau}^2+\left(m_{\chiz }^2-m_{\lstau}^2\right)^2} \bigg\{\left\{\SP^2 \left(2 \LM^2+\LP^2\right) + \SM^2
\left(\LM^2+2 \LP^2\right)\right\} s\, m_{\lstau}^2 \right. \notag \\ 
&\ \quad  + \left\{\SP^2 \left(3 \LM^2 + 2 \LP^2\right) + \SM^2 \left(2 \LM^2 + 3 \LP^2\right)\right\} \left(m^{2}_{\chiz }-m^{2}_{\lstau}\right)^2 \bigg\} \notag \\
&\ \quad -  \frac{2}{-s+2 m_{\chiz }^2-2 m_{\lstau}^2} \log \left[\frac{-\sqrt{s \left(s-4 m_{\chiz }^2\right)} +  \left(s-2 m_{\chiz }^2+2 m_{\lstau}^2\right)}{\sqrt{s \left(s-4 m_{\chiz }^2\right)} +  \left(s-2 m_{\chiz }^2+2 m_{\lstau}^2\right)}\right] \notag \\
&\ \quad \times \bigg\{\left\{\SP^2 \left(3 \LM^2+2 \LP^2\right) + \SM^2 \left(2 \LM^2 + 3 \LP^2\right)\right\} \left(m^{2}_{\chiz } - m^{2}_{\lstau}\right)^2 \notag \\
&\ \quad + s \left\{-\left(\SP^2 \LM^2 + \SM^2\LP^2\right) m_{\chiz }^2+\left(\SP^2 \left(2 \LM^2+\LP^2\right)+\SM^2 \left(\LM^2+2 \LP^2\right)\right) m_{\lstau}^2\right\}\bigg\}\Bigg\}
\intertext{And the denominator is}
&\ \qquad \qquad \qquad 128 \pi \, s\, \mwfr \cbfr \swfr \cwfr \left(s-4 m_{\chiz }^2\right)
\end{align}

\subsection{$\chiz   \lstau \rightarrow \gamma \tau/\mu$}

In the limit $m_{\tau} \rightarrow 0$ the cross-section for $\chiz\ \lstau \rightarrow \gamma\ \tau$ is calculated. This process involves $s$-channel $\tau$ mediation and  $t$-channel $\lstau$ exchange. The numerator is 
\begin{align}
 \left(\SP^2+\SM^2\right) e^{4} &\left\{ \log\left[\frac{ m_{\chiz }^2 -  \left(s+m_{\lstau}^2\right) - 
\sqrt{m_{\chiz }^4+\left(-s+m_{\lstau}^2\right)^2-2 m_{\chiz }^2 \left(s+m_{\lstau}^2\right)}}{ m_{\chiz }^2 -  \left(s+m_{\tilde{\tau
}_1}^2\right) + \sqrt{m_{\chiz }^4+\left(-s+m_{\lstau}^2\right)^2 - 2 m_{\chiz }^2 \left(s+m_{\lstau}^2\right)}}\right] \right. \notag \\
&\quad \times s \left(m_{\chiz }^2-3 m_{\lstau}^2\right) + \left(s-2 m_{\chiz }^2+2 m_{\lstau}^2\right) \sqrt{m_{\chiz }^4+\left(-s+m_{\tilde{\tau
}_1}^2\right)^2-2 m_{\chiz }^2 \left(s+m_{\lstau}^2\right)} \Bigg\}
\intertext{And the denominator is}
&\ 32\pi\, s\, \mwsq \cbsq  \swsq \cwsq \left\{m_{\chiz
}^4+\left(-s+m_{\lstau}^2\right)^2-2 m_{\chiz }^2 \left(s+m_{\lstau}^2\right)\right\}
\end{align}
Whereas in the limit $m_{\mu} \rightarrow 0$ the cross-section for $\chiz\ \lstau \rightarrow \gamma\ \mu$ is calculated. This process involves $s$-channel $\mu$ mediation and  $t$-channel $\lstau$ exchange. The numerator is 
\begin{align}
 \left(\LP^2+\LM^2\right) e^{4} &\left\{ \log\left[\frac{ m_{\chiz }^2 -  \left(s+m_{\lstau}^2\right) - 
\sqrt{m_{\chiz }^4+\left(-s+m_{\lstau}^2\right)^2-2 m_{\chiz }^2 \left(s+m_{\lstau}^2\right)}}{ m_{\chiz }^2 -  \left(s+m_{\tilde{\tau
}_1}^2\right) + \sqrt{m_{\chiz }^4+\left(-s+m_{\lstau}^2\right)^2 - 2 m_{\chiz }^2 \left(s+m_{\lstau}^2\right)}}\right] \right. \notag \\
&\quad \times s \left(m_{\chiz }^2-3 m_{\lstau}^2\right) + \left(s-2 m_{\chiz }^2+2 m_{\lstau}^2\right) \sqrt{m_{\chiz }^4+\left(-s+m_{\tilde{\tau
}_1}^2\right)^2-2 m_{\chiz }^2 \left(s+m_{\lstau}^2\right)} \Bigg\}
\intertext{And the denominator is}
&\ 32\pi\, s\, \mwsq \cbsq  \swsq \cwsq \left\{m_{\chiz
}^4+\left(-s+m_{\lstau}^2\right)^2-2 m_{\chiz }^2 \left(s+m_{\lstau}^2\right)\right\}
\end{align}


\bibliographystyle{JHEP}

\bibliography{flav_coan_bib}

\providecommand{\href}[2]{#2}\begingroup\raggedright\begin{thebibliography}{10}

\bibitem{Jungman:1995df}
G.~Jungman, M.~Kamionkowski, and K.~Griest, {\it {Supersymmetric dark matter}},
   {\em Phys. Rept.} {\bf 267} (1996) 195--373,
  [\href{http://xxx.lanl.gov/abs/hep-ph/9506380}{{\tt hep-ph/9506380}}].

\bibitem{Goldberg:1983nd}
H.~Goldberg, {\it {Constraint on the Photino Mass from Cosmology}},  {\em Phys.
  Rev. Lett.} {\bf 50} (1983) 1419.

\bibitem{Ellis:1983ew}
J.~R. Ellis, J.~S. Hagelin, D.~V. Nanopoulos, K.~A. Olive, and M.~Srednicki,
  {\it {Supersymmetric relics from the big bang}},  {\em Nucl. Phys.} {\bf
  B238} (1984) 453--476.

\bibitem{Chankowski:1998za}
P.~H. Chankowski, J.~R. Ellis, K.~A. Olive, and S.~Pokorski, {\it {Cosmological
  fine tuning, supersymmetry, and the gauge hierarchy problem}},  {\em
  Phys.Lett.} {\bf B452} (1999) 28--38,
  [\href{http://xxx.lanl.gov/abs/hep-ph/9811284}{{\tt hep-ph/9811284}}].

\bibitem{wmap7}
D.~Larson, J.~Dunkley, G.~Hinshaw, E.~Komatsu, M.~Nolta, {\em et.~al.}, {\it
  {Seven-Year Wilkinson Microwave Anisotropy Probe (WMAP) Observations: Power
  Spectra and WMAP-Derived Parameters}},  {\em Astrophys.J.Suppl.} {\bf 192}
  (2011) 16, [\href{http://xxx.lanl.gov/abs/1001.4635}{{\tt arXiv:1001.4635}}].

\bibitem{ArkaniHamed:2006mb}
N.~Arkani-Hamed, A.~Delgado, and G.~Giudice, {\it {The Well-tempered
  neutralino}},  {\em Nucl.Phys.} {\bf B741} (2006) 108--130,
  [\href{http://xxx.lanl.gov/abs/hep-ph/0601041}{{\tt hep-ph/0601041}}].

\bibitem{Baer:2003wx}
H.~Baer, C.~Balazs, A.~Belyaev, T.~Krupovnickas, and X.~Tata, {\it {Updated
  reach of the CERN LHC and constraints from relic density, $b \to s \gamma$
  and a($\mu$) in the mSUGRA model}},  {\em JHEP} {\bf 0306} (2003) 054,
  [\href{http://xxx.lanl.gov/abs/hep-ph/0304303}{{\tt hep-ph/0304303}}].

\bibitem{Djouadi:2006be}
A.~Djouadi, M.~Drees, and J.-L. Kneur, {\it {Updated constraints on the minimal
  supergravity model}},  {\em JHEP} {\bf 0603} (2006) 033,
  [\href{http://xxx.lanl.gov/abs/hep-ph/0602001}{{\tt hep-ph/0602001}}].

\bibitem{cmv}
L.~Calibbi, Y.~Mambrini, and S.~Vempati, {\it {SUSY-GUTs, SUSY-seesaw and the
  neutralino dark matter}},  {\em JHEP} {\bf 0709} (2007) 081,
  [\href{http://xxx.lanl.gov/abs/0704.3518}{{\tt arXiv:0704.3518}}].

\bibitem{utpala-term}
U.~Chattopadhyay, D.~Das, A.~Datta, and S.~Poddar, {\it {Non-zero trilinear
  parameter in the mSUGRA model: Dark matter and collider signals at Tevatron
  and LHC}},  {\em Phys.Rev.} {\bf D76} (2007) 055008,
  [\href{http://xxx.lanl.gov/abs/0705.0921}{{\tt arXiv:0705.0921}}].

\bibitem{barger}
V.~Barger, D.~Marfatia, and A.~Mustafayev, {\it {Neutrino sector impacts SUSY
  dark matter}},  {\em Phys.Lett.} {\bf B665} (2008) 242--251,
  [\href{http://xxx.lanl.gov/abs/0804.3601}{{\tt arXiv:0804.3601}}].

\bibitem{gomez-lola-kang}
M.~Gomez, S.~Lola, P.~Naranjo, and J.~Rodriguez-Quintero, {\it {WMAP Dark
  Matter Constraints on Yukawa Unification with Massive Neutrinos}},  {\em
  JHEP} {\bf 0904} (2009) 043, [\href{http://xxx.lanl.gov/abs/0901.4013}{{\tt
  arXiv:0901.4013}}].

\bibitem{Kang:2009pj}
S.~K. Kang, A.~Kato, T.~Morozumi, and N.~Yokozaki, {\it {Threshold corrections
  to the radiative breaking of electroweak symmetry and neutralino dark matter
  in supersymmetric seesaw model}},  {\em Phys.Rev.} {\bf D81} (2010) 016011,
  [\href{http://xxx.lanl.gov/abs/0909.2484}{{\tt arXiv:0909.2484}}].

\bibitem{Biggio:2010me}
C.~Biggio and L.~Calibbi, {\it {Phenomenology of SUSY $SU(5)$ with Type I+Iii
  Seesaw}},  {\em JHEP} {\bf 10} (2010) 037,
  [\href{http://xxx.lanl.gov/abs/1007.3750}{{\tt arXiv:1007.3750}}].

\bibitem{Esteves:2010ff}
J.~N. Esteves, J.~C. Romao, M.~Hirsch, F.~Staub, and W.~Porod, {\it
  {Supersymmetric Type-Iii Seesaw: Lepton Flavour Violating Decays and Dark
  Matter}},  {\em Phys. Rev.} {\bf D83} (2011) 013003,
  [\href{http://xxx.lanl.gov/abs/1010.6000}{{\tt arXiv:1010.6000}}].

\bibitem{olivereview}
J.~Ellis, A.~Mustafayev, and K.~A. Olive, {\it {Resurrecting No-Scale
  Supergravity Phenomenology}},  {\em Eur.Phys.J.} {\bf C69} (2010) 219--233,
  [\href{http://xxx.lanl.gov/abs/1004.5399}{{\tt arXiv:1004.5399}}].

\bibitem{olive1}
K.~Kadota, K.~A. Olive, and L.~Velasco-Sevilla, {\it {A Sneutrino NLSP in the
  nu CMSSM}},  {\em Phys.Rev.} {\bf D79} (2009) 055018,
  [\href{http://xxx.lanl.gov/abs/0902.2510}{{\tt arXiv:0902.2510}}].

\bibitem{BHS}
R.~Barbieri, L.~J. Hall, and A.~Strumia, {\it {Violations of lepton flavor and
  CP in supersymmetric unified theories}},  {\em Nucl.Phys.} {\bf B445} (1995)
  219--251, [\href{http://xxx.lanl.gov/abs/hep-ph/9501334}{{\tt
  hep-ph/9501334}}].

\bibitem{Calibbi:2006nq}
L.~Calibbi, A.~Faccia, A.~Masiero, and S.~K. Vempati, {\it {Lepton Flavour
  Violation from Susy-Guts: Where Do We Stand for Meg, Prism / Prime and a
  Super Flavour Factory}},  {\em Phys. Rev.} {\bf D74} (2006) 116002,
  [\href{http://xxx.lanl.gov/abs/hep-ph/0605139}{{\tt hep-ph/0605139}}].

\bibitem{fnsoftterms}
E.~Dudas, S.~Pokorski, and C.~A. Savoy, {\it {Soft scalar masses in
  supergravity with horizontal U(1)-x gauge symmetry}},  {\em Phys.Lett.} {\bf
  B369} (1996) 255--261, [\href{http://xxx.lanl.gov/abs/hep-ph/9509410}{{\tt
  hep-ph/9509410}}].

\bibitem{Dudas:1996fe}
E.~Dudas, C.~Grojean, S.~Pokorski, and C.~A. Savoy, {\it {Abelian flavor
  symmetries in supersymmetric models}},  {\em Nucl.Phys.} {\bf B481} (1996)
  85--108, [\href{http://xxx.lanl.gov/abs/hep-ph/9606383}{{\tt
  hep-ph/9606383}}].

\bibitem{Barbieri:1997tu}
R.~Barbieri, L.~J. Hall, and A.~Romanino, {\it {Consequences of a U(2) flavor
  symmetry}},  {\em Phys.Lett.} {\bf B401} (1997) 47--53,
  [\href{http://xxx.lanl.gov/abs/hep-ph/9702315}{{\tt hep-ph/9702315}}].

\bibitem{Kobayashi:2002mx}
T.~Kobayashi, H.~Nakano, H.~Terao, and K.~Yoshioka, {\it {Flavor violation in
  supersymmetric theories with gauged flavor symmetries}},  {\em
  Prog.Theor.Phys.} {\bf 110} (2003) 247--267,
  [\href{http://xxx.lanl.gov/abs/hep-ph/0211347}{{\tt hep-ph/0211347}}].

\bibitem{Chankowski:2005qp}
P.~H. Chankowski, K.~Kowalska, S.~Lavignac, and S.~Pokorski, {\it {Update on
  fermion mass models with an anomalous horizontal U(1) symmetry}},  {\em
  Phys.Rev.} {\bf D71} (2005) 055004,
  [\href{http://xxx.lanl.gov/abs/hep-ph/0501071}{{\tt hep-ph/0501071}}].

\bibitem{Antusch:2008jf}
S.~Antusch, S.~F. King, M.~Malinsky, and G.~G. Ross, {\it {Solving the SUSY
  Flavour and CP Problems with Non-Abelian Family Symmetry and Supergravity}},
  {\em Phys.Lett.} {\bf B670} (2009) 383--389,
  [\href{http://xxx.lanl.gov/abs/0807.5047}{{\tt arXiv:0807.5047}}].

\bibitem{Scrucca:2007pj}
C.~A. Scrucca, {\it {Soft masses in superstring models with anomalous U(1)
  symmetries}},  {\em JHEP} {\bf 0712} (2007) 092,
  [\href{http://xxx.lanl.gov/abs/0710.5105}{{\tt arXiv:0710.5105}}].

\bibitem{susylr}
J.~Esteves, J.~Romao, M.~Hirsch, A.~Vicente, W.~Porod, {\em et.~al.}, {\it {LHC
  and lepton flavour violation phenomenology of a left-right extension of the
  MSSM}},  {\em JHEP} {\bf 1012} (2010) 077,
  [\href{http://xxx.lanl.gov/abs/1011.0348}{{\tt arXiv:1011.0348}}].

\bibitem{Esteves:2011gk}
J.~Esteves, J.~Romao, M.~Hirsch, W.~Porod, F.~Staub, {\em et.~al.}, {\it {Dark
  matter and LHC phenomenology in a left-right supersymmetric model}},  {\em
  JHEP} {\bf 1201} (2012) 095, [\href{http://xxx.lanl.gov/abs/1109.6478}{{\tt
  arXiv:1109.6478}}].

\bibitem{Hisano:1995nq}
J.~Hisano, T.~Moroi, K.~Tobe, M.~Yamaguchi, and T.~Yanagida, {\it {Lepton
  Flavor Violation in the Supersymmetric Standard Model with Seesaw Induced
  Neutrino Masses}},  {\em Phys. Lett.} {\bf B357} (1995) 579--587,
  [\href{http://xxx.lanl.gov/abs/hep-ph/9501407}{{\tt hep-ph/9501407}}].

\bibitem{Masina:2002mv}
I.~Masina and C.~A. Savoy, {\it {Sleptonarium (Constraints on the CP and
  Flavour Pattern of Scalar Lepton Masses)}},  {\em Nucl. Phys.} {\bf B661}
  (2003) 365--393, [\href{http://xxx.lanl.gov/abs/hep-ph/0211283}{{\tt
  hep-ph/0211283}}].

\bibitem{Paradisi:2005fk}
P.~Paradisi, {\it {Constraints on SUSY Lepton Flavour Violation by Rare
  Processes}},  {\em JHEP} {\bf 10} (2005) 006,
  [\href{http://xxx.lanl.gov/abs/hep-ph/0505046}{{\tt hep-ph/0505046}}].

\bibitem{Hisano:2002iy}
J.~Hisano, R.~Kitano, and M.~M. Nojiri, {\it {Slepton Oscillation at Large
  Hadron Collider}},  {\em Phys. Rev.} {\bf D65} (2002) 116002,
  [\href{http://xxx.lanl.gov/abs/hep-ph/0202129}{{\tt hep-ph/0202129}}].

\bibitem{Hisano:2008ng}
J.~Hisano, M.~M. Nojiri, and W.~Sreethawong, {\it {Discriminating
  Electroweak-Ino Parameter Ordering at the Lhc and Its Impact on Lfv
  Studies}},  {\em JHEP} {\bf 06} (2009) 044,
  [\href{http://xxx.lanl.gov/abs/0812.4496}{{\tt arXiv:0812.4496}}].

\bibitem{Griest:1990kh}
K.~Griest and D.~Seckel, {\it {Three exceptions in the calculation of relic
  abundances}},  {\em Phys.Rev.} {\bf D43} (1991) 3191--3203.

\bibitem{Belanger:2010gh}
G.~Belanger, F.~Boudjema, P.~Brun, A.~Pukhov, S.~Rosier-Lees, {\em et.~al.},
  {\it {Indirect search for dark matter with micrOMEGAs2.4}},  {\em
  Comput.Phys.Commun.} {\bf 182} (2011) 842--856,
  [\href{http://xxx.lanl.gov/abs/1004.1092}{{\tt arXiv:1004.1092}}].

\bibitem{lucanpb}
M.~Ciuchini, A.~Masiero, P.~Paradisi, L.~Silvestrini, S.~Vempati, {\em
  et.~al.}, {\it {Soft SUSY breaking grand unification: Leptons versus quarks
  on the flavor playground}},  {\em Nucl.Phys.} {\bf B783} (2007) 112--142,
  [\href{http://xxx.lanl.gov/abs/hep-ph/0702144}{{\tt hep-ph/0702144}}].

\bibitem{Nakamura:2010zzi}
{\bf Particle Data Group} Collaboration, K.~Nakamura {\em et.~al.}, {\it
  {Review of Particle Physics}},  {\em J. Phys.} {\bf G37} (2010) 075021.

\bibitem{lhcbounds}
\url{https://twiki.cern.ch/twiki/bin/view/AtlasPublic},
  \url{https://twiki.cern.ch/twiki/bin/view/CMSPublic/PhysicsResultsHIG}.

\bibitem{Ellis:2002iu}
J.~R. Ellis, T.~Falk, K.~A. Olive, and Y.~Santoso, {\it {Exploration of the
  MSSM with Non-Universal Higgs Masses}},  {\em Nucl. Phys.} {\bf B652} (2003)
  259--347, [\href{http://xxx.lanl.gov/abs/hep-ph/0210205}{{\tt
  hep-ph/0210205}}].

\bibitem{nonunivHiggs}
H.~Baer, A.~Mustafayev, S.~Profumo, A.~Belyaev, and X.~Tata, {\it {Direct,
  indirect and collider detection of neutralino dark matter in SUSY models with
  non-universal Higgs masses}},  {\em JHEP} {\bf 0507} (2005) 065,
  [\href{http://xxx.lanl.gov/abs/hep-ph/0504001}{{\tt hep-ph/0504001}}].

\bibitem{Ellis:2008eu}
J.~R. Ellis, K.~A. Olive, and P.~Sandick, {\it {Varying the Universality of
  Supersymmetry-Breaking Contributions to MSSM Higgs Boson Masses}},  {\em
  Phys.Rev.} {\bf D78} (2008) 075012,
  [\href{http://xxx.lanl.gov/abs/0805.2343}{{\tt arXiv:0805.2343}}].

\bibitem{Ellis:2007by}
J.~R. Ellis, S.~King, and J.~Roberts, {\it {The Fine-Tuning Price of Neutralino
  Dark Matter in Models with Non-Universal Higgs Masses}},  {\em JHEP} {\bf
  0804} (2008) 099, [\href{http://xxx.lanl.gov/abs/0711.2741}{{\tt
  arXiv:0711.2741}}].

\bibitem{Roszkowski:2009sm}
L.~Roszkowski, R.~Ruiz~de Austri, R.~Trotta, Y.-L.~S. Tsai, and T.~A. Varley,
  {\it {Global fits of the Non-Universal Higgs Model}},  {\em Phys.Rev.} {\bf
  D83} (2011) 015014, [\href{http://xxx.lanl.gov/abs/0903.1279}{{\tt
  arXiv:0903.1279}}].

\bibitem{Das:2010kb}
D.~Das, A.~Goudelis, and Y.~Mambrini, {\it {Exploring SUSY light Higgs boson
  scenarios via dark matter experiments}},  {\em JCAP} {\bf 1012} (2010) 018,
  [\href{http://xxx.lanl.gov/abs/1007.4812}{{\tt arXiv:1007.4812}}].

\bibitem{upcoming}
D.~Chowdhury and S.~K. Vempati {\em [In Preparation]}.

\bibitem{Hinchliffe:2000np}
I.~Hinchliffe and F.~Paige, {\it {Lepton flavor violation at the CERN LHC}},
  {\em Phys.Rev.} {\bf D63} (2001) 115006,
  [\href{http://xxx.lanl.gov/abs/hep-ph/0010086}{{\tt hep-ph/0010086}}].

\bibitem{Allanach:2008ib}
B.~Allanach, J.~Conlon, and C.~Lester, {\it {Measuring Smuon-Selectron Mass
  Splitting at the CERN LHC and Patterns of Supersymmetry Breaking}},  {\em
  Phys.Rev.} {\bf D77} (2008) 076006,
  [\href{http://xxx.lanl.gov/abs/0801.3666}{{\tt arXiv:0801.3666}}].

\bibitem{calibbi2}
A.~J. Buras, L.~Calibbi, and P.~Paradisi, {\it {Slepton mass-splittings as a
  signal of LFV at the LHC}},  {\em JHEP} {\bf 1006} (2010) 042,
  [\href{http://xxx.lanl.gov/abs/0912.1309}{{\tt arXiv:0912.1309}}].

\bibitem{bartl}
A.~Bartl, K.~Hidaka, K.~Hohenwarter-Sodek, T.~Kernreiter, W.~Majerotto, {\em
  et.~al.}, {\it {Test of lepton flavor violation at LHC}},  {\em Eur.Phys.J.}
  {\bf C46} (2006) 783--789,
  [\href{http://xxx.lanl.gov/abs/hep-ph/0510074}{{\tt hep-ph/0510074}}].

\bibitem{Nojiri:1994it}
M.~M. Nojiri, {\it {Polarization of $\tau$ lepton from scalar $\tau$ decay as a
  probe of neutralino mixing}},  {\em Phys.Rev.} {\bf D51} (1995) 6281--6291,
  [\href{http://xxx.lanl.gov/abs/hep-ph/9412374}{{\tt hep-ph/9412374}}].

\bibitem{Nojiri:1996fp}
M.~M. Nojiri, K.~Fujii, and T.~Tsukamoto, {\it {Confronting the minimal
  supersymmetric standard model with the study of scalar leptons at future
  linear e+ e- colliders}},  {\em Phys.Rev.} {\bf D54} (1996) 6756--6776,
  [\href{http://xxx.lanl.gov/abs/hep-ph/9606370}{{\tt hep-ph/9606370}}].

\bibitem{Guchait:2002xh}
M.~Guchait and D.~Roy, {\it {Using $\tau$ polarization as a distinctive SUGRA
  signature at LHC}},  {\em Phys.Lett.} {\bf B541} (2002) 356--361,
  [\href{http://xxx.lanl.gov/abs/hep-ph/0205015}{{\tt hep-ph/0205015}}].

\bibitem{Hamaguchi:2004df}
K.~Hamaguchi, Y.~Kuno, T.~Nakaya, and M.~M. Nojiri, {\it {A Study of late
  decaying charged particles at future colliders}},  {\em Phys.Rev.} {\bf D70}
  (2004) 115007, [\href{http://xxx.lanl.gov/abs/hep-ph/0409248}{{\tt
  hep-ph/0409248}}].

\bibitem{Godbole:2008it}
R.~Godbole, M.~Guchait, and D.~Roy, {\it {Using Tau Polarization to probe the
  Stau Co-annihilation Region of mSUGRA Model at LHC}},  {\em Phys.Rev.} {\bf
  D79} (2009) 095015, [\href{http://xxx.lanl.gov/abs/0807.2390}{{\tt
  arXiv:0807.2390}}].

\bibitem{Brignole:2004ah}
A.~Brignole and A.~Rossi, {\it {Anatomy and phenomenology of mu-tau lepton
  flavor violation in the MSSM}},  {\em Nucl.Phys.} {\bf B701} (2004) 3--53,
  [\href{http://xxx.lanl.gov/abs/hep-ph/0404211}{{\tt hep-ph/0404211}}].

\bibitem{suseflav_docu}
D.~Chowdhury, R.~Garani, and S.~K. Vempati, {\it {SUSEFLAV: Program for
  supersymmetric mass spectra with seesaw mechanism and rare lepton flavor
  violating decays}},  \href{http://xxx.lanl.gov/abs/1109.3551}{{\tt
  arXiv:1109.3551}}.

\bibitem{Pierce:1996zz}
D.~M. Pierce, J.~A. Bagger, K.~T. Matchev, and R.-j. Zhang, {\it {Precision
  Corrections in the Minimal Supersymmetric Standard Model}},  {\em Nucl.
  Phys.} {\bf B491} (1997) 3--67,
  [\href{http://xxx.lanl.gov/abs/hep-ph/9606211}{{\tt hep-ph/9606211}}].

\bibitem{Heinemeyer:1999be}
S.~Heinemeyer, W.~Hollik, and G.~Weiglein, {\it {The Mass of the Lightest MSSM
  Higgs Boson: a Compact Analytical Expression at the Two-Loop Level}},  {\em
  Phys. Lett.} {\bf B455} (1999) 179--191,
  [\href{http://xxx.lanl.gov/abs/hep-ph/9903404}{{\tt hep-ph/9903404}}].

\bibitem{Pukhov:1999gg}
A.~Pukhov {\em et.~al.}, {\it {Comphep: a Package for Evaluation of Feynman
  Diagrams and Integration over Multi-Particle Phase Space. User's Manual for
  Version 33}},  \href{http://xxx.lanl.gov/abs/hep-ph/9908288}{{\tt
  hep-ph/9908288}}.

\bibitem{Barate:2003sz}
{\bf LEP Working Group for Higgs boson searches} Collaboration, R.~Barate {\em
  et.~al.}, {\it {Search for the Standard Model Higgs Boson at Lep}},  {\em
  Phys. Lett.} {\bf B565} (2003) 61--75,
  [\href{http://xxx.lanl.gov/abs/hep-ex/0306033}{{\tt hep-ex/0306033}}].

\bibitem{Frere:1983ag}
J.~M. Frere, D.~R.~T. Jones, and S.~Raby, {\it {Fermion Masses and Induction of
  the Weak Scale by Supergravity}},  {\em Nucl. Phys.} {\bf B222} (1983) 11.

\bibitem{AlvarezGaume:1983gj}
L.~Alvarez-Gaume, J.~Polchinski, and M.~B. Wise, {\it {Minimal Low-Energy
  Supergravity}},  {\em Nucl.Phys.} {\bf B221} (1983) 495. Revised version.

\bibitem{Claudson:1983et}
M.~Claudson, L.~J. Hall, and I.~Hinchliffe, {\it {Low-Energy Supergravity:
  False Vacua and Vacuous Predictions}},  {\em Nucl.Phys.} {\bf B228} (1983)
  501.

\bibitem{Nihei:2002sc}
T.~Nihei, L.~Roszkowski, and R.~Ruiz~de Austri, {\it {Exact cross-sections for
  the neutralino slepton coannihilation}},  {\em JHEP} {\bf 0207} (2002) 024,
  [\href{http://xxx.lanl.gov/abs/hep-ph/0206266}{{\tt hep-ph/0206266}}].

\end{thebibliography}\endgroup

\end{document}